%% file: main.tex
\documentclass[fleqn,usenatbib]{mnras}

\usepackage{newtxtext,newtxmath}

\usepackage[T1]{fontenc}

\DeclareRobustCommand{\VAN}[3]{#2}
\let\VANthebibliography\thebibliography
\def\thebibliography{\DeclareRobustCommand{\VAN}[3]{##3}\VANthebibliography}

\usepackage{booktabs}  
\usepackage{multirow}
\usepackage{subfig}
\usepackage{graphicx}	
\usepackage{amsmath}	
\usepackage{color}
\usepackage{acro} 
\usepackage{lscape}

\DeclareRobustCommand{\ion}[2]{\textup{#1\,\textsc{\lowercase{#2}}}}

\newcommand{\kms}{km~s\ensuremath{^{-1}}}
\newcommand{\ha}{H$\alpha$}
\newcommand{\hb}{H$\beta$}
\newcommand{\msun}{${\rm M}_{\odot}$}

\newcommand{\oiii}{[\ion{O}{iii}]}
\newcommand{\nii}{[\ion{N}{ii}]}
\newcommand{\sii}{[\ion{S}{ii}]}
\newcommand{\otheroiii}{[\ion{O}{iii}]$\lambda$4959}
\newcommand{\fulloiii}{[\ion{O}{iii}]$\lambda$5007}
\newcommand{\fullnii}{[\ion{N}{ii}]$\lambda$6584}
\newcommand{\fullsii}{[\ion{S}{ii}]$\lambda\lambda$6716,6731}

\newcommand{\lbol}{${L}_\mathrm{bol}$}
\newcommand{\npn}{${N}_\mathrm{PN}$}
\newcommand{\mpn}{${m}_{5007}$}
\newcommand{\hii}{\ion{H}{ii}}
\newcommand{\nbpt}{[\ion{N}{II}]-BPT}
\newcommand{\sbpt}{[\ion{S}{II}]-BPT}

\DeclareAcronym{CS}{short = CS, long = central star}
\DeclareAcronym{PN}{
    short = PN,
    short-plural = e,
    long = planetary nebula,
    long-plural = e,
}
\DeclareAcronym{PSF}{short = PSF, long = point spread function}
\DeclareAcronym{LSF}{short = LSF, long = line spread function}
\DeclareAcronym{FWHM}{short = FWHM, short-plural-form = FWHM, long = full width at half maximum, long-plural-form = full widths at maximum}
\DeclareAcronym{iFTS}{short = iFTS, long = imaging Fourier transform spectrometer}
\DeclareAcronym{FOV}{short = FOV, short-plural-form = FOV, long = field of view, long-plural-form = fields of view}
\DeclareAcronym{SNR}{short = SNR, long = supernova remnant}
\DeclareAcronym{S/N}{short = S/N, long = signal to noise ratio}
\DeclareAcronym{AGB}{short = AGB, long = asymptotic giant branch}
\DeclareAcronym{CMD}{short = CMD, long = colour magnitude diagram}
\DeclareAcronym{LOS}{short = LOS, short-plural-form = LOS, long = line of sight, long-plural-form = lines of sight}
\DeclareAcronym{PNS}{short = PN.S, long = Planetary Nebulae Spectrograph}
\DeclareAcronym{CDI}{short = CDI, long = counter-dispersed imaging}
\DeclareAcronym{ETG}{short = ETG, long = early-type galaxy, long-plural-form = early-type galaxies}
\DeclareAcronym{LTG}{short = LTG, long = late-type galaxy, long-plural-form = late-type galaxies}
\DeclareAcronym{IFS}{short = IFS, long = integral-field spectrograph}
\DeclareAcronym{PNLF}{short = PNLF, long = planetary nebula luminosity function}
\DeclareAcronym{SIGNALS}{short = SIGNALS, long = {Star formation, Ionized Gas, and Nebular Abundances Legacy Survey}, first-style = long}
\DeclareAcronym{CFHT}{short = CFHT, long = Canada–France–Hawaii Telescope}
\DeclareAcronym{ILS}{short = ILS, long = instrumental line shape}
\DeclareAcronym{HST}{short = {\it HST}, long = {\it Hubble Space Telescope}}
\DeclareAcronym{WFC3}{short = WFC3, long = Wide Field Camera 3}
\DeclareAcronym{ACS}{short = ACS, long = Advanced Camera for Surveys}
\DeclareAcronym{WFC}{short = WFC, long = Wide Field Channel}
\DeclareAcronym{TRGB}{short = TRGB, long = tip of the red giant branch}
\DeclareAcronym{LG}{short = LG, long = local group}

\title[SIGNALS PNe]{An automated method for planetary nebula detection with SIGNALS: first applications to NGC~4214 and NGC~4449}

\author[N.\ Yang et al.]{Nancy Yang,$^{1}$\thanks{E-mail: nancy.yang@physics.ox.ac.uk}
Johanna Hartke,$^{2,3,4}$
Martin Bureau,$^1$
Chiara Spiniello,$^{1,5}$
Louis-Simon Guit\'e,$^{6,7}$
\newauthor 
Guy Flint,$^1$
Magda Arnaboldi,$^5$
Ana In\'es Ennis,$^{8,9}$
R.\ Pierre Martin,$^{10}$
Thomas Martin,$^{11,12}$
\newauthor 
Carmelle Robert,$^{11}$ 
Laurie Rousseau-Nepton,$^{13,14}$
Lucas M.\ Valenzuela$^{15}$
and S\'ebastien Vicens-Mouret$^{11}$\\
$^1$Sub-department of Astrophysics, Department of Physics, University of Oxford, Denys Wilkinson Building, Keble Road, Oxford, OX1 3RH, United Kingdom\\
$^2$Finnish Centre for Astronomy with ESO (FINCA), University of Turku, FI-20014 Turun yliopisto, Finland\\
$^3$Tuorla Observatory, Department of Physics and Astronomy, University of Turku, FI-20014 Turun yliopisto, Finland\\
$^4$Turku Collegium for Science, Medicine and Technology (TCSMT), University of Turku, FI-20014 Turun yliopisto, Finland\\
$^5$European Southern Observatory, Karl-Schwarzschild-Straße 2, 85748 Garching bei München, Germany\\
$^6$Université Paris-Saclay, Université Paris Cité, CEA, CNRS, AIM, 91191, Gif-sur-Yvette, France\\
$^7$Département de Physique, Université de Montréal, Montréal, QC, H3C 3J7, Canada\\
$^8$Perimeter Institute for Theoretical Physics, Waterloo, Ontario N2L 2Y5, Canada\\
$^9$Waterloo Centre for Astrophysics, University of Waterloo, Waterloo, Ontario, N2L 3G1, Canada\\
$^{10}$Department of Physics \& Astronomy, University of Hawaii at Hilo, Hilo, 96720, USA\\
$^{11}$Département de Physique, de génie physique et d’optique, Université Laval, G1V 0A6, Québec, Canada\\
$^{12}$C\'egep Garneau, 1660 Boulevard de l’Entente, Qu\'ebec, QC G1S 4S3, Canada\\
$^{13}$David A.\ Dunlap Department of Astronomy and Astrophysics, University of Toronto, 50 St-George Street, Toronto, Ontario M5S 3H4, Canada\\
$^{14}$Dunlap Institute for Astronomy and Astrophysics, 50 St-George Street, Toronto, Ontario M5S 3H4, Canada\\
$^{15}$Universitäts-Sternwarte, Fakultät für Physik, Ludwig-Maximilians-Universität München, Scheinerstr.\ 1, 81679 München, Germany}

\date{Accepted XXX. Received YYY; in original form ZZZ}

\pubyear{2025}

\begin{document}
\label{firstpage}
\pagerange{\pageref{firstpage}--\pageref{lastpage}}
\maketitle

\begin{abstract}
Utilising the optical imaging Fourier transform spectrograph SITELLE, the Star-formation, Ionized Gas and Nebular Abundances Legacy Survey (SIGNALS) is designed to study the connection between star-forming regions and their environments. Targeting $31$ local star-forming galaxies, its data products also lend themselves to planetary nebula (PN) surveys. We present here a new pipeline to find PNe using automated emission-line diagnostics and morphology tests, that is able to distinguish PNe from contaminants with an accuracy similar to that of past visual methods. We also perform thorough completeness tests using mock PNe inserted into the data cubes with full spectra. We apply these tools to a pilot sample of two dwarf irregular galaxies from the SIGNALS survey, NGC~4214 and NGC~4449, with other galaxies to follow. For these two galaxies, we identify $25$ PNe (including $6$ new discoveries) and $23$ PNe (including $13$ new discoveries), respectively, and calculate PN luminosity function distances of $3.09^{+0.25}_{-0.46}$ and $3.91^{+0.33}_{-0.52}$~Mpc, respectively, the latter consistent with previous estimates. We also calculate the bolometric PN specific frequency of our galaxies ($\alpha_\mathrm{bol}$), as well as a newly defined $V$-band PN specific frequency ($\alpha_\mathrm{V}$) based solely on the galaxies' total luminosities in that band. 
\end{abstract}




\section{Introduction}
\label{sec:pne_intro}

\Acp{PN} are the progeny of low- to intermediate-mass ($\approx1$ -- $8$~\msun) stars at the end of their \ac{AGB} phases \citep[e.g.][]{kwok_PNe_origins_1978}. \Acp{PN} have varying lifetimes of $\sim10,000$~yr before their ionised shells dissipate. They are almost always brightest in the \fulloiii\ emission line, although exceptions exist, such as very low-excitation \acp{PN} with cool central stars that emit more in the \ha\ line \citep{frew_parker_PNe_2010}. The exact strength of the \fulloiii\ line depends on the metallicity, initial mass, and temperature of each star, but a high-metallicity \ac{PN} can emit up to $15\%$ of its \ac{CS}'s luminosity in the \fulloiii\ line alone \citep{dopita_15percent_1992}. The bright \fulloiii\ emission allows \acp{PN} to be detected out to distances of $\approx100$~Mpc \citep{gerhard_coma_2005}. \Acp{PN} can also be detected in galaxy haloes, where the stellar continua are faint. They are therefore popular tools to trace the kinematics of galaxies to large galactocentric radii \citep[e.g.][]{arnaboldi_ngc1316_1998, spiniello_pne_2018, hartke_halo_2020, bhattacharya_kinematics_2023}. 

\Acp{PN} also serve as a secondary distance indicator, through the \ac{PNLF}. \citet{ciardullo_PNLF_1989} showed that there is a lower limit to the absolute magnitudes of observed \acp{PN} (i.e.\ there is a maximum \ac{PN} brightness) which is nearly universal across galaxies, regardless of morphological type. By fitting a measured \ac{PNLF}, this lower limit can be determined and a distance modulus inferred. First described in \citet{ciardullo_PNLF_1989} and adapted for a varying faint-end \ac{PNLF} slope by \citet{longobardi_planetary_2013}, the \ac{PNLF} is described by
\begin{equation} \label{eq:pnlf}
    N(M_{5007})=c_1\,e^{c_2\,M_{5007}}\left(1-e^{3(M^*_{5007}-M_{5007})}\right)\,,
\end{equation}
where $M_{5007}$ is the absolute magnitude of the \fulloiii\ line, $c_1$ is a normalisation constant, $c_2$ is the faint-end slope index and $M^*_{5007}$ is the absolute magnitude of the bright-end cut-off. The \mpn\ (apparent) magnitude approximates the $V$-band (apparent) magnitude one would observe if the total \fulloiii\ line emission was distributed over the $V$ band, and can be obtained through the integrated \fulloiii\ flux, ${F}_{5007}$, via the Jacoby relation \citep{jacoby_m5007_1989}:
\begin{equation}
    {m}_{5007}=-2.5\,\log\left(\frac{{F}_{5007}}{\mathrm{erg~s}^{-1}~\mathrm{cm}^{-2}}\right)-13.74\,.
\end{equation}

\Ac{PNLF} distances can be accurately derived (uncertainties $<10\%$) for distances of up to $\approx20$~Mpc \citep{ciardullo_pnlf_2012, jacoby_towards_2024} and have been found to be of comparable precision and accuracy to those derived from Cepheids and \ac{TRGB} methods (see \citealt{congiu_sculptor_2025} for a recent comparison). 

While the \ac{PNLF} is effective as a distance indicator, the physics behind the universality of the bright-end cut-off remains unclear. Naively, older stellar populations should have fainter \acp{PN}, as older stars exit the main sequence at lower masses and lower progenitor masses produce fainter \acp{PN} \citep{marigo_evolution_2004}. Yet, observations have shown that the bright-end cut-off is always at the same absolute magnitude, so there must be some mechanism(s) for old stellar populations to produce bright \acp{PN} \citep{jacoby_m5007_1989}. Recent simulations suggest that these bright \acp{PN} may form within older stellar populations of high metallicity, as a result of metal-rich stars \citep{valenzuela_picsI_2025}. 
Because metal-rich stars evolve more slowly than metal-poor stars of the same mass, they form \acp{PN} later and with a higher initial mass, leading to brighter \acp{PN}.
Metallicity is also predicted to affect the absolute magnitude cut-off, such that $M^*_{5007}$ increases with decreasing metallicities \citep{ciardullo_pnlf_2010}. In turn, one could constrain the metallicity of a galaxy by assuming a distance and fitting for $M^*_{5007}$. 

Another way to investigate stellar populations with \acp{PN} is to use the bolometric specific frequency of \acp{PN}, that is the number of \acp{PN} per unit bolometric luminosity, defined as
\begin{equation}\label{eq:alpha}
    \alpha\equiv\frac{N_\mathrm{PN}}{L_\mathrm{bol}}\,,
\end{equation}
where \npn\ is the total number of \acp{PN} in a galaxy and \lbol\ is the galaxy's total bolometric luminosity.

In practice, $N_{\mathrm{PN}}$ can only be determined over a limited range of magnitudes, down from the bright cut-off. A more practical definition is thus given by
\begin{equation}\label{eq:alpha_int}
    \alpha_{\Delta m} = \frac{1}{L_{\mathrm{bol}}} \int^{m^*_{5007}+\Delta m}_{m^*_{5007}} N(m)\mathrm{d}m\,,
\end{equation}
where $N(m)$ is the number of \acp{PN} between (apparent) magnitudes $m$ and $m+\mathrm{d}m$ and $\Delta m=2.5$ is commonly used in the literature.

This specific frequency $\alpha$ has long been shown observationally to correlate with galaxy colour \citep{peimbert_alpha_1990, ciardullo_alpha_1991, ciardullo_binaries_2005, buzzoni_planetary_2006}, such that redder galaxies produce fewer \acp{PN} (per unit luminosity). However, stellar population modelling produces the opposite prediction, i.e.\ redder galaxies should have more \acp{PN} \citep{buzzoni_planetary_2006}.

The specific frequency can also vary as a function of galactocentric radius within a galaxy \citep{hui_5128_1993, longobardi_planetary_2013, hartke_m49_2017, bhattacharya_survey_2019}. 

Traditionally, \ac{PN} candidates have been identified using the on-band/off-band method, for which follow-up spectroscopy is required to confirm the \ac{PN} classifications and measure their velocities. Slitless spectroscopy instruments like the Planetary Nebula Spectrograph (PN.S; \citealt{douglas_PNS_2002}) offer a more efficient solution, providing \ac{PN} diagnostics and accurate velocity measurements from a single observation \citep[e.g.][]{coccato_pns_2009, pulsoni_epns_2023}.

A further improvement is offered by \acp{IFS}: \acp{PN} no longer require dedicated surveys but can be found in \ac{IFS} data cubes as by-products. Indeed, since the first \ac{PN} study using \ac{IFS} \citep{sarzi_sauron_2011}, there have been many more (e.g.\ \citealt{kreckel_PNLF_NGC628_2017, scheuermann_spirals_2022}; but see also the reviews by \citealt{roth_proceedings_2025} and \citealt{hartke_review_2025}).


Nevertheless, the effectiveness of most \acp{IFS} is limited by their generally small \acp{FOV}, often necessitating numerous pointings for single targets and thus large mosaics \citep{congiu_sculptor_2025}. SITELLE \citep{drissen_sitelle_2019} is an \ac{iFTS} with a very large \ac{FOV} ($11\arcmin\times11\arcmin$) mounted at the \ac{CFHT}. Its spectral coverage and \ac{FOV} lend themselves perfectly to \ac{PN} studies. In particular, the wide \ac{FOV} allows to study \acp{PN} in the outskirts of galaxies, where they stand out easily from the faint stellar continua.

The \ac{SIGNALS} (\ac{SIGNALS}; \citealt{rousseau-nepton_signals_2019}) targeted $31$ local (distances $\approx1$ -- $12$~Mpc) star-forming galaxies with SITELLE. Most past \ac{PN} surveys have focused on \acp{ETG}, for their lack of contaminating \oiii-sources \citep[e.g.][]{coccato_pns_2009, cortesi_S0_2013, pulsoni_survey_2018}.
A \ac{PN} survey of the \ac{SIGNALS} galaxies will allow to probe \ac{PN} population (and associated \ac{PNLF}) changes with e.g.\ galaxy morphology (from dwarf irregular galaxies to grand-design spirals), stellar mass and stellar metallicity. So far, the \ac{PN} populations of only M~31 \citep{martin_m31_2018} and NGC~4214 \citep{vicens-mouret_planetary_2023} have been catalogued using \ac{SIGNALS} data. To expand these studies to the entire survey, a more automated pipeline for \ac{PN} detection is required. This is the primary goal of this work. 

This paper thus sets out in detail the automated methods we developed to construct \ac{PN} catalogues of \ac{SIGNALS} galaxies. For illustrative purposes, we also applied those methods to the galaxies NGC~4214 and NGC~4449. We describe the data and calibration methods in Section~\ref{sec:data}. Section~\ref{sec:methods} covers \ac{PN} detection, the elimination of contaminants and the construction of mock \ac{PN} catalogues to estimate the completeness of observations. We present our results for NGC~4214 and NGC~4449 separately in Section~\ref{sec:results}, comparing their \ac{PNLF} distances and $\alpha$ parameters to those of previous works. We discuss our results in Section~\ref{sec:discussion} and summarise them in Section~\ref{sec:conclusions}. 

Throughout this paper, we refer to \fulloiii, \fullnii\ and \fullsii\ simply as \oiii, \nii\ and \sii.


\section{Data and calibrations}
\label{sec:data}
\subsection{SITELLE}
\label{sec:sitelle}

As mentioned above, SITELLE is an \ac{iFTS} with an $11\arcmin\times11\arcmin$ \ac{FOV} (sampled with $0\farcs32\times0\farcs32$ spaxels) mounted at the Cassegrain focus of the \ac{CFHT} \citep{drissen_sitelle_2019}. The design is that of an off-axis Michelson interferometer. A beam splitter directs the light onto two different mirrors: the fixed mirror and the scanning mirror. The two beams meet again at the beam splitter, where they interfere and form two new beams, each recorded with its own camera. The scanning mirror can be moved in small steps, changing the optical path difference between the two beams. 

An interferogram cube is obtained by a series of short exposures ($\approx1$~min each), in between which the scanning mirror moves by a few microns. The spectral resolution can be fine-tuned to any resolution in the range $1$ to $10,000$ by changing the number of steps of the scanning mirror. A Fourier transform then returns a data cube in which two dimensions are spatial and the third is spectral (wave number). This results in an instrumental \ac{LSF} that is a sinc as a function of wave number \citep{martin_sincgauss_2016} rather than the more typical Gaussian as a function of wavelength. Emission (and absorption) lines broadened by the Doppler effect have a Gaussian shape, which when observed with SITELLE is convolved into a "sincgauss" shape. 

Early on during the commissioning of SITELLE, the image quality was found to degrade towards the edges of the \ac{FOV}, where the \ac{PSF} becomes elongated. As mentioned in \citet{martin_m31_2018} and \citet{drissen_sitelle_2019}, the extent of this distortion changes from image to image, so it cannot be corrected. Fortunately, however, more recent data suffer less from this issue, thanks to an update to the SITELLE optics \citep{sitelle_optics_update}.

\subsection{Data selection} \label{sec:data_selection}

\ac{SIGNALS} \citep{rousseau-nepton_signals_2019} is a survey that uses SITELLE's strengths to quantify the impact of environment on star-forming regions, and is therefore optimised to study emission-line objects. \ac{SIGNALS} uses SITELLE's SN1 (wavelength range $365$ -- $385$~nm, spectral resolution $R=1000$), SN2 ($480$ -- $520$~nm, $R=1000$) and SN3 ($651$ -- $685$~nm, $R=5000$) filters. It combines large spectral coverage (allowing to observe important emission lines for \hii-region and \ac{PN} studies, such as \oiii, \hb, \ha\ and \nii) with high spectral resolution (allowing to measure precise \hii-region and \ac{PN} velocities).

The selection criteria of SIGNALS were driven by the need to observe a large number of \hii\ regions in a variety of galactic environments, resulting in the observations of $31$ star-forming galaxies at distances $D\leq12$~Mpc. While the survey is not optimised to reveal \acp{PN}, a previous study by \citet{vicens-mouret_planetary_2023} utilised \ac{SIGNALS} data to identify $15$ new \acp{PN} in the starburst galaxy NGC~4214. To both illustrate and benchmark our new \ac{PN} detection pipeline, we thus re-analyse the NGC~4214 \ac{SIGNALS} data here. Using \ac{HST} data, \citet{annibali_ngc4449_2017} discovered $28$ (potential) \acp{PN} in the central region of another similar starburst galaxy, NGC~4449, $5$ of which they obtained full spectra for. We thus also present the \ac{SIGNALS} data of NGC~4449 here, to add to this catalogue and benchmark our work. 

While $\approx75$\% of the stellar populations of NGC~4214 are over $\approx8$~Gyr old, it is currently undergoing a new burst of star formation \citep{williams_sfh-4214_2011}. NGC~4449 has been continuously forming stars over the lifetime of the galaxy, with a peak of star formation taking place $5$ -- $20$~Myr ago \citep{sacchi_sfh-ngc4449_2018}.

We obtained fully calibrated and reduced data cubes from the Canadian Astronomical Data Centre\footnote{https://www.cadc-ccda.hia-iha.nrc-cnrc.gc.ca/}. The SITELLE data reduction pipeline is as described in \citet{martin_orbs_2015}, and Table~\ref{tab:obs} presents an overview of the observations. We assume distances to our galaxies based on listed values from the NASA/IPAC Extragalactic Database, using only those derived from primary distance indicators. Where available, distances from the following methods were taken into account: Cepheids, \ac{CMD}, \ac{TRGB} and Carbon stars. We only use distances with reported uncertainties.

We note that while \ac{SIGNALS} uses the SN1 filter, only the data cubes obtained with the SN2 and SN3 filters are used in this work. We also note that while the NGC~4214 data were obtained with a lower spectral resolution in SN3 than the usual \ac{SIGNALS}' resolution, this does not hinder our ability to detect \acp{PN}. 

\input{tables/obs}

\subsection{Flux calibration}
\label{sec:flux_cal}

While the \ac{SIGNALS} data cube fluxes are calibrated using standard stars in the aforementioned pipeline, there is still some uncertainty in the zero points. However, we can use narrow-band \ac{HST} data to re-calibrate these zero points \citep{martin_m31_2018}. 

Because for \acp{PN} we care most about the \oiii\ line emission, we focus our calibration on SITELLE's SN2 filter. 

\begin{figure*}
    \centering
    \subfloat[NGC~4214]{%
        \includegraphics[width=.48\linewidth]{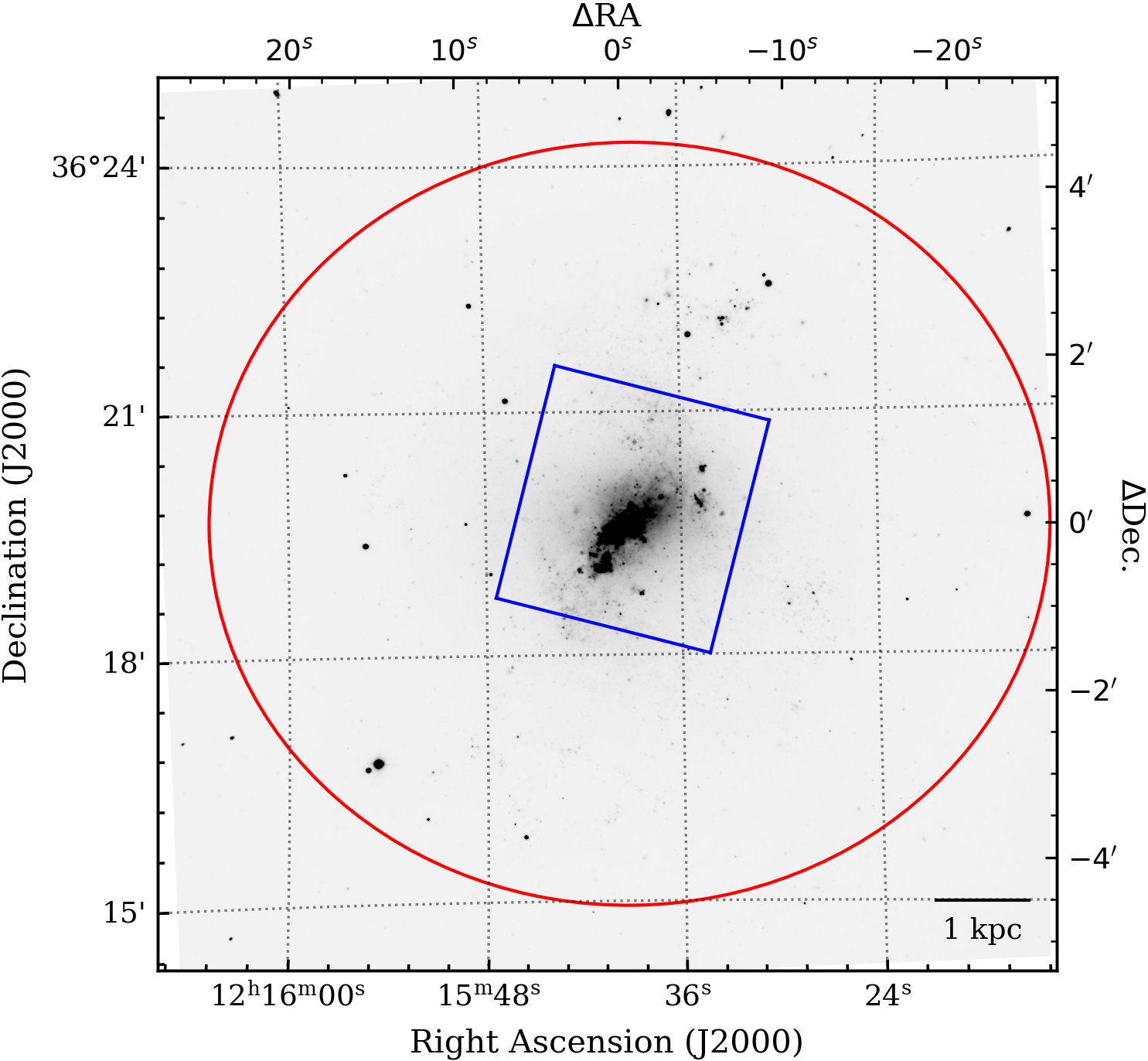}%
    } \hfill
    \subfloat[NGC~4449]{%
        \includegraphics[width=.48\linewidth]{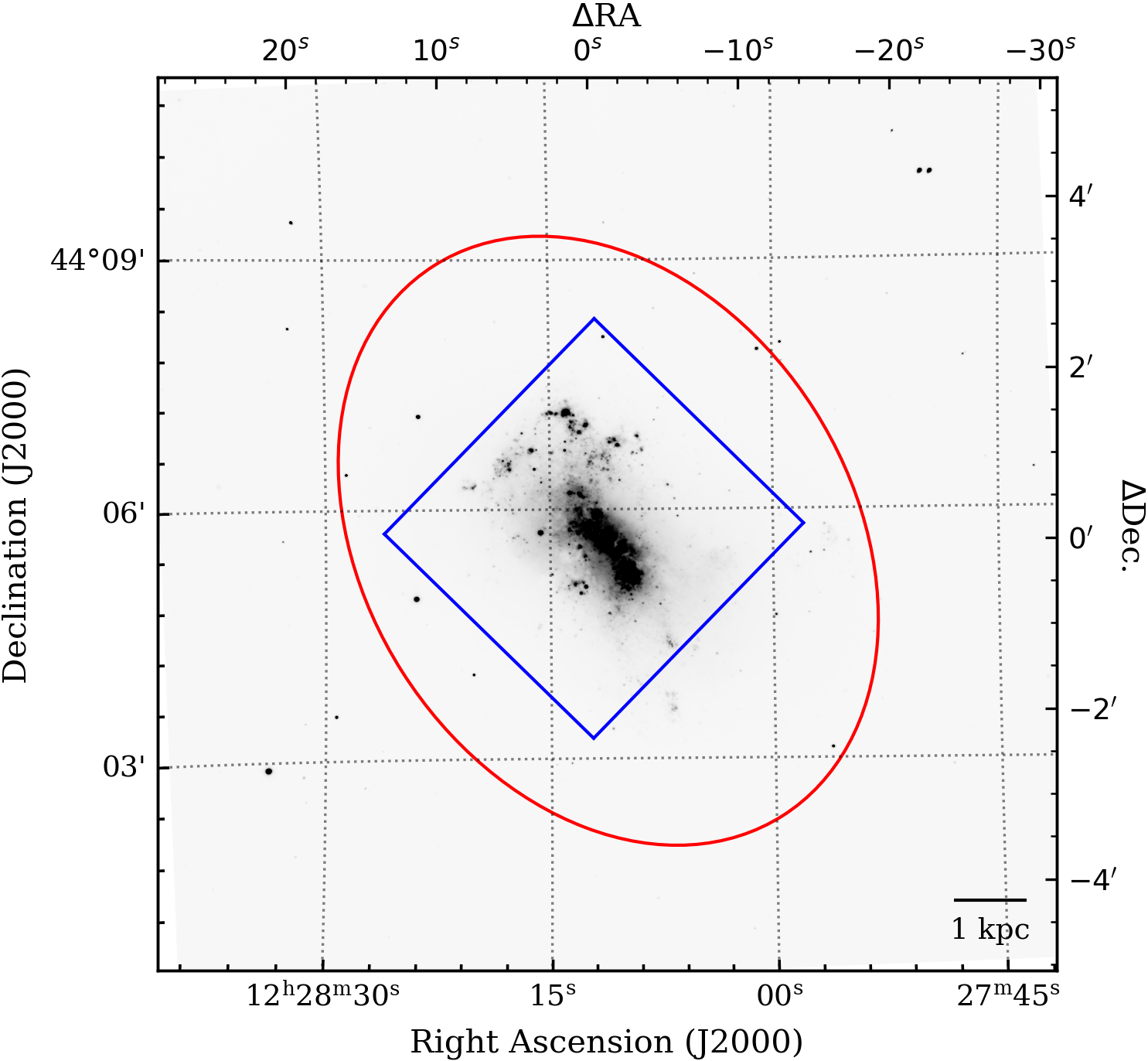}%
    }
    \caption{SITELLE SN2 deep images of NGC~4214 (left) and NGC~4449 (right). Blue rectangles are overlaid to indicate the \acp{FOV} of the corresponding \ac{HST} images, taken with WFC3 (left) and ACS (right). The red ellipses represent the apertures used by \citet{cook_spitzerphoto_2014}, whose total $B$ and $V$ magnitudes we adopt.}
    \label{fig:FOV}
\end{figure*}

We obtained fully calibrated and reduced \ac{HST} images taken in narrow-band filters around \oiii, which are fully encompassed by the SN2 filter, from the Barbara A.\ Mikulski Archive for Space Telescopes\footnote{https://mast.stsci.edu}. For NGC~4214, we use data taken with \ac{HST}'s \ac{WFC3} in the UVIS channel using the F502N filter (program ID: 11360, PI: R.\ O'Connell). These data were also used by \citet{dopita_ngc4214_2010} to identify \ac{PN} candidates. The \ac{WFC3} \ac{FOV} is $160\arcsec\times160\arcsec$, much smaller than SITELLE's \ac{FOV}, but yielding an overlap area sufficient for calibration (see Fig.~\ref{fig:FOV}). For NGC~4449, we use data taken with \ac{HST}'s \ac{ACS} \ac{WFC} in a similar F502N filter (program ID: 10522, PI: D.\ Calzetti). These data were also used in the \ac{PN} survey of \citet{annibali_ngc4449_2017}.

For each galaxy, we integrate the SN2 datacubes over the wave number range of the appropriate F502N filter, scaling each slice by the transmission at that wave number. This simulates an image taken with SITELLE in the F502N filter. We account for the different pixel sizes by re-binning the \ac{HST} images to match the SITELLE pixel size. Finally, we convolve the original \ac{HST} images with a Gaussian to match the seeing of the SITELLE data in the relevant SN2 cube.

\begin{figure*}
    \centering
    \subfloat[NGC~4214]{%
        \includegraphics[width=.45\linewidth]{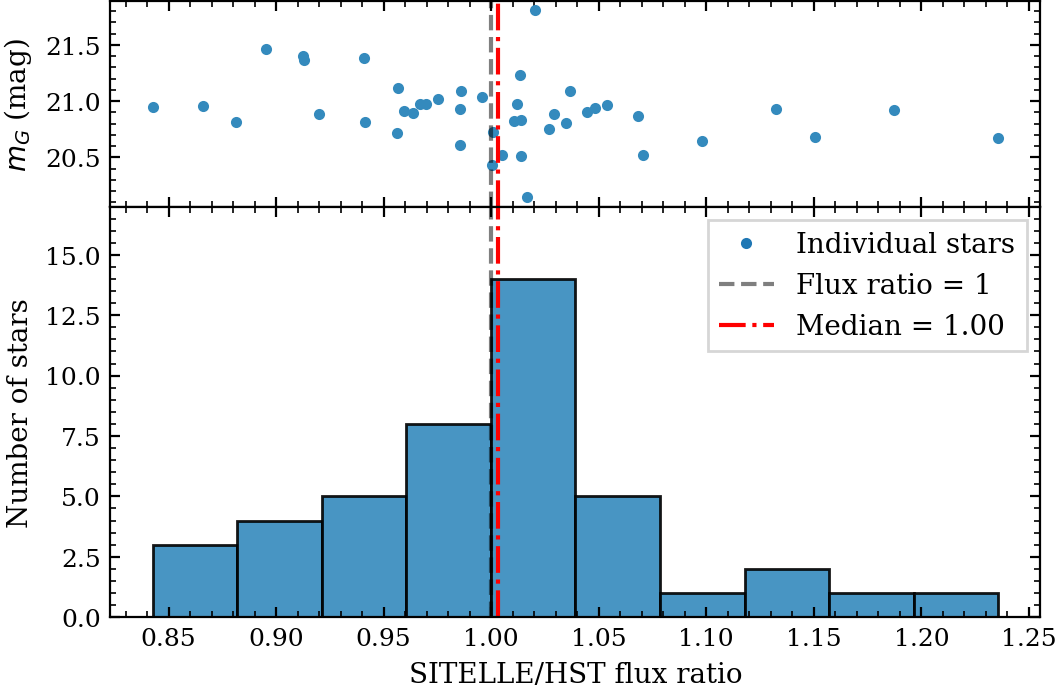}%
    } \hfill
    \subfloat[NGC~4449]{%
        \includegraphics[width=.44\linewidth]{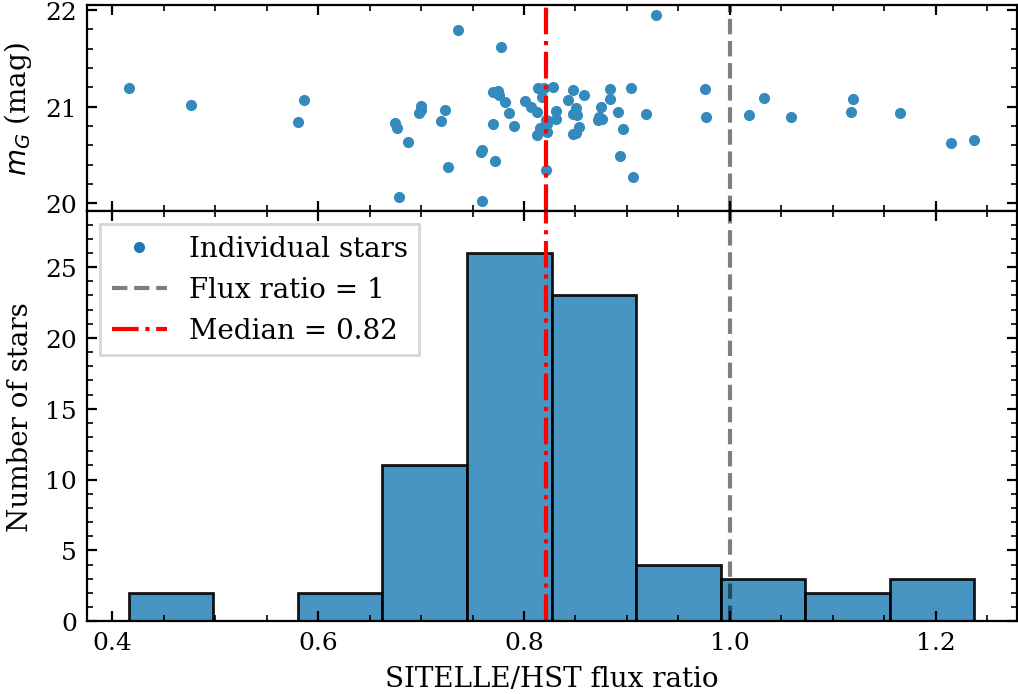}%
    }
    \caption{Flux correction factors of NGC~4214 (bottom-left) and NGC~4449 (bottom-right), determined by taking the flux ratios of stars measured using the SITELLE SN2 data cube (integrated over a narrowband \ac{HST} filter) and an \ac{HST} image. The sigma-clipped medians are indicated by red dot-dashed vertical lines, ratios of $1$ by grey dashed vertical lines. The apparent {\it GAIA} $G$-band magnitudes of the stars used are shown in the top parts of the plots.
    }
    \label{fig:HST}
\end{figure*}

We then perform aperture photometry on stars from the {\it GAIA} Data Release 3 catalogue \citep{gaia_dr3_2023} twice, once on the SITELLE slice and once on the re-binned \ac{HST} image. To avoid saturated stars, we select the stars to have apparent $G$-band magnitudes $m_G>20$, leaving $\approx 50$ stars in the FOV of each galaxy. We extract the fluxes using $8$-pixel apertures, subtracting the background flux measured within annuli of radii $10$ -- $12$~pixels, and compute the ratio of the \ac{HST} and SITELLE fluxes. We then define the correction factor to be the three sigma-clipped median of all the flux ratios. As shown in Fig.~\ref{fig:HST}, the \ac{HST} and SITELLE flux calibrations of NGC~4214 are in very good agreement, but while there is some scatter in the flux ratios of NGC~4449, there is also a clear offset between the two flux calibrations. We assume the uncertainties of our flux calibrations to be the sigma-clipped standard deviations of the flux ratios, and thus recover (re-)calibration factors of $1.00\pm0.08$ and $0.82\pm0.14$ from the \ac{HST} data of NGC~4214 and NGC~4449, respectively. These flux calibrations are optimised for the \oiii\ line, but we also apply them to other emission lines for consistency. This should not influence the line ratios used in our chosen diagnostic diagrams.

While the galaxies in this paper do have archival \ac{HST} data in the relevant filters, this is not the case for most SIGNALS sample galaxies. Since the survey data products were first released, a more widely applicable flux calibration method has been under development that uses {\it GAIA} spectra \citep{gaia_dr3_2023}. This method will be described in detail in an upcoming paper \citet[][in prep.]{vicens_flux_inpre}. For comparison, the new method using {\it GAIA} spectra yields (preliminary) calibration factors of $0.997\pm0.032$ in SN2 and $0.954\pm0.024$ in SN3 for NGC~4214, and $0.924\pm0.038$ in SN2 and $0.856\pm0.020$ in SN3 for NGC~4449, instead of the ones we derived using {\it HST} data. The {\it GAIA}-based flux calibration has the main advantage of being applicable to all \ac{SIGNALS} galaxies, while only a few galaxies have \ac{HST} data in the relevant filters. In this work we will use the established calibration method using the \ac{HST} data as described here. However, future catalogues will likely be published using the calibrated data from \citet[][in prep.]{vicens_flux_inpre}. 

We also correct our line fluxes for Milky Way extinction assuming $R_V=3.1$, $E_{B-V}$ values taken at the central coordinates of each galaxy from the \citet{schlegel_dustmap_1998} dust maps, and a \citet{fitzpatrick_extinctionlaw_1999} extinction law.

\subsection{Sky-line velocity map}
\label{sec:skyline_velcal}

All velocities quoted in this paper have had a barycentric velocity correction applied, based on the start time of the observations.

As described in detail in \citet{martin_m31_2018}, SITELLE's wavelength calibration relies on a high spectral resolution data cube of a laser source, taken with the telescope pointing at the zenith. However, due to temperature variations and the changing direction of the gravity vector, laser cubes obtained for different pointings show (absolute) velocity calibration errors of up to $25$~\kms\ \citep{flagey_eagle_2020}. In addition, the exact wavelength of the laser used for the calibration has some uncertainty, as it is dependent on the observing conditions. The data reduction pipeline assumes the manufacturer-stated wavelength of $543.5$~nm, but a $0.1$~nm error leads to a $55$~\kms\ offset \citep{martin_reduction_2021}. 

\begin{figure*}
    \centering
    \subfloat[NGC~4214]{%
        \includegraphics[width=.4\linewidth]{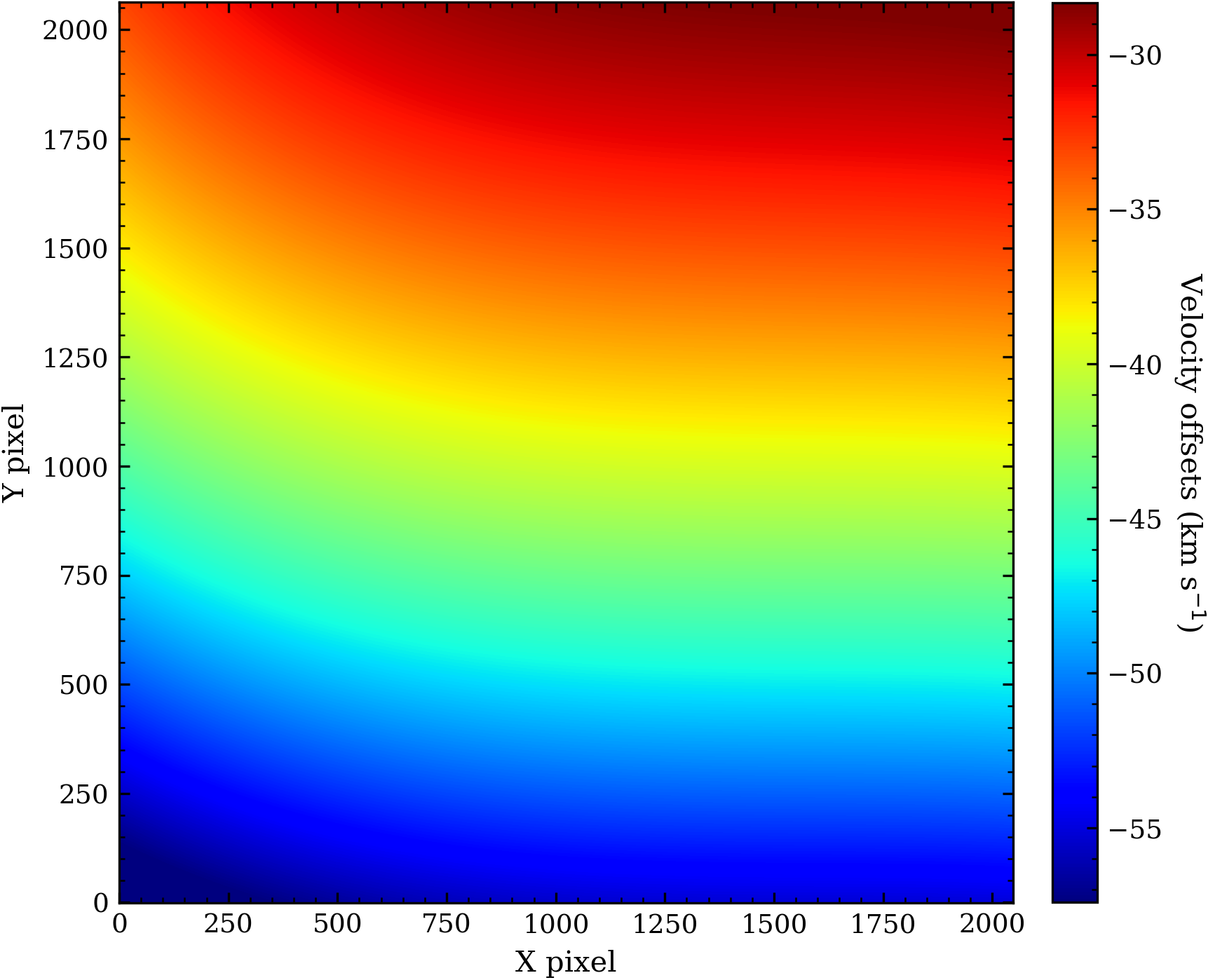}%
    } \hfill
    \subfloat[NGC~4449]{%
        \includegraphics[width=.4\linewidth]{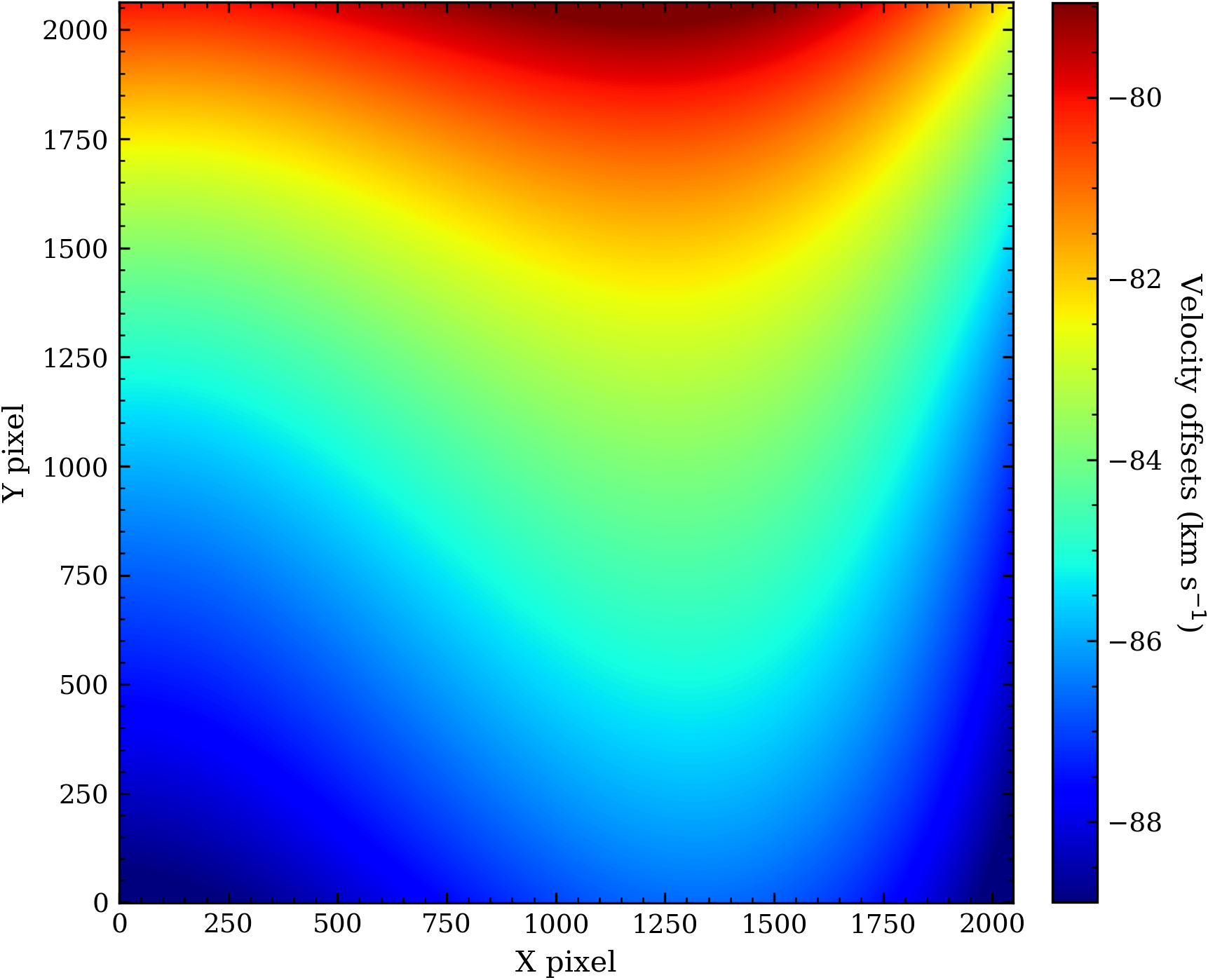}%
    }
    \caption{Fitted map of velocity offsets across the SITELLE \ac{FOV}, as measured from sky lines in the SN3 filter, for NGC~4214 (left) and NGC~4449 (right). The images cover the same FOV and have the same orientations as those in Fig.~\ref{fig:FOV}.}
    \label{fig:sky_vel}
\end{figure*}

To achieve a better accuracy, a fit to the Meinel OH sky lines can be performed in the SN3 filter. Indeed, these should have zero velocity, allowing to map the velocity offset across the SITELLE FOV. As the bright central regions of galaxies do not have clear sky lines, we can not constrain the velocity offsets there by direct fits to the spectra. Nevertheless, we can constrain the velocity offsets at the edges of the \ac{FOV} and fit a model to those measurements (as a function of position), thus allowing to predict the offset in the central regions. The details of the model and the initial parameters of the fit can be found in \citet{martin_m31_2018}, and they have been implemented in \texttt{Outils de R\'{e}duction de Cubes Spectraux} (\texttt{ORCS}; \citealt{martin_orbs_2015}). Here we sampled each cube on a $40\times40$ grid, masking the $5\%$ brightest pixels. The resulting sky velocity maps are presented in Fig.~\ref{fig:sky_vel}. The standard deviation of the residuals from the fit is $\approx4$~\kms\ in both cases, and these have been propagated into our final velocity uncertainties. 

The SN2 data cubes do not have any detectable sky line. For \acp{PN} which do not have a reliable emission line in SN3, an additional velocity calibration step for the SN2 velocities is presented in Section~\ref{sec:SN2_vel}.


\section{Planetary nebula pipeline} \label{sec:methods}

\begin{figure*}
    \centering
    \includegraphics[width=1.\linewidth]{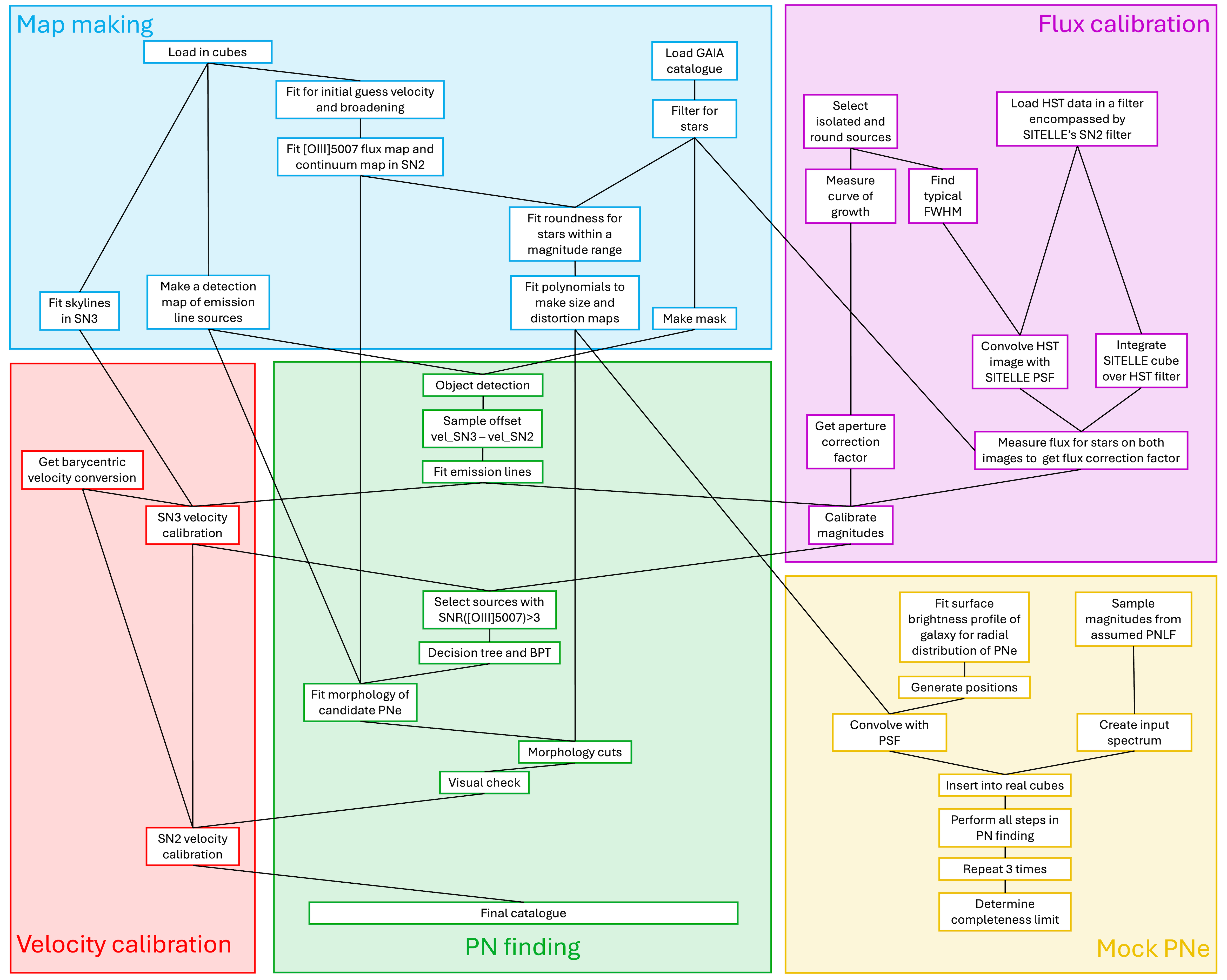}
    \caption{Schematic illustration of the \ac{SIGNALS} \ac{PN} identification pipeline.
    }
    \label{fig:flowchart}
\end{figure*}

The construction of our \ac{PN} catalogues can be broken down into two main steps: i) finding \ac{PN} candidates and ii) distinguishing \acp{PN} from contaminants. In this section, we describe these two steps in detail. We start with the creation of a two-dimensional (2D) image from which we detect emission-line sources. The emission lines are then fitted and diagnostic diagrams are used to weed out \hii\ regions and \acp{SNR}. Finally, we remove sources whose morphologies are inconsistent with a point-source. See Fig.~\ref{fig:flowchart} for an overview of our \ac{PN} detection pipeline,  including map-making and calibration steps.

\subsection{Source identification}
\label{sec:photutils}

To identify emission-line objects, we created a so-called detection map using an inbuilt function of \texttt{ORCS} \citep{martin_orbs_2015}, as used in \citet{martin_m31_2018}. The process is similar to median filtering. At each spaxel, the median spectrum of a $9\times9$ pixel$^2$ "background" spatial box (excluding the central $3\times3$ pixel$^2$) is first subtracted from the median spectrum of a $3\times3$ pixel$^2$ spatial box, both centred on the spaxel under consideration.
The highest flux at any wave number of this background-subtracted spectrum is then assigned to the spaxel. This procedure therefore reveals sources that are brighter than their surroundings in at least one (wave-number) frame. By applying this procedures to our SN2 data cubes, we highlight sources that are bright in either \fulloiii, \otheroiii\ or \hb. The box sizes quoted here are the default options, chosen to fit the typical seeing of SITELLE observations ($\approx1\arcsec$ or $\approx3$ pixels). While the sizes can be changed, they are restricted to odd numbers of pixels and therefore cannot be precisely fine-tuned. 

We search for objects in the SN2 detection map using the DAOFIND algorithm \citep{stetson_daophot_1987} as implemented in python by \citet{bradley_photutils_2022}. This finds local maxima with peak fluxes above a certain threshold flux. The routine \texttt{DAOStarFinder} also returns two different roundness measures for each detected source, but we opted not to use those in the following steps as both measures turned out to be correlated with magnitude. We therefore proceed with the source centroids only, and (re-)fit their roundnesses later. 

\begin{figure*}
    \centering
    \includegraphics[width=1\linewidth]{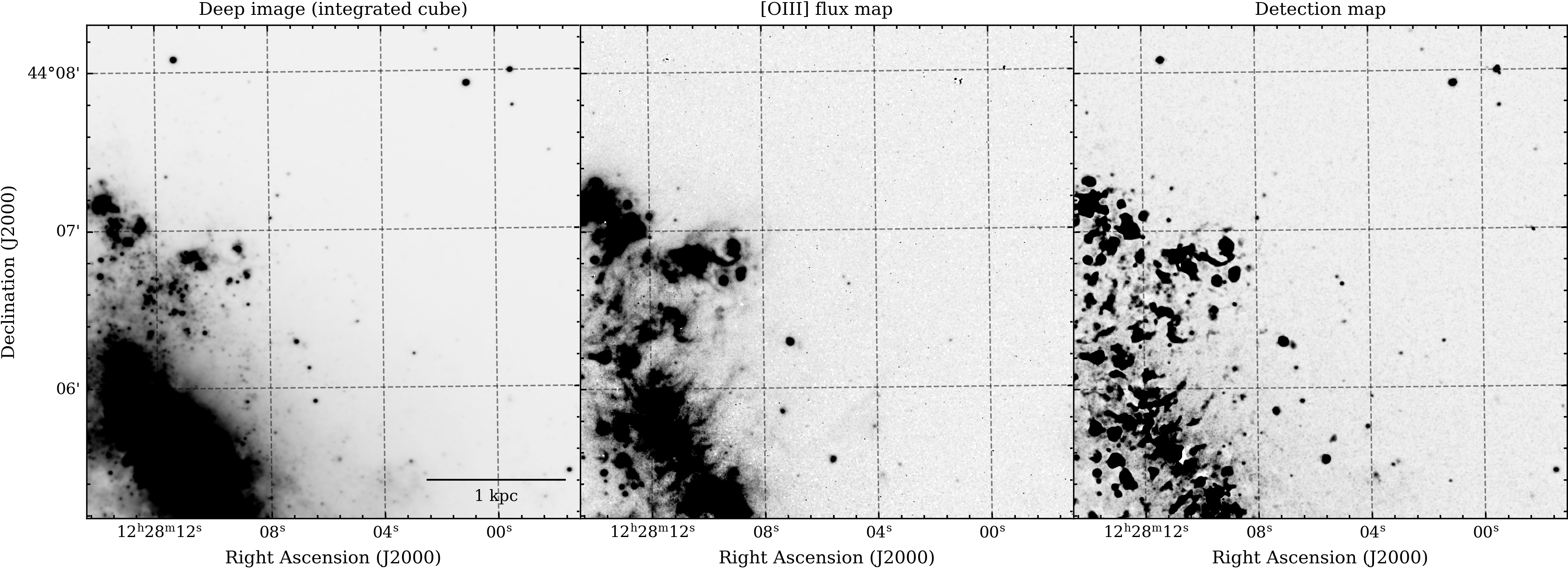}
    \caption{Inner section of the deep image (left), \oiii\ map (centre) and detection map (right) of NGC~4449. The single black pixels in the central panel show spaxels for which the emission-line fit failed. As expected, these are more common toward the edge of the map.}
    \label{fig:maps}
\end{figure*}

Running the algorithm on the detection maps yields $\approx5000$ sources for each galaxy. Most of these sources are associated with \hii\ regions, but using the detection map is nevertheless preferable to using a simple map of the \fulloiii\ emission, as \hii\ regions are very bright and extended in \oiii. The detection map is better at revealing point sources, which is extremely useful when searching for \acp{PN}; any source bright and compact in \oiii\ should also be prominent in the detection map. To illustrate this, we show a comparison of (part of) the deep image (constructed by integrating over all frames of the interference data cube before Fourier transforming), the \oiii\ map and the detection map of NGC~4449 in Fig.~\ref{fig:maps}.

\subsection{Source spectra and photometry}
\label{sec:fitting}

We use \texttt{ORCS} \citep{martin_orbs_2015} for emission-line fitting, which relies on the \texttt{curve\_fit} function of the \texttt{SciPy} library \citep{scipy}. For each emission-line source identified in the detection map, we extract a spectrum from a circular aperture of $3$ pixels in radius, from which we subtract the median spectrum from a background annulus of inner radius $8$ and outer radius $10$ pixels. The typical \ac{FWHM} is $\approx3$ pixels, but we purposefully keep the aperture small to minimise contamination from other sources \citep{soemitro_crowded_2023}. This is the same aperture used by \citet{vicens-mouret_planetary_2023} and increases the signal-to-noise ratio ($S/N$; \citealt{howel_photometry_1989}). The small aperture does introduce the need for an aperture correction, which we determine by constructing the curves of growth of round, isolated sources. As the \acp{PSF} of the SN2 and SN3 data cubes are different, we measure this correction separately for each data cube. For NGC~4214, the aperture correction factors used are $1.35\pm0.11$ for SN2 and $1.34\pm0.12$ for SN3. For NGC~4449, they are $1.35\pm0.19$ for SN2 and $1.40\pm0.20$ for SN3.

When fitting emission lines, \texttt{ORCS} has the option to fit the velocity and velocity dispersion of each line separately. We did not use this option, however, as we assume \ac{PN} emission lines are from a single source (superpositions are unlikely in these low-mass galaxies). This has the added benefit of yielding more robust velocity measurements, as all the lines can be used simultaneously to determine a single velocity and velocity dispersion. 

\begin{figure*}
    \centering
    \includegraphics[width=0.9\linewidth]{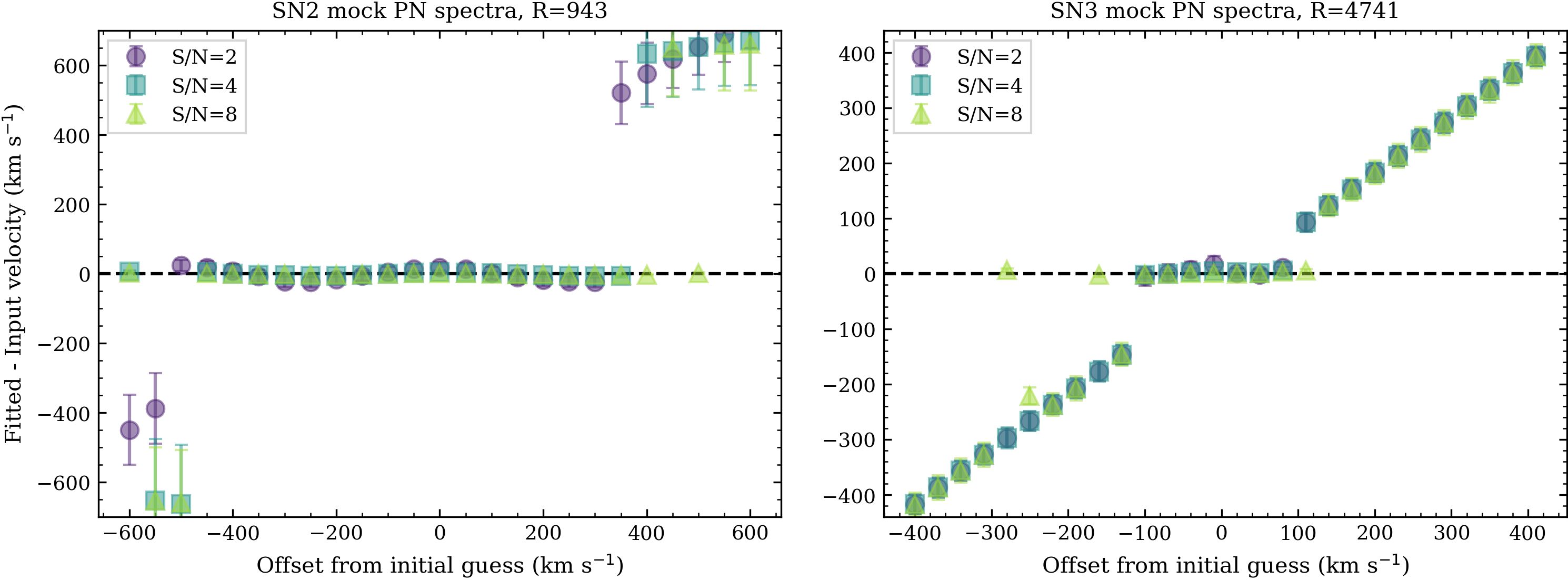}
    \caption{Fitting accuracy of mock spectra as a function of initial velocity guess and signal-to-noise ratio.}
    \label{fig:initial_vel}
\end{figure*}

Good initial guesses of the velocities of the sources are required for the fits to be accurate. 
To illustrate this, we generated mock PN spectra  with varying background noise. We set the spectral resolution to match that of the datacubes of NGC~4449, which is the SIGNALS standard. We then fitted these spectra with a range of initial velocity guesses, to test the limits within which the fits still converge to the correct velocity. We show the results in Fig.~\ref{fig:initial_vel}. For SN2, the fitting procedure yields the correct velocity as long as the initial guess is within $\approx300$~\kms\ of the truth, which is a wider range than the velocity dispersion of the galaxies studied. For SN3, the allowed range is smaller and an initial guess within $\approx100$~\kms\ of the truth is required. This is due to the SN3 cube having a higher spectral resolution, which is SIGNALS standard practice. In contrast to the strong dependence on spectral resolution, there is only a small dependence on signal-to-noise ratio. 

For the galaxies in this paper, we select a suitable initial velocity guess by fitting the emission lines of the $30$ brightest sources identified using \texttt{DAOStarFinder}. These fits are checked by eye to ensure a good fit to the appropriate lines and yield: (1) an initial guess for the velocity and (2) a typical offset of the fitted velocities $\Delta v=v_\mathrm{SN3}-v_\mathrm{SN2}$. The two galaxies in this paper do not have a large velocity gradient, so we assume that the \acp{PN} will also have velocities within $300$~\kms\ of the sampled velocity. Under this assumption, the initial velocity guess should allow lines in the SN2 filter to be fitted correctly for all our candidate \acp{PN}. The resulting $v_\mathrm{SN2}$ can then be used as the initial guess for the emission lines in the SN3 filter, by applying $\Delta v$. To account for the uncertainty in $v_\mathrm{SN2}$ and to allow for some variation of $\Delta v$, we also re-fit SN3 with initial velocities $100$~\kms\ above and below the initial guess of $\Delta v+v_\mathrm{SN2}$. From these, we select the best fit based on the reduced $\chi^2$. 
While for these galaxies we performed three SN3 fits per source, this number can be increased as necessary to allow for a greater range of velocities.

After fitting, we apply an \oiii\ luminosity cut, keeping only sources with a reliable \oiii\ detection, defined as an \oiii\ flux $S/N>3$. As \acp{PN} are generally brightest in the \oiii\ emission line, we would risk eliminating many \acp{PN} by requiring a similarly strong detection of any other emission line. We thus define other emission lines as reliably detected if they have flux $S/N>1$, and adopt $1\sigma$ flux upper limits for non-detections.

\subsection{Contaminant elimination}
\subsubsection{Emission-line ratio diagnostics}
\label{sec:bpt}

The next step is to distinguish \acp{PN} from other emission-line sources. The most common contaminants are \hii\ regions and \acp{SNR}, although background intermediate-redshift galaxies are also occasionally detected in \oiii. While \acp{PN} can have emission-line ratios very similar to those of some of these contaminants \citep{frew_parker_PNe_2010}, in most cases only they occupy a specific area in the top-left corner of a typical emission-line ratio diagnostic diagram (so-called BPT diagrams; \citealt{baldwin_bpt_1981}). 

We use the demarcation line of \citet{kauffmann_2003_bpt} on the \nbpt\ and \citet{kewley_bpt_2001} on the \sbpt\  to filter out \hii\ regions. The empirical \citet{sabin_bpt_2013} demarcation line is used on both diagrams to filter out \acp{SNR}. As many of our sources are undetected in one or more emission lines, we have allowed for and evaluated every combination of detections and non-detections in a decision tree, shown in Fig.~\ref{fig:tree}. 

For most candidates, both the \nbpt\ and \sbpt\ diagrams will be considered in some form. The use of two diagnostic diagrams means that we have two intermediate classifications that may not always match, and that need to be consolidated into one final conclusion. To avoid confusion, we only discuss the intermediate classifications in this section and consolidate the results in Section~\ref{sec:catalogue}. 

\begin{figure*}
    \centering
    \subfloat[NGC~4214]{%
        \includegraphics[width=0.9\linewidth]{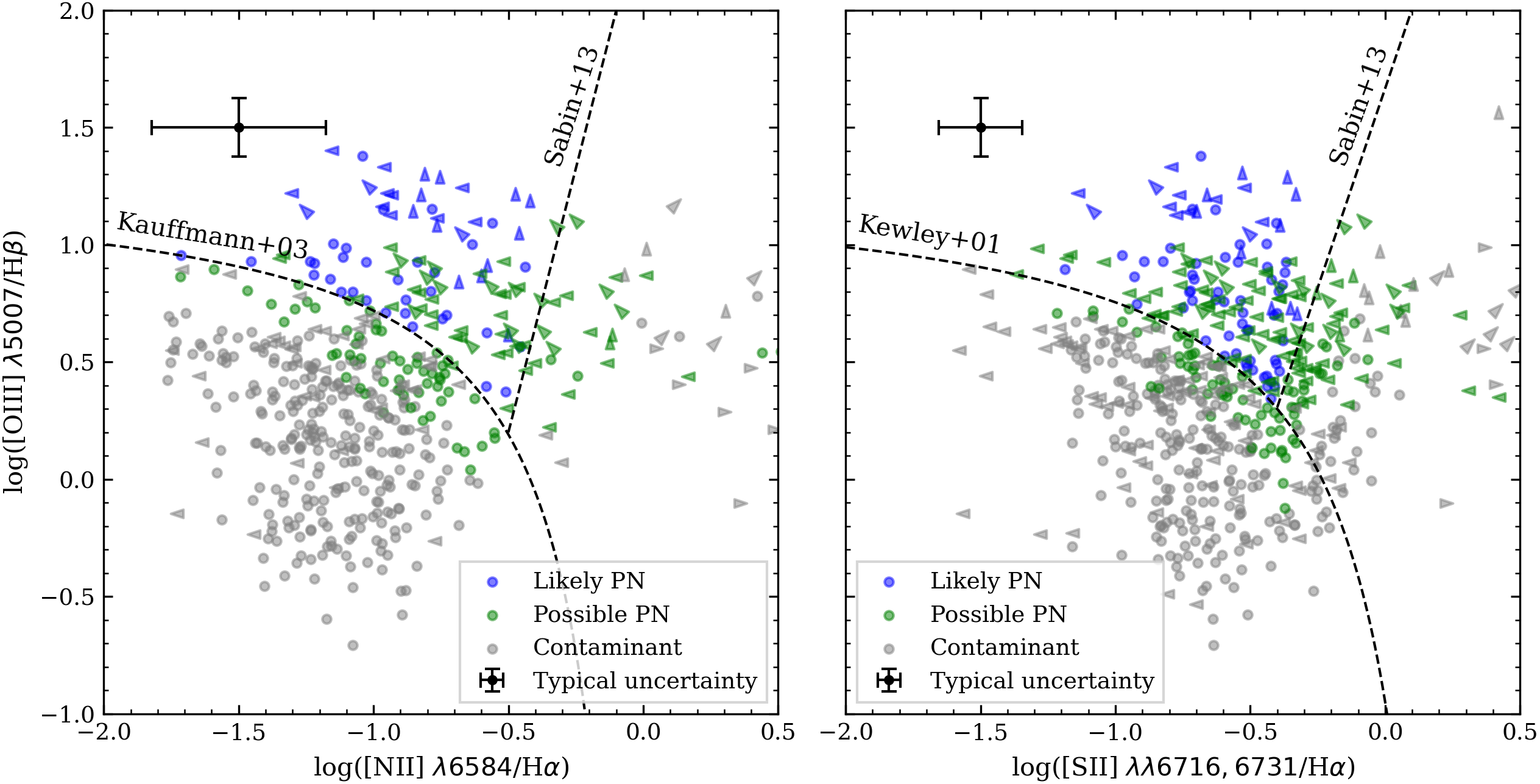}%
    }\\
    \subfloat[NGC~4449]{%
        \includegraphics[width=0.9\linewidth]{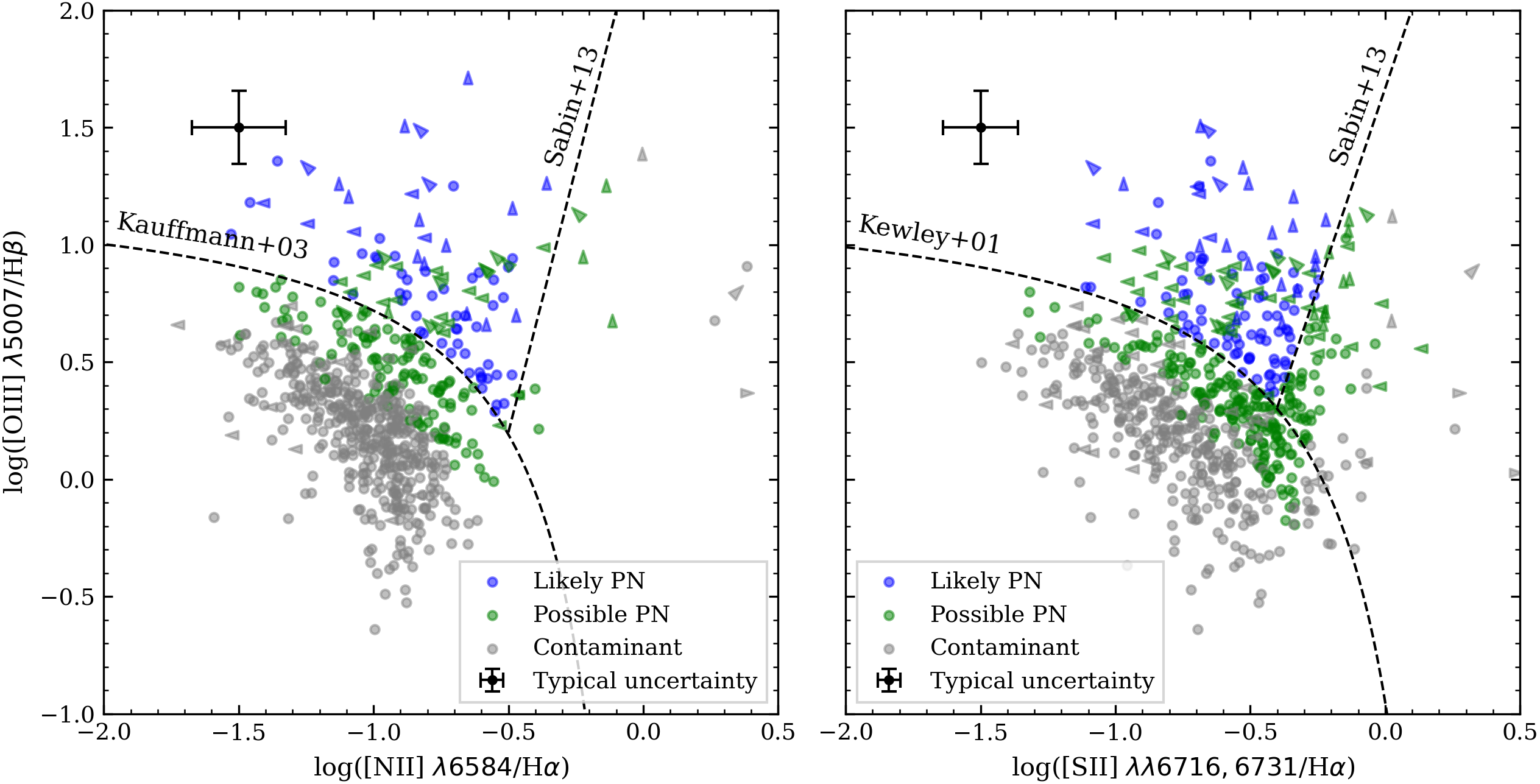}%
    }
    \caption{Classification of the emission-line sources of NGC~4214 (top) and NGC~4449 (bottom), based on the \nbpt\ diagram (left) and the \sbpt\ diagram (right). Triangular markers indicate limits and their directions, in cases of non-detection of at least one of the emission lines.
    }
    \label{fig:bpt}
\end{figure*}

Fig.~\ref{fig:bpt} shows the BPT diagrams for the emission-line sources of both galaxies, with colours indicating the intermediate classifications. Let us consider the case of a source with all emission lines detected. Sources located within the \ac{PN}-only region are considered likely \ac{PN} candidates. These are indicated by blue symbols in Fig.~\ref{fig:bpt}. If a source could be located within the \ac{PN}-only region when considering its emission-line ratio uncertainties, it is then considered a possible \ac{PN} candidate. These are indicated by green symbols in Fig.~\ref{fig:bpt}. A source not satisfying either criterion is rejected as not a \ac{PN} (i.e.\ a contaminant, plotted in grey in Fig.~\ref{fig:bpt}). 

\begin{figure}
    \centering
    \includegraphics[width=1\linewidth]{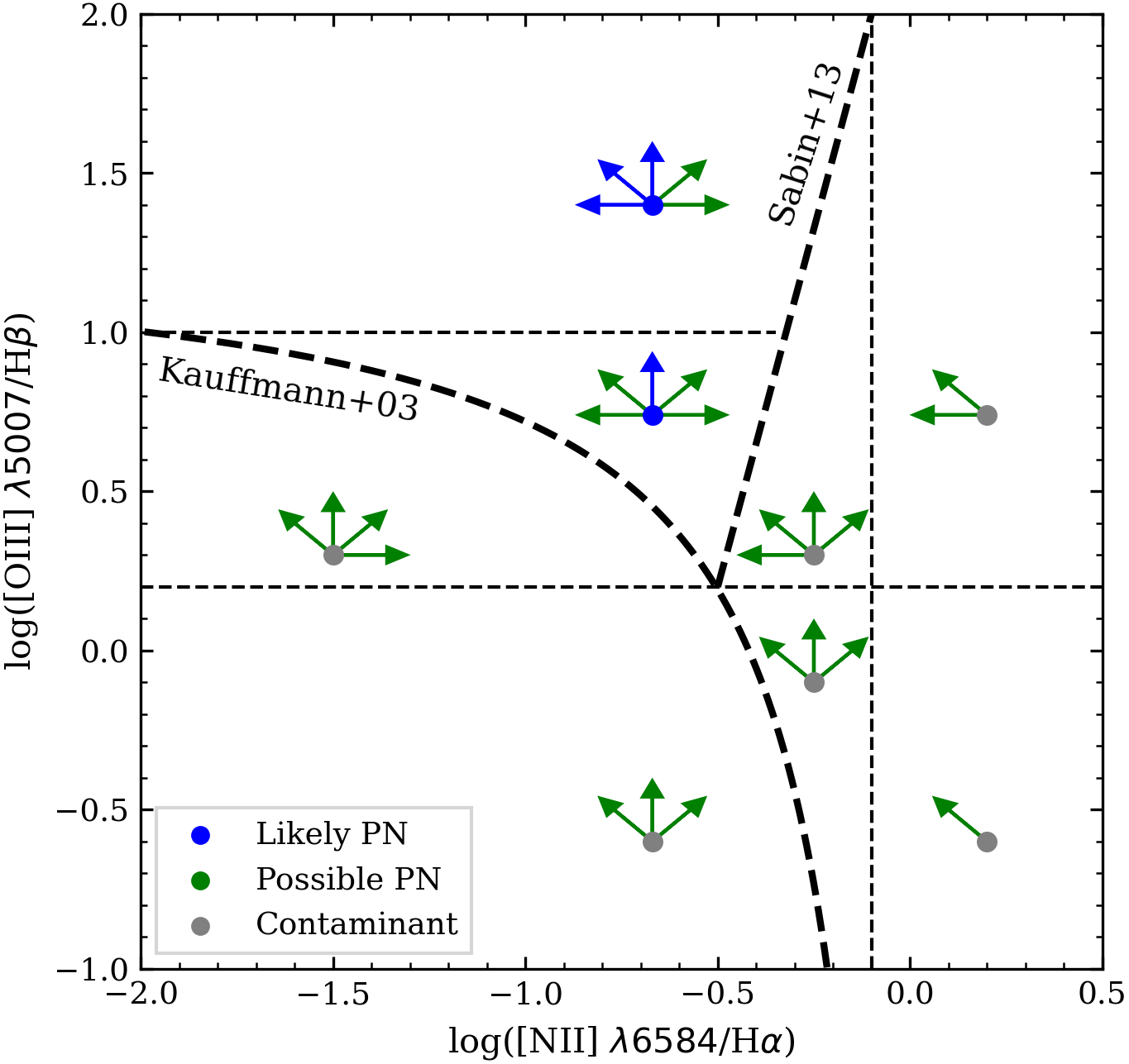}
    \caption{\nbpt\ diagnostic diagram, illustrating how a source is classified based on its location and emission-line ratio lower/upper limits. Diagonally pointing arrows indicate an upper limit along two axes. 
    The horizontal line at $\log([\ion{O}{iii}]/\mathrm{H}\beta)=0.2$ shows the intersection of the \protect\citet{kauffmann_2003_bpt} and \protect\citet{sabin_bpt_2013} demarcation lines. The horizontal line at $\log([\ion{O}{iii}]/\mathrm{H}\beta)=1$ and the vertical line at $\log([\ion{N}{ii}]/\mathrm{H}\alpha)=-0.1$ are there because we only consider the demarcation lines as far as the boundaries plotted here. 
    There is no down-pointing arrow as we require each emission-line source to have a reliable \oiii\ detection.}
    \label{fig:limits}
\end{figure}

Also plotted in Fig.~\ref{fig:bpt}, but using triangles, are sources with emission-line flux upper limits. For these, we consider whether the direction of the limit makes a source point away from, or toward, the PN-only region. In doing so, we consider the demarcation lines (and the regions they delineate) to be defined only as far as the emission-line flux ratio ranges plotted in Fig.~\ref{fig:bpt}. For example, the PN-only region does not extend to infinity above the \citet{sabin_bpt_2013} line. The intermediate classifications of sources with upper limits are illustrated in Fig.~\ref{fig:limits} for the \nbpt; they are analogous for the \sbpt. 

A direct consequence of our classification criteria, as indicated in the decision tree (Fig.~\ref{fig:tree}), is that we classify an emission-line source as a possible \ac{PN} if it has a reliable \oiii\ detection but no other emission-line detection. This will almost always be the case for the faintest \acp{PN} detectable. As a consequence, we can only ever label these as possible \acp{PN}, rather than likely \acp{PN}.

Another special case is an emission-line source with reliable \oiii\ and \nii\ detections but no other emission-line detection, as this is characteristic of Type~I \acp{PN} \citep{peimbert_typeI_1983, torres-peimbert_typeI_1997}. For these objects, we first consider their position on the \nbpt\ as usual. If they are not in the \ac{PN}-only region, we still classify them as possible \acp{PN} if they have $\log($\nii/\oiii) $\geq-0.3$. For the galaxies presented in this paper, there is no Type~I \ac{PN}.

\subsubsection{Source morphology}
\label{sec:roundness}

Spatially unresolved objects like \acp{PN} should be spatially identical to the \ac{PSF}, and should therefore appear as circular Gaussian-shaped objects. As \hii\ regions and \acp{SNR} are usually spatially extended, the extents and shapes of emission-line sources are of great help to remove potential contaminants remaining after applying the aforementioned emission-line ratio diagnostics. However, the SITELLE \ac{PSF} varies across the \ac{FOV}, objects being more distorted toward the edges. A point source located far from the data cube centre could therefore appear more extended and elongated, making it more difficult to distinguish it from an intrinsically extended source. We have therefore attempted to map the \ac{PSF} variations across the \ac{FOV}, so that we can still apply compactness and shape criteria to determine whether a \ac{PN} candidate (based on emission-line ratios) is indeed a spatially unresolved \ac{PN}. 

We map the variations of the \ac{PSF} in SN2 by quantifying the departures from perfect 2D Gaussians of sources across the \ac{FOV}. We use all the stars from the {\it GAIA} Data Release 3 \citep{gaia_dr3_2023} catalogue that are located within the SITELLE \ac{FOV} and have $m_G>17$ (to avoid saturated stars). We filter for stars using \texttt{classprob\_dsc\_combmod\_star} $>0.995$, which indicates the probability of any source being a single star.

\begin{figure*}
    \centering
    \subfloat[NGC~4214]{%
        \includegraphics[width=.45\linewidth]{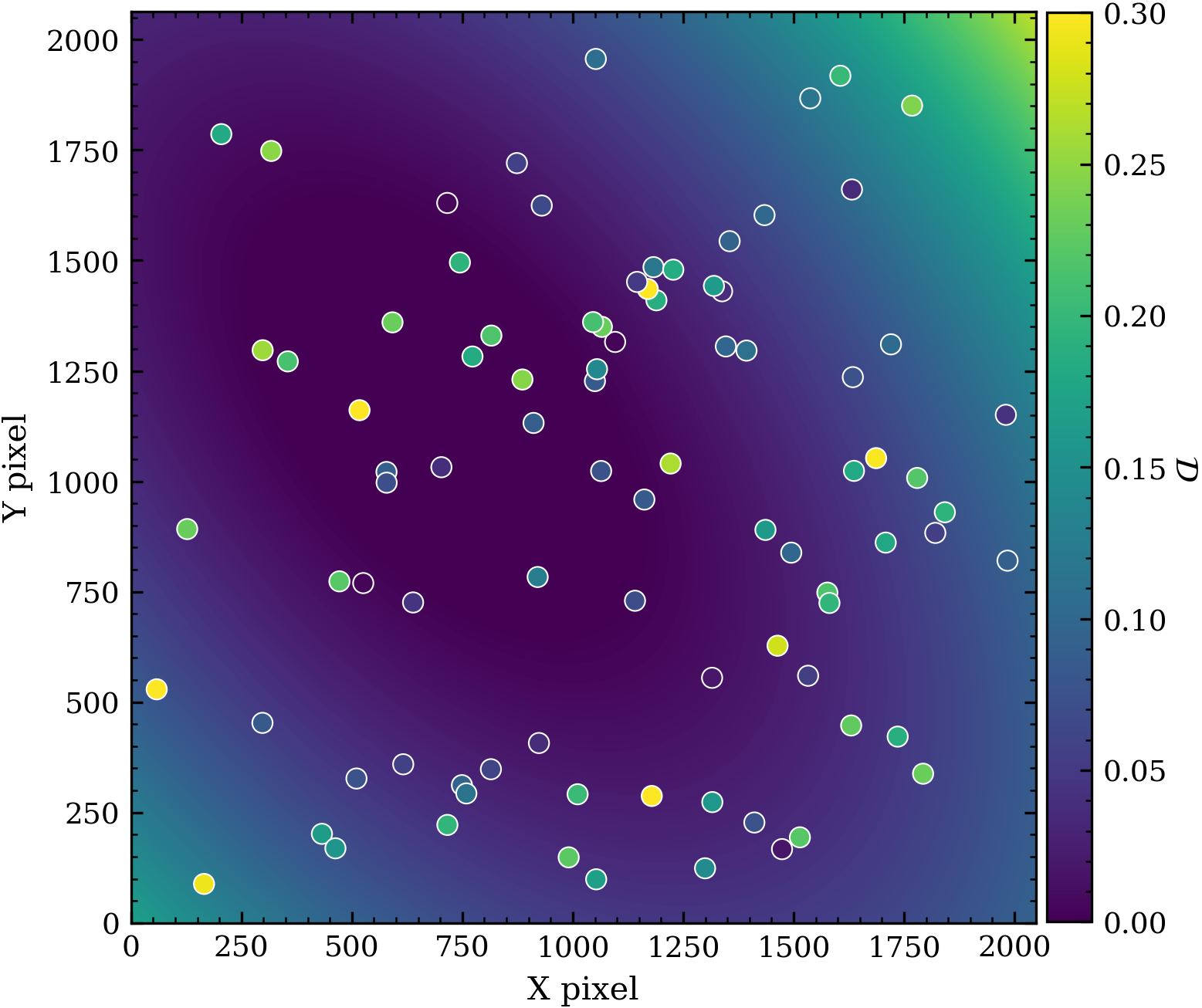}%
    } \hfill
    \subfloat[NGC~4449]{%
        \includegraphics[width=.45\linewidth]{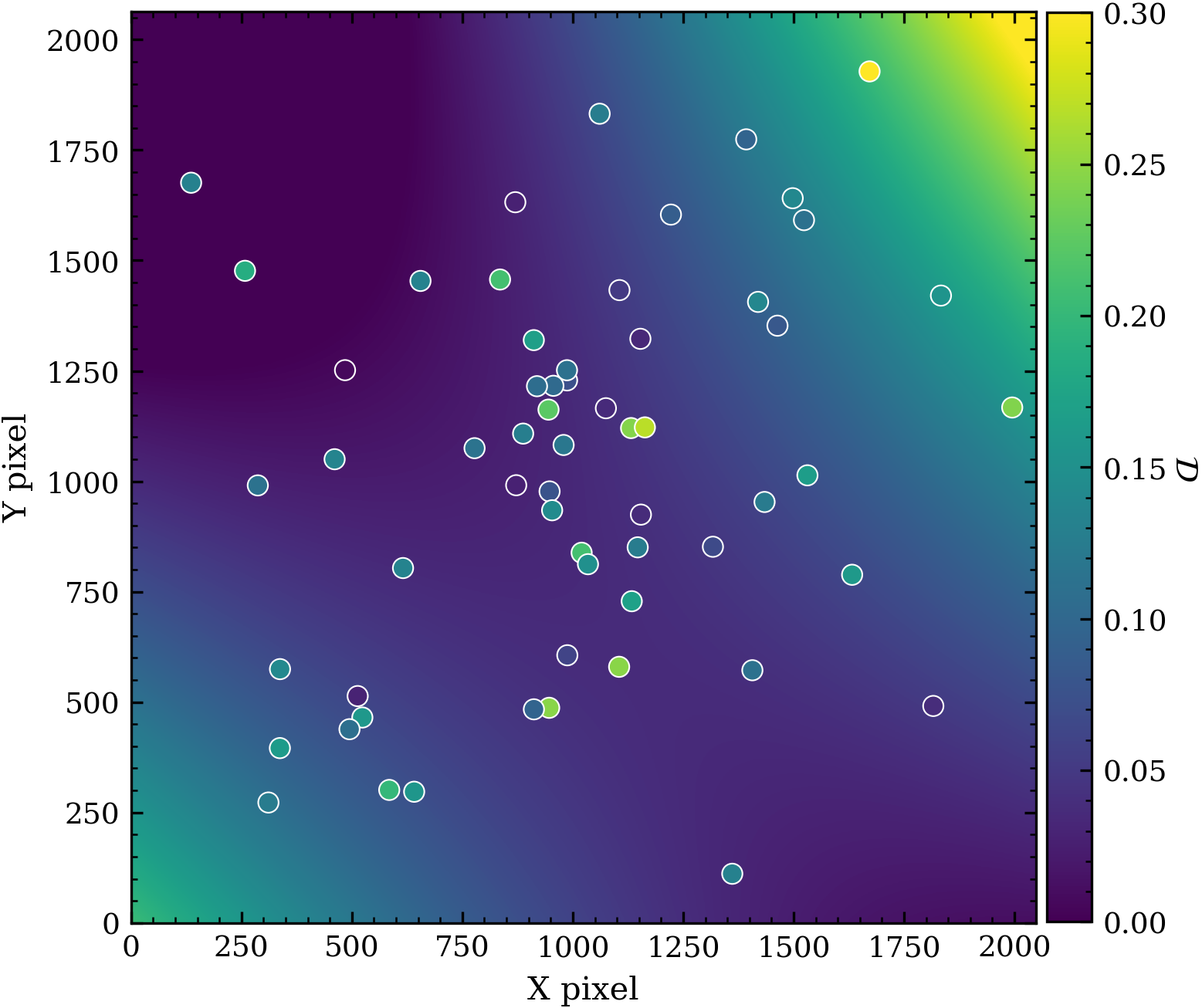}%
    }
    \caption{Distortion maps of NGC~4214 (left) and NGC~4449 (right), fitted to stars. The stars used for the fit are plotted as filled circles, with a colour matching that of the background map and colour table. Lighter colours indicate higher asymmetric distortions.
    }
    \label{fig:distortion}
\end{figure*}

We then fit the stars' shapes on the SN2 continuum map using 2D Gaussians, and define a distortion parameter $\mathcal{D}\equiv1-\frac{\sigma_{\mathrm b}}{\sigma_{\mathrm a}}$, where $\sigma$ is the standard deviation of the best-fitting Gaussian and a and b refer to the best-fitting Gaussian major and minor axis, respectively. A perfectly round source has $\mathcal{D}=0$ while an elongated source has $0<\mathcal{D}<1$. We then fit a polynomial surface to all the measurements across the \ac{FOV}, yielding the 2D distortion map ($\mathcal{D}_\mathrm{fit}$) shown in Fig.~\ref{fig:distortion} for each of NGC~4214 and NGC~4449. Each map informs us on how elongated an intrinsically round source appears in our data. 

Having fitted the {\it GAIA} stars, we then carry out 2D Gaussian fits of all the \oiii\ emission-line sources. The $S/N$ of \acp{PN} in the deep images are not sufficient to carry out reliable fits, as they are only emitting across a very limited number of frames in each data cube. We therefore fit the emission-line sources using the \oiii\ flux maps and the detection maps, and calculate a distortion parameter $\mathcal{D}_\mathrm{source}$ from each map. Each can then be compared to the prediction of the polynomial fit at the source location. We classify sources as round if $\mathcal{D}_\mathrm{source}-\mathcal{D}_\mathrm{fit}\leq0.3$ for at least one of the two maps.

\begin{figure*}
    \centering
    \subfloat[NGC~4214]{%
        \includegraphics[width=.45\linewidth]{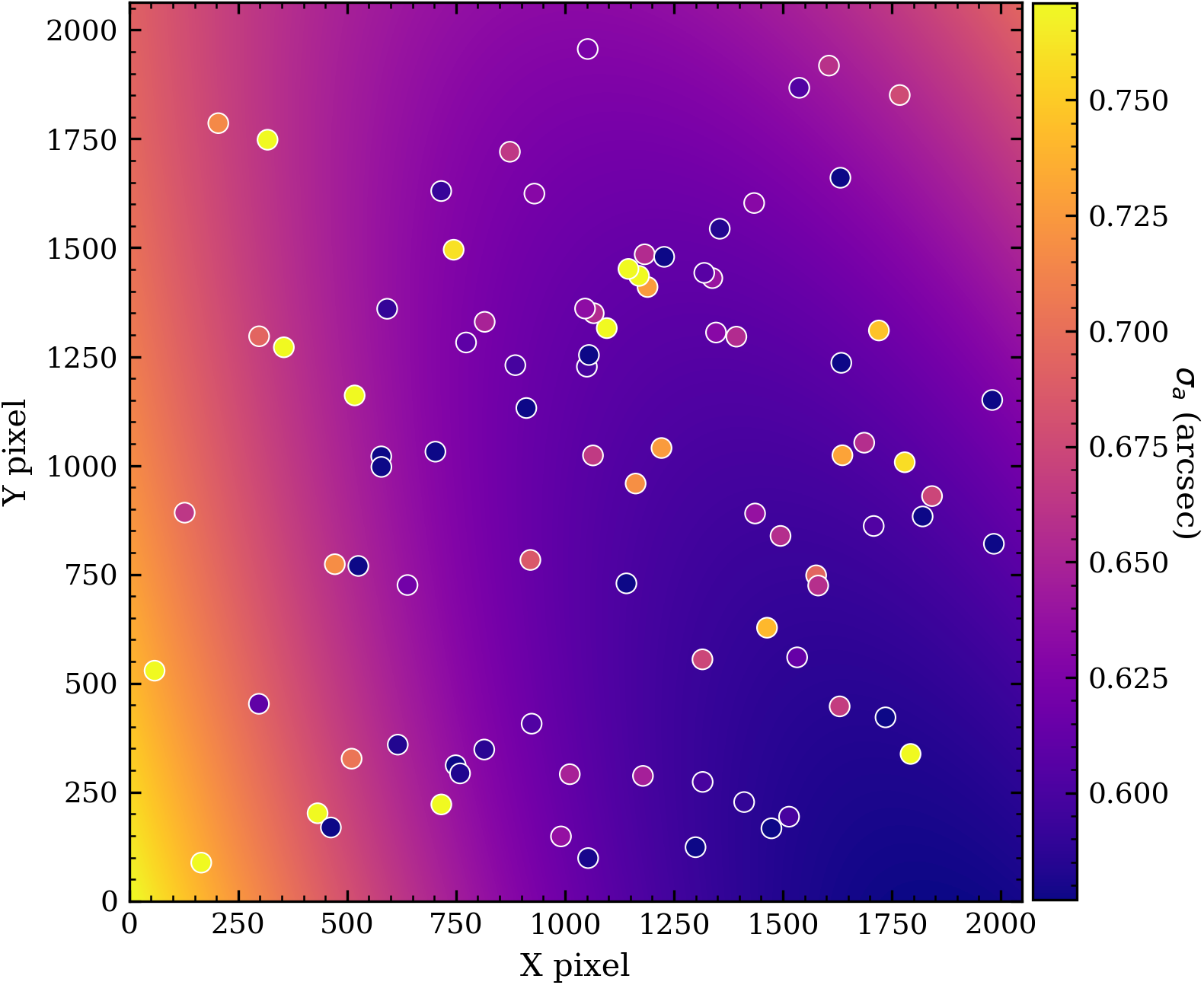}%
    } \hfill
    \subfloat[NGC~4449]{%
        \includegraphics[width=.45\linewidth]{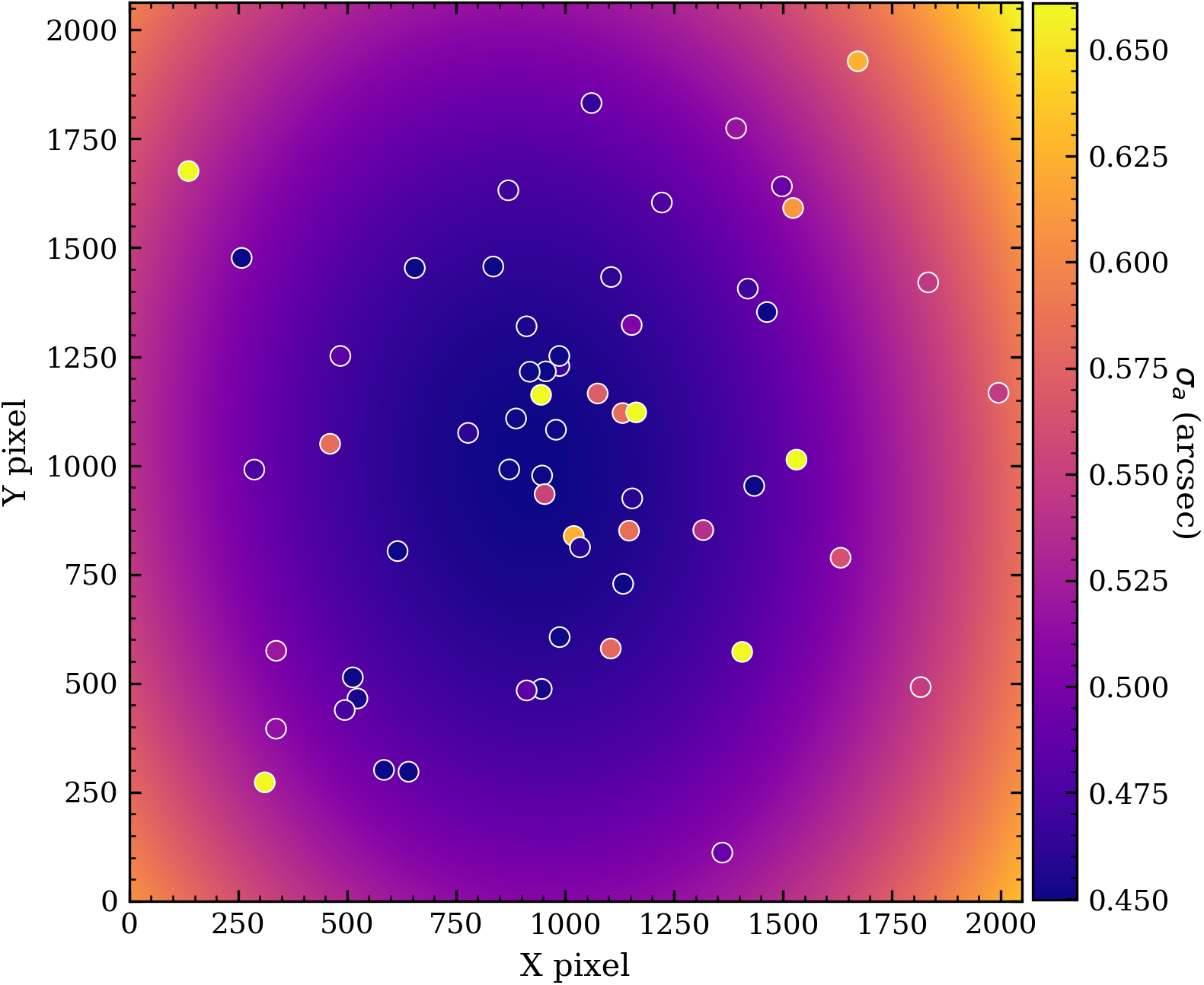}%
    }
    \caption{$\sigma_\mathrm{a}$ maps of NGC~4214 (left) and NGC~4449 (right), fitted to stars. The stars used for the fit are plotted as filled circles, with a colour matching that of the background map and colour table. Lighter colours indicate larger $\sigma_a$ and therefore more severe size distortions.
    }
    \label{fig:semimajor}
\end{figure*}

\acp{PN} should also be spatially unresolved. We thus fit another polynomial surface, this time simply to the semi-major axes ($\sigma_\mathrm{a}$) of the stars (see Fig.~\ref{fig:semimajor}). We classify a \ac{PN} candidate as unresolved if $\sigma_\mathrm{a,source}/\sigma_\mathrm{a,fit}\leq1.3$. 

\subsubsection{Final catalogue} \label{sec:catalogue}
We visually inspect all the sources that are (1) round, unresolved and classified as a likely \ac{PN} on at least one of the two BPT diagrams, and (2) round, unresolved and classified as a possible \ac{PN} on both diagnostic diagrams (or on the only one used). Objects from category (1) that pass the visual inspection are labelled bona fide \acp{PN}, while those from category (2) are labelled  possible \acp{PN}. The final classifications of all objects that are at least possible \acp{PN} are shown in Tables~\ref{tab:NGC4214} and \ref{tab:NGC4449} for NGC~4214 and NGC~4449, respectively.

\subsection{Velocity calibration}
\label{sec:SN2_vel}

While the SN3 initial velocities were based on the uncalibrated SN2 velocities (Section~\ref{sec:fitting}), the resulting fitted SN3 velocities are usually of a higher precision and can be calibrated using the skyline method described in Section~\ref{sec:skyline_velcal}. The exceptions to this are sources that have $S/N<3$ in the \ha\ line, for which the SN2-based velocities generally remain more reliable. Nevertheless, the SN2 filter has a much lower spectral resolution and, more importantly, it lacks (detected) sky lines required for a velocity calibration. We thus calibrate the SN2 velocities of these \acp{PN} by finding the average offset between the uncalibrated SN2 velocities and the fully calibrated SN3 velocities of nearby objects. This method is very similar to that used by \citet{vicens-mouret_planetary_2023}, but we nevertheless explain it in full here as a few modifications were made.

First, we compile a catalogue of sources that are reliably detected (flux $S/N>3$) in both \oiii\ and \ha, so that these sources have velocities estimated using both the SN2 ($v_\mathrm{SN2}$) and the SN3 ($v_\mathrm{SN3}$) filters. These sources are not required to have \ac{PN}-like emission or be point sources. As we can use the sky-line calibration for SN3, we treat the calibrated SN3 velocities as the ground truths. For each \ac{PN} that does not have a velocity derived using the SN3 filter, we search for objects within a radius of $150$~pixels, iteratively increasing that radius by $100$~pixels at a time, if necessary, until at least $20$ sources are found. We then calculate the median $v_\mathrm{SN3}-v_\mathrm{SN2}$ of these sources, and use that as a correction factor for the \ac{PN} SN2 velocity. For error propagation, the standard deviation of all the $v_\mathrm{SN3}-v_\mathrm{SN2}$ measurements is taken as the uncertainty of the correction factor. 

\begin{figure}
    \centering
    \includegraphics[width=0.9\linewidth]{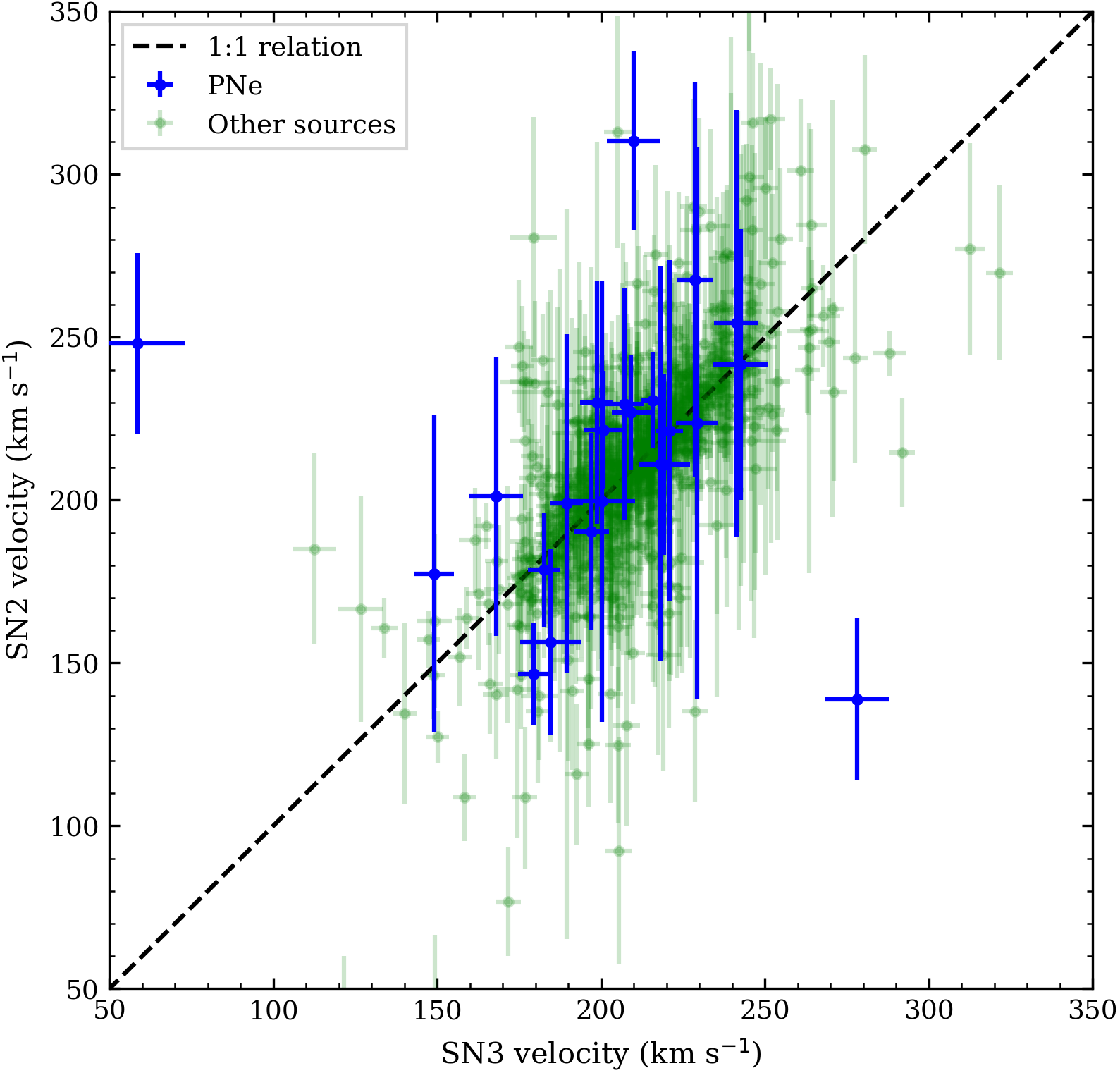}
    \caption{Comparison of the SN2 (cross-matching) and SN3 (sky-line) velocity calibration methods for sources in NGC~4449.
    }
    \label{fig:velocity_check}
\end{figure}

This method assumes that the extra correction to the SN2-derived velocities varies smoothly across the \ac{FOV},
but we can test its accuracy by also applying it to sources which do have SN3 (sky line-based) velocities. The outcome of this is shown in Fig.~\ref{fig:velocity_check} for NGC~4449, revealing that the corrected SN2 and SN3 velocities of the sources follow the $1:1$ line, with a standard deviation of $\approx 20~\mathrm{km~s}^{-1}$. The velocities of sources detected only in SN2 may therefore be slightly less reliable. In this work, we quote the SN3 velocities for sources with $S/N>3$ in the \ha\ line, SN2 velocities otherwise.

\subsection{Mock catalogue and recovery tests}
\label{sec:mocks}

To better understand our recovery rate and to define a completeness limit for our \ac{PN} catalogues, we generate mock populations of \acp{PN}. As \acp{PN} are the progeny of stars, their distribution should to first order reflect that of the underlying stellar light. We therefore start by performing surface photometry of each galaxy, and use that to dictate where the mock \acp{PN} are inserted spatially. We first mask out sources identified as stars by {\it GAIA}, as these do not belong to the galaxy. To avoid being biased by emission-line regions, we {use a map of the galaxy (stellar) continuum emission in SN2.} We then carry out surface photometry, fitting the surface brightness in elliptical annuli of constant ellipticity, position angle and centre. We use $12$ annuli, increasing the semi-major and semi-minor axes logarithmically.
The relative surface brightness of each annulus determines the number of mock \acp{PN} it receives. Within each elliptical annulus, the positions of the mock \acp{PN} are randomised. To test the \ac{PN} recovery as a function of apparent magnitude, we also randomly sample the magnitudes from an assumed Ciardullo-like \ac{PNLF} \citep{ciardullo_PNLF_1989}. In doing so, we need to assume a bright-end cut-off and therefore a distance to each galaxy, but this does not affect the completeness limit inferred, which is at a much fainter magnitude. The distance used for each galaxy is as listed in Table~\ref{tab:obs}.

To insert each \ac{PN} in a data cube, we create a 2D Gaussian spatial model using \texttt{astropy}, with input parameters as determined from foreground stars. For each mock \ac{PN}, we take the fitted $\sigma_\mathrm{a}$ and $\mathcal{D}$ at its input position, using the maps created in Section~\ref{sec:roundness}. 
 
The spectra of the mock \acp{PN} are scaled versions of each other, in the sense that the \oiii\ flux matches that of the sampled \mpn. We select input emission-line ratios, such that a source with these emission lines would be located in the appropriate region of each BPT diagram, and assign the same ratios to each mock \ac{PN}. \texttt{ORCS} is then used to simulate SN2 and SN3 spectra at spectral resolutions matching that of the real data cubes, assuming the same velocity offsets and broadening as the real \acp{PN}. 

Finally, the mock \acp{PN} are inserted into the real data cubes, to best recreate the conditions under which real \acp{PN} are searched for (e.g.\ same noise and contaminant sources). To ensure sufficient numbers of mock \acp{PN} (and thus reliable statistics) but avoid source confusion, for each galaxy we create $3$ unique sets of mock data cubes with $\approx300$ \acp{PN} in each. 

\begin{figure*}
    \centering
    \subfloat[NGC~4214]{%
        \includegraphics[width=1\linewidth]{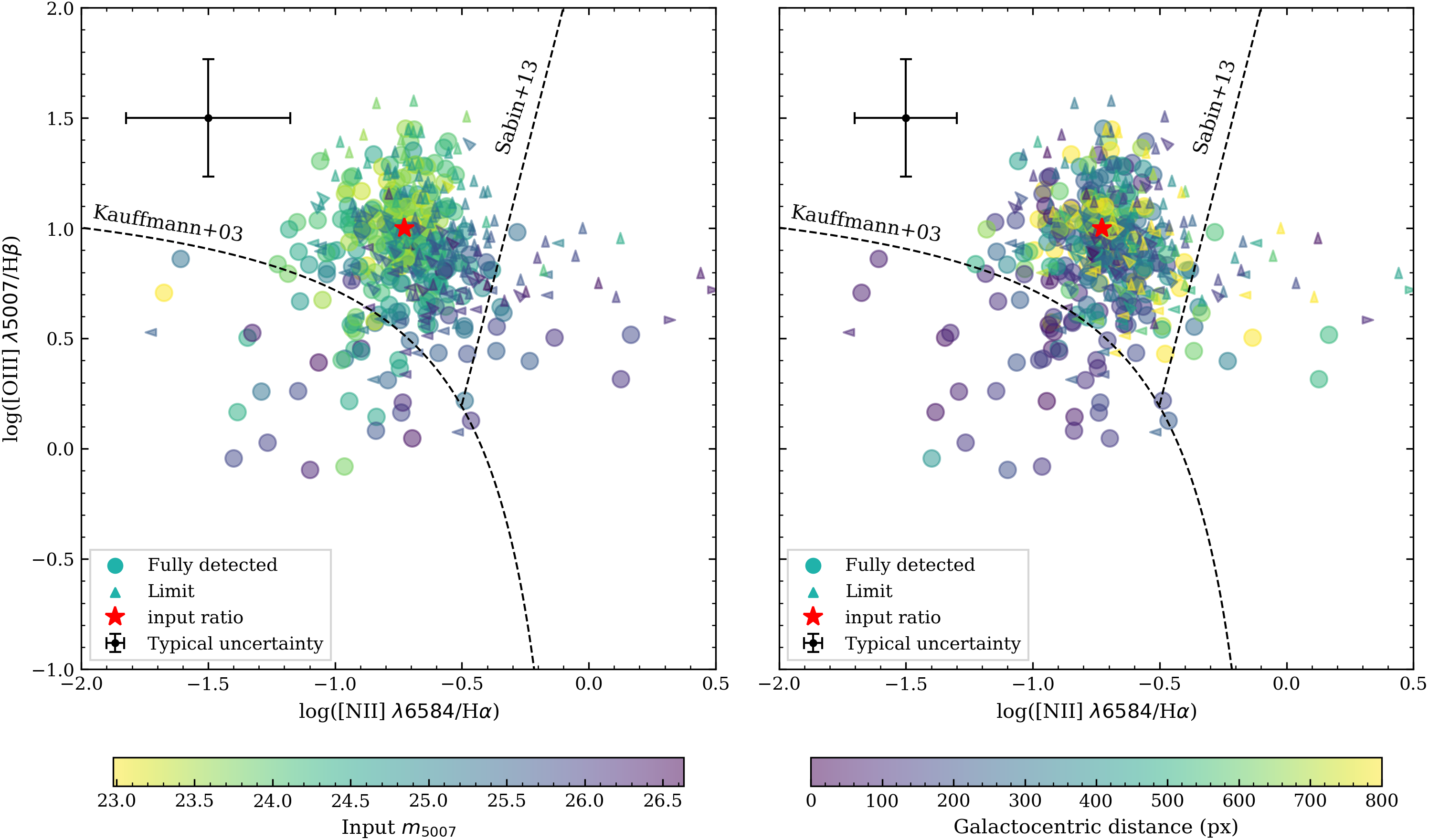}%
    } \\
    \subfloat[NGC~4449]{%
        \includegraphics[width=1\linewidth]{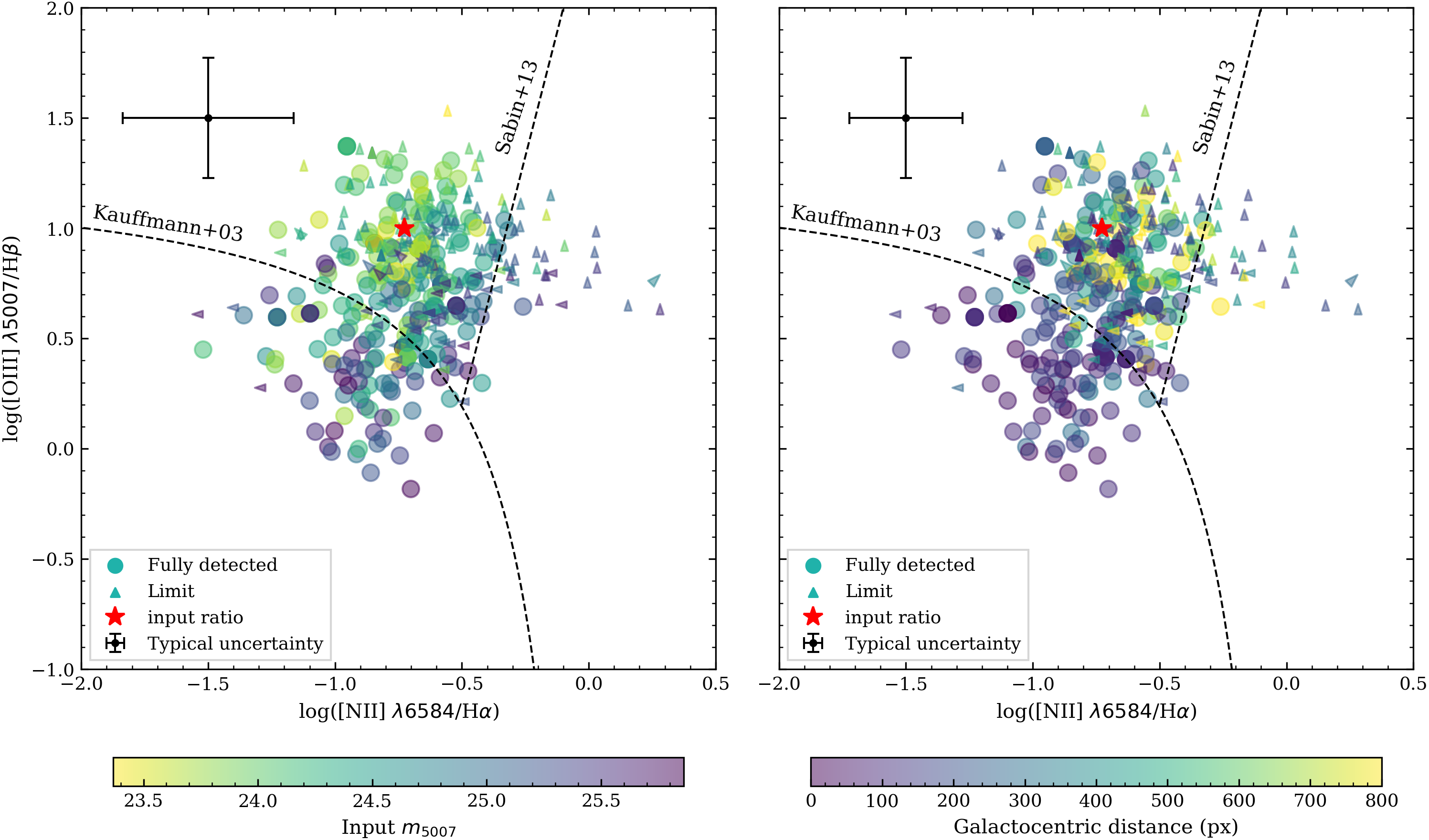}%
    }
    \caption{\nbpt\ diagrams of the mock \acp{PN} generated for NGC~4214 (top) and NGC~4449 (bottom). The mock \acp{PN} are colour-coded by input apparent magnitude (left) and galactocentric distance (right). The input emission-line ratio is indicated by the red star. 
    }
    \label{fig:mock_bpt}
\end{figure*}

Using the same methods as for the real catalogue, we create a detection map using \texttt{ORCS}, run our source detection algorithms using \texttt{photutils} and carry out emission-line fitting with the same parameters as for the real data. We then cross-match the input mock \acp{PN} with the recovered sources. Fig.~\ref{fig:mock_bpt} shows how our mock \acp{PN} populate the \nbpt\ emission-line ratio diagnostic diagram, where the input emission-line ratios are indicated by the red star. 
The colour coding of the data points shows that the scatter of the mock \acp{PN} about the input emission-line ratios increases with faintness (larger magnitudes) and proximity to the galaxy centre (increased contamination due to bright \hii\ regions). 

\begin{figure*}
    \captionsetup[subfigure]{labelformat=empty}
    \centering
    \subfloat[NGC~4214
    ]{%
        \includegraphics[width=.48\linewidth]{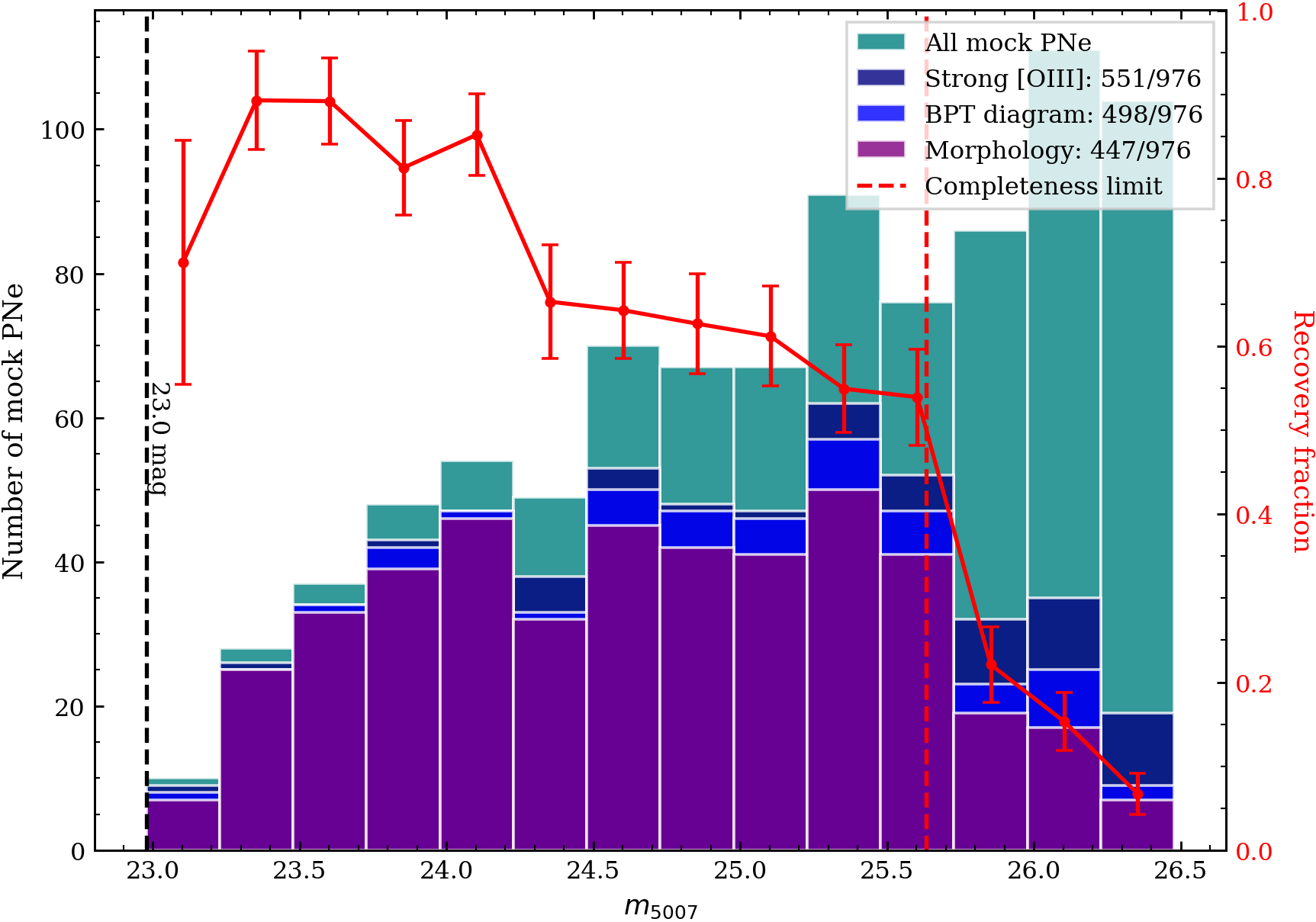}%
    } \hfill
    \subfloat[
    ]{%
        \includegraphics[width=.5\linewidth]{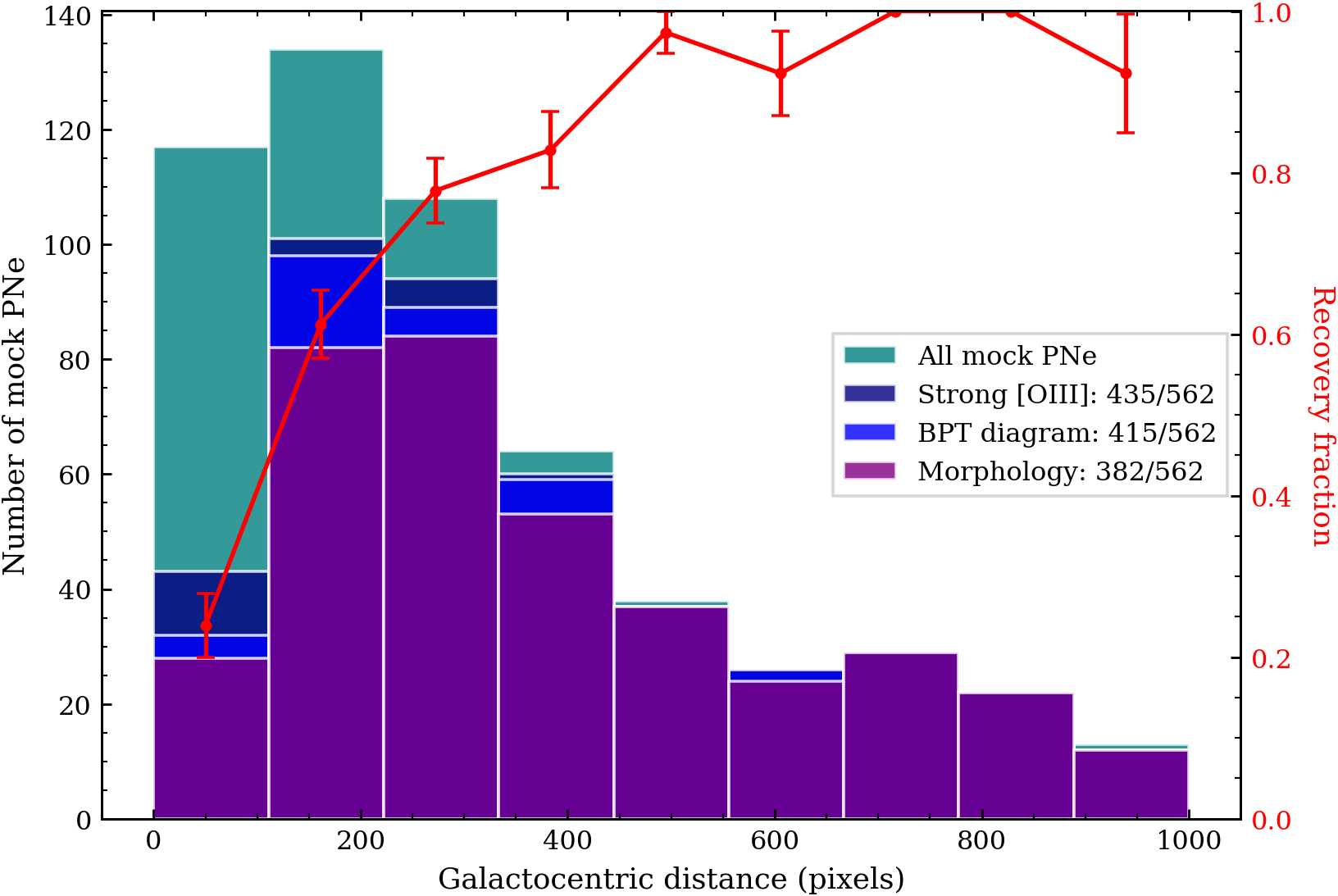}%
    }\\
    \subfloat[NGC~4449
    ]{%
        \includegraphics[width=.48\linewidth]{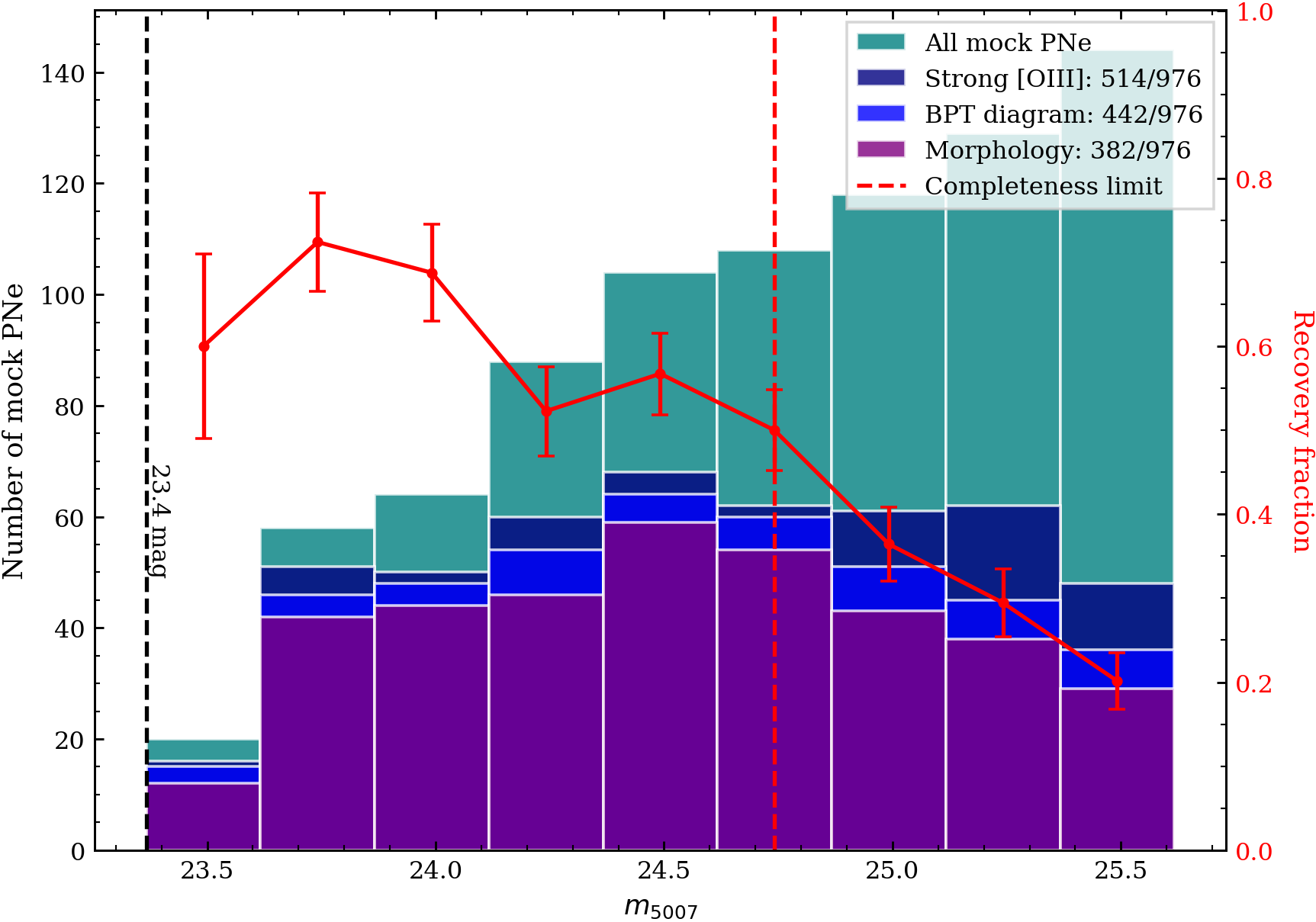}%
    } \hfill
    \subfloat[
    ]{%
        \includegraphics[width=.5\linewidth]{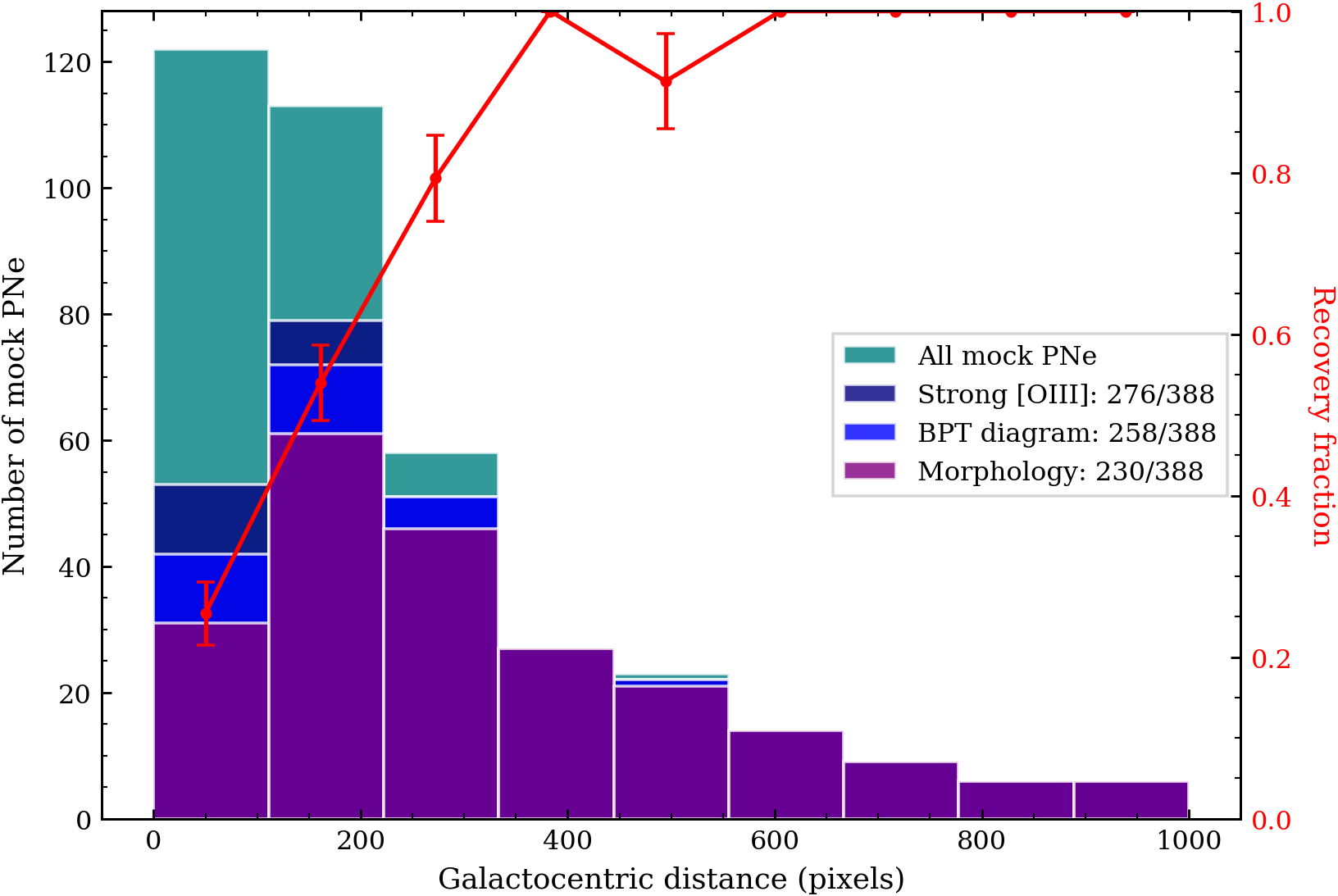}%
    }
    \caption{Recovery rates of mock \acp{PN} for NGC~4214 (top) and NGC~4449 (bottom). Left: \acp{PNLF} of the mock \acp{PN}, overlaid with the recovery fractions. Colours indicate the pipeline step at which the mock \acp{PN} fail to be recovered. From top to bottom in the legend: all generated mock \acp{PN}, recovered mock \acp{PN} with $S/N(\mathrm{\oiii})>3$, recovered mock \acp{PN} which also have the correct emission-line ratios on the BPT diagram(s) and recovered mock \acp{PN} which additionally are point sources. The vertical black dashed lines indicate the bright-end cut-offs used and the vertical red dashed lines the derived completeness limits. The recovery fractions decrease toward fainter magnitudes as expected, but even the brighter bins do not have a perfect recovery due to the presence of contaminant sources. Right: Mock \acp{PN} binned by galactocentric distance, overlaid  with the recovery fractions, for \acp{PN} brighter than the completeness limit only. The recovery rates increase toward the galaxy outskirts, where there are fewer contaminant sources.}
    \label{fig:mock_recovery}
\end{figure*}

The mock \ac{PN} roundnesses were determined from the mock detection frames and the mock \oiii\ flux maps using only the automated method (no visual inspection). We use the same distortion maps and size maps as fitted before ($\mathcal{D}_\mathrm{fit}$ and $\sigma_\mathrm{a}$) to determine whether sources are sufficiently round. We also keep the size cuts the same. The total recovery rates and corresponding \acp{PNLF} are shown in the left panels of Fig.~\ref{fig:mock_recovery}. We consider the samples complete in magnitude bins for which the recovery rate is at least $50\%$. This yields completeness limits of ${m}_{5007}=25.5$ for NGC~4214 and ${m}_{5007}=24.7$ for NGC~4449. 

The strong dependence on magnitude present in the left panels of Fig.~\ref{fig:mock_recovery} is expected, but we can also investigate the spatial dependences of the completeness of our samples by considering only the subsets of mock \acp{PN} brighter than the completeness limits. As illustrated in the right panels of Fig.~\ref{fig:mock_recovery}, the recovery rates are low in the galaxy centres, due to the bright and extended contaminant emission-line regions there. By having the mock \ac{PN} positions follow the stellar continuum light, many are indeed located in the galaxy central regions where they are harder to detect. If we had populated the data cubes randomly spatially, we would have significantly overestimated the completeness of our samples.

\subsection{PNLF and the $\alpha$ parameter}

The planetary nebulae luminosity function specifies the number of \acp{PN} within a certain \mpn\ range. It has empirically been shown to function as a distance indicator through its bright-end cut-off, applicable to both early- and late-type galaxies. It is described by Eq.~\ref{eq:pnlf}, but we use the simplified version that keeps the faint-end slope index fixed at $c_2=0.307$ and we assume an absolute magnitude of the bright-end cut-off $M^*_{5007}=-4.47$ \citep{ciardullo_pnlf_2012}. The only parameter we fit for is the apparent magnitude of the bright-end cut-off ($m^*_{5007}$), for which we use the method of maximum likelihood \citep{ciardullo_PNLF_1989} as implemented by \citet{scheuermann_spirals_2022}. By treating the \ac{PNLF} as a probability distribution, we avoid binning the \acp{PN} in magnitude. For the initial guess of the distance modulus we assume the distance listed in Table~\ref{tab:obs} for each galaxy, which is the mean of the distances listed in the NASA/IPAC Extragalactic Database. For NGC~4214 and NGC~4449, these come from a combination of methods including TRGB.

The specific frequency of \acp{PN}, also referred to as the $\alpha$ parameter, is defined in Eqs.~\ref{eq:alpha} and \ref{eq:alpha_int}. Although $\alpha$ is traditionally defined using a bolometric luminosity, it has become common practice to use $B$ or $V$ band photometry with a bolometric correction, as described in \cite{buzzoni_planetary_2006}. We also follow this convention and calculate $\alpha_\mathrm{bol}$ parameters based on $V$-band photometry from \cite{cook_spitzerphoto_2014}, using the integrated magnitudes obtained from their infrared apertures as they most closely match the areas over which we detect \acp{PN} (see Fig.~\ref{fig:FOV}). We assume distances as listed in Table.~\ref{tab:obs}, rather than the ones derived from PNLF fitting. In addition to $\alpha_\mathrm{bol}$, we provide an $\alpha_V$ parameter for each galaxy, which simply uses the $V$-band luminosity without a bolometric correction. 

The number of \acp{PN} we recover is only a subset of the entire population of \acp{PN}, even at bright magnitudes. This needs to be accounted for when calculating the $\alpha$ parameter. For each galaxy, we determine the recovery rate using the mock catalogue of \acp{PN}. This recovery rate as a function of magnitude is then used to scale the number of recovered \acp{PN} to what should be more representative of the full population. To facilitate comparisons between surveys of varying depth, the $\alpha$ parameter is often defined up to $0.5$ or $2.5$ magnitudes from $m^*_{5007}$. It should be noted that with a completeness threshold defined at $50\%$, the \ac{PN} catalogue of NGC~4449 is only complete up to $0.5$ magnitudes from the fitted bright-end cut-off.


\section{Results}
\label{sec:results}

\input{tables/alpha}

\subsection{NGC~4214}
In NGC~4214, we identify $628$ sources with reliably detected \oiii\ emission. Of these, $279$ (potentially) have emission-line ratios in the appropriate regions of the BPT diagnostic diagrams and $121$ are initially classified as round using our automated methods. The $72$ sources that are in both of those categories were then visually inspected for roundness, leaving only $11$ bona fide and $14$ possible \acp{PN}. Most of the final contaminants that were manually removed fall within the bright centre of the galaxy, where we cannot make out whether the source is a distinct point source or not. Other contaminants are associated with extended sources, where it is unclear if they are part of the larger source or superposed. We have kept our catalogue as clean as possible and removed any source that does not appear to be a point source.

\begin{figure*}
    \captionsetup[subfigure]{labelformat=empty}
    \centering
    \subfloat[]{%
        \includegraphics[width=.55\linewidth]{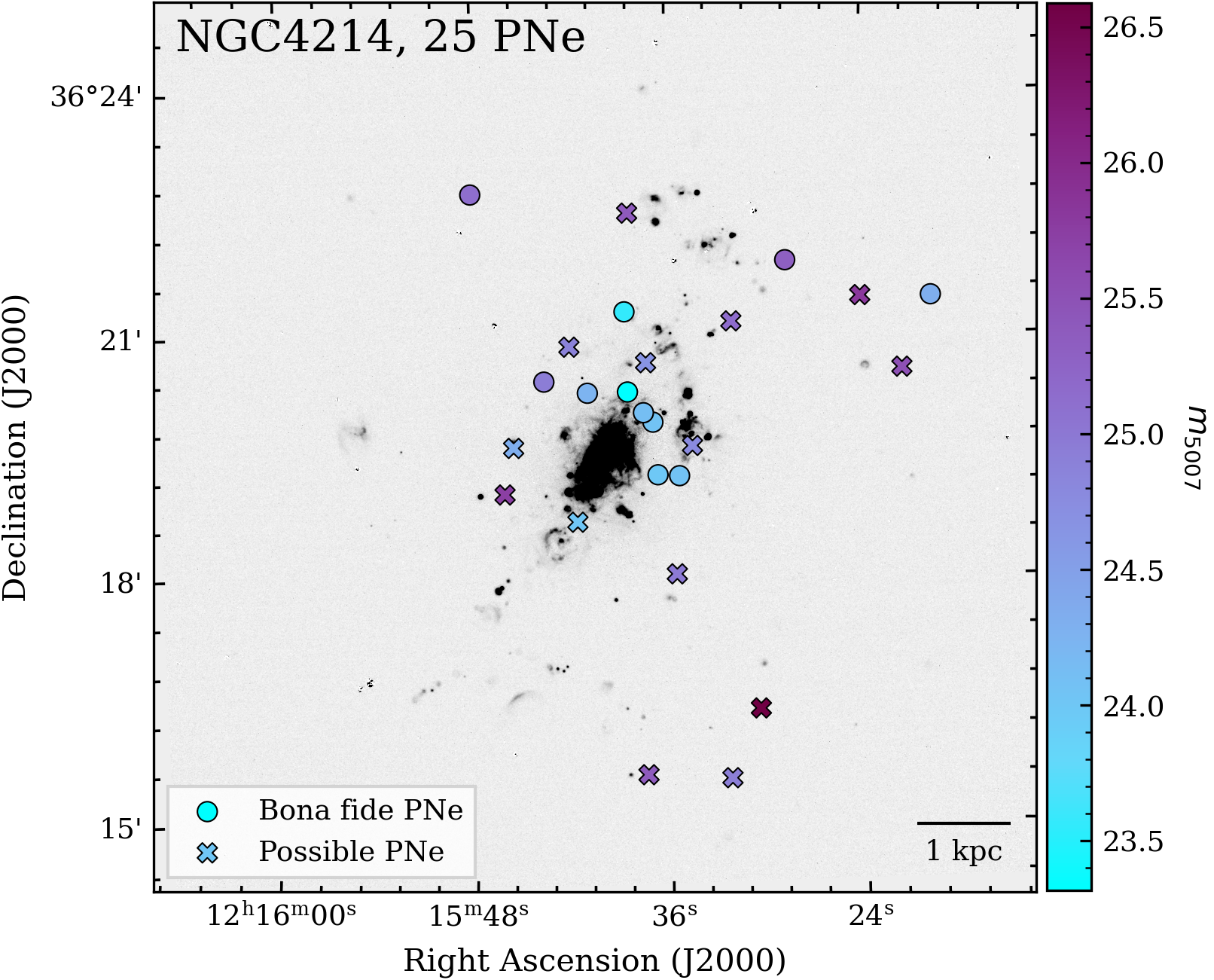}%
    }
    \subfloat[
    ]{%
        \includegraphics[width=.45\linewidth]{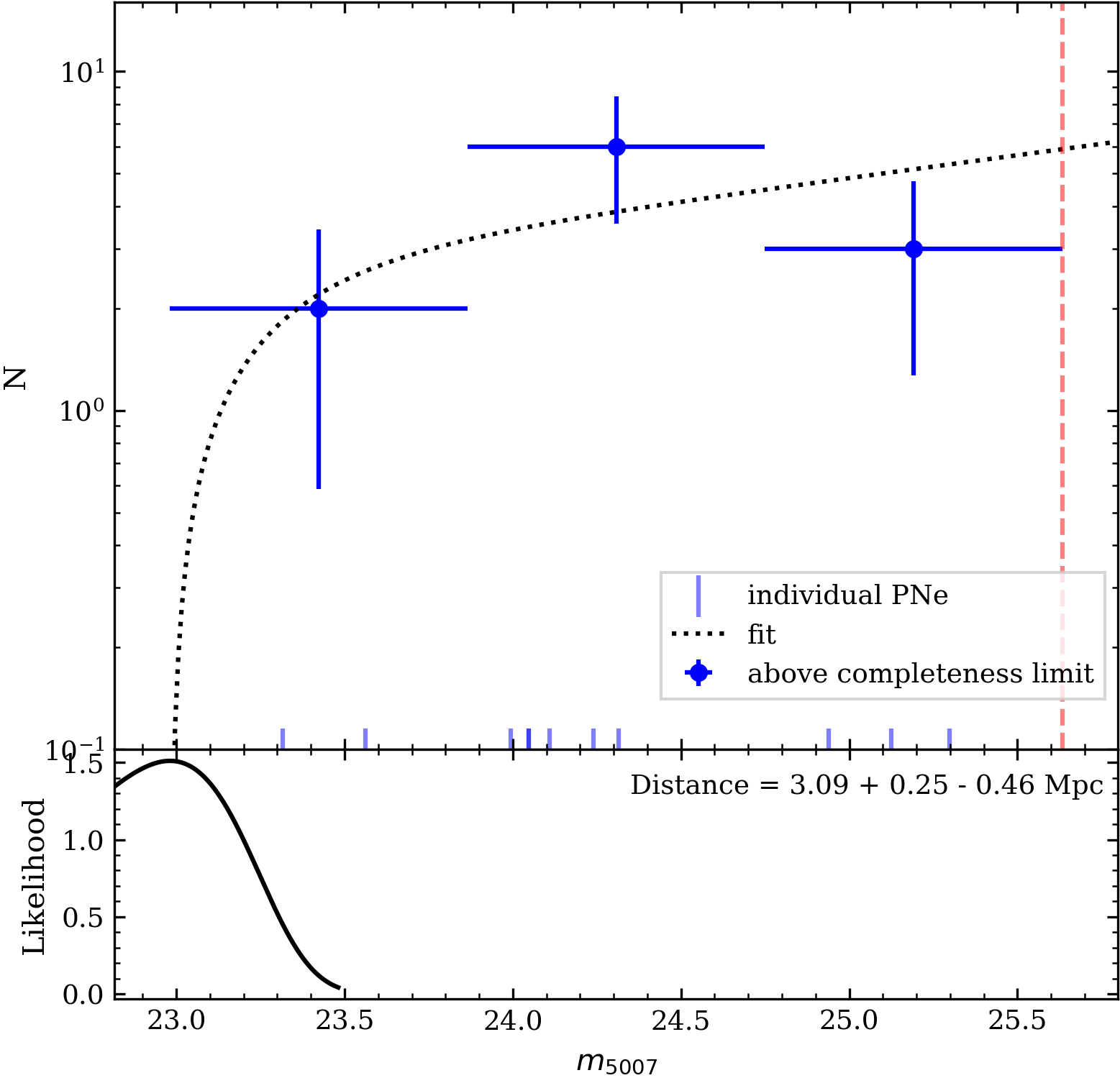}%
    }
    \caption{Left: bona fide and possible \acp{PN} of NGC~4214, overlaid on the \oiii\ flux map. Right: \ac{PNLF} of NGC~4214, computed using only the bona fide \acp{PN}. The black dotted line shows the best-fitting \ac{PNLF}. The vertical red dashed line indicates the completeness limit. The likelihood of the apparent bright-end cut-off is plotted underneath.}
    \label{fig:pne_4214}
\end{figure*}

The final \ac{PN} catalogue of NGC~4214 is presented in Table~\ref{tab:NGC4214}. The \ac{PN} spatial distribution is shown in the left panel of Fig.~\ref{fig:pne_4214}. There is a notable lack of \ac{PN} in the brightest part of the galaxy, as expected from the mock \ac{PN} recovery test (Fig.~\ref{fig:mock_recovery}). 

Of the $17$ \ac{PN} candidates identified by \citet{dopita_ngc4214_2010} using narrow-band \ac{HST} imaging, we do not recover D9 -- D12 and D16 on the detection map. It should however be noted that \citet{dopita_ngc4214_2010} suspected D10 and D13 to be compact high-excitation \hii\ regions rather than \acp{PN}. All the other candidates that are detected also have strong enough \oiii\ emission. D13 and D15 do not have the right emission line ratios, while D2 and D8 are rejected because of their morphologies. In the end, we classify $6$ of the $17$ \citet{dopita_ngc4214_2010} \ac{PN} candidates as bona fide \acp{PN} and a further $2$ as possible \acp{PN}. Our conclusions are very similar to those of \citet{vicens-mouret_planetary_2023}, who recovered the same $8$ \acp{PN} but also included D2 and D15 in their catalogue.

All of the $25$ \acp{PN} identified with the \ac{SIGNALS} data by eye by \citet{vicens-mouret_planetary_2023}, including the $10$ that are in common with \citet{dopita_ngc4214_2010}, are also recovered on the detection map. However, VDR1 and VDR10 are not reliably detected in the \oiii\ emission line ($S/N\leq3$), while VDR4 and VDR14 are not in the appropriate region of either BPT diagnostic diagram. Aside from these 4 mentioned here, and D2 and D15 mentioned before, we we stil classify $10$ of the $25$ \citet{vicens-mouret_planetary_2023} \acp{PN} as bona fide \acp{PN} and a further $9$ as possible \acp{PN}. 

Overall, in addition to the $19$ \acp{PN} we recover from \citet{dopita_ngc4214_2010} and \citet{vicens-mouret_planetary_2023}, we also identify $6$ new ones. One of these, NGC4214-Y6, has a velocity in strong disagreement with those in the rest of the catalogue. We list it in our catalogue regardless, as the velocities may not be accurate, but we caution that it may be a contaminant.

\begin{figure}
    \centering
    \includegraphics[width=1\linewidth]{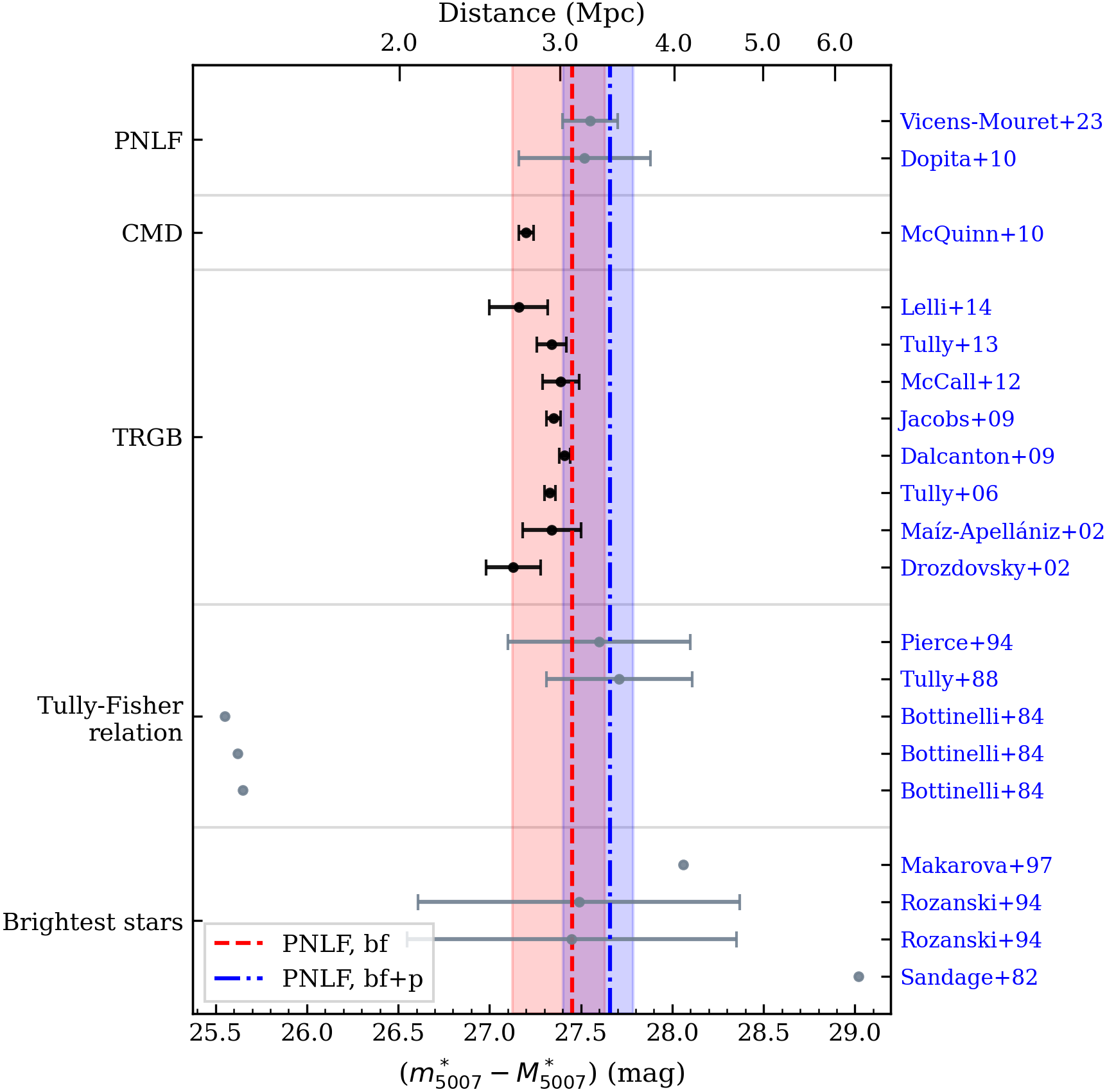}
    \caption{Comparison of our NGC~4214 PNLF distances to literature measurements. The red dashed line indicates our fit to the bona fide \acp{PN} only, the blue dot-dashed line to both bona fide and possible \acp{PN}. Shaded areas represent $1\sigma$ uncertainties. References are listed on the right, methods on the left (CMD: colour-magnitude diagram). Distance measurements with black markers were used in calculating the assumed distance to NGC~4214, as used throughout this paper.}
    \label{fig:distances_4214}
\end{figure}

A PNLF fit to the bona fide \ac{PN} sample, shown in the right panel of Fig.~\ref{fig:pne_4214}, yields a distance modulus of $27.45^{+0.18}_{-0.33}$~mag and thus a distance of $3.09^{+0.25}_{-0.46}$~Mpc, where the $1\sigma$ uncertainties are as sampled from the likelihood shown in the bottom-right panel of Fig~\ref{fig:pne_4214}. If we instead also include the possible \ac{PN} sample, we get a distance modulus of $27.66^{+0.13}_{-0.26}$~mag and a distance of $3.40^{+0.20}_{-0.40}$~Mpc. Both of these distances are in agreement with the previous PNLF measurements of \citet{dopita_ngc4214_2010} ($3.19\pm0.36$~Mpc) and \citet{vicens-mouret_planetary_2023} ($3.23^{+0.18}_{-0.25}$~Mpc). We warn however that due to the nature of the \ac{PNLF}, incompleteness can easily lead to an overestimation of the distance. We have plotted where our distance measurements lie compared to a more comprehensive collection of literature measurements based on different distance indicators in Fig.~\ref{fig:distances_4214}. 
We also list the corresponding $\alpha$ parameters in Table~\ref{tab:alpha}. 

\nocite{tully_distance_1988, pierce_distance_1994, bottinelli_distance_1984, sandage_distance_1982, makarova_distance_1997, rozanski_distance_1994, dalcanton_distance_2009, mccall_distance_2012, jacobs_distance_2009, maiz-apellaniz_distance_2002, tully_distance_2013, tully_distance_2006, lelli_distances_2014, drozdovsky_distance_2002, dopita_ngc4214_2010, mcquinn_distance_2010}

\subsection{NGC~4449}

In NGC~4449, we identify $606$ sources with  reliably detected \oiii\ emission. Of these, $247$ (potentially) have \ac{PN}-like emission-line ratios and $95$ are classified as round using our automated methods. The $48$ objects that are in both categories were visually inspected, leaving $9$ bona fide and $14$ possible \acp{PN}. Even more so than for NGC4214, the majority of the contaminants which were manually removed are in the central-most area, where we cannot distinguish any point source. Some objects are round but are too large to be point sources. 

\begin{figure*}
    \captionsetup[subfigure]{labelformat=empty}
    \centering
    \subfloat[
    ]{%
        \includegraphics[width=.55\linewidth]{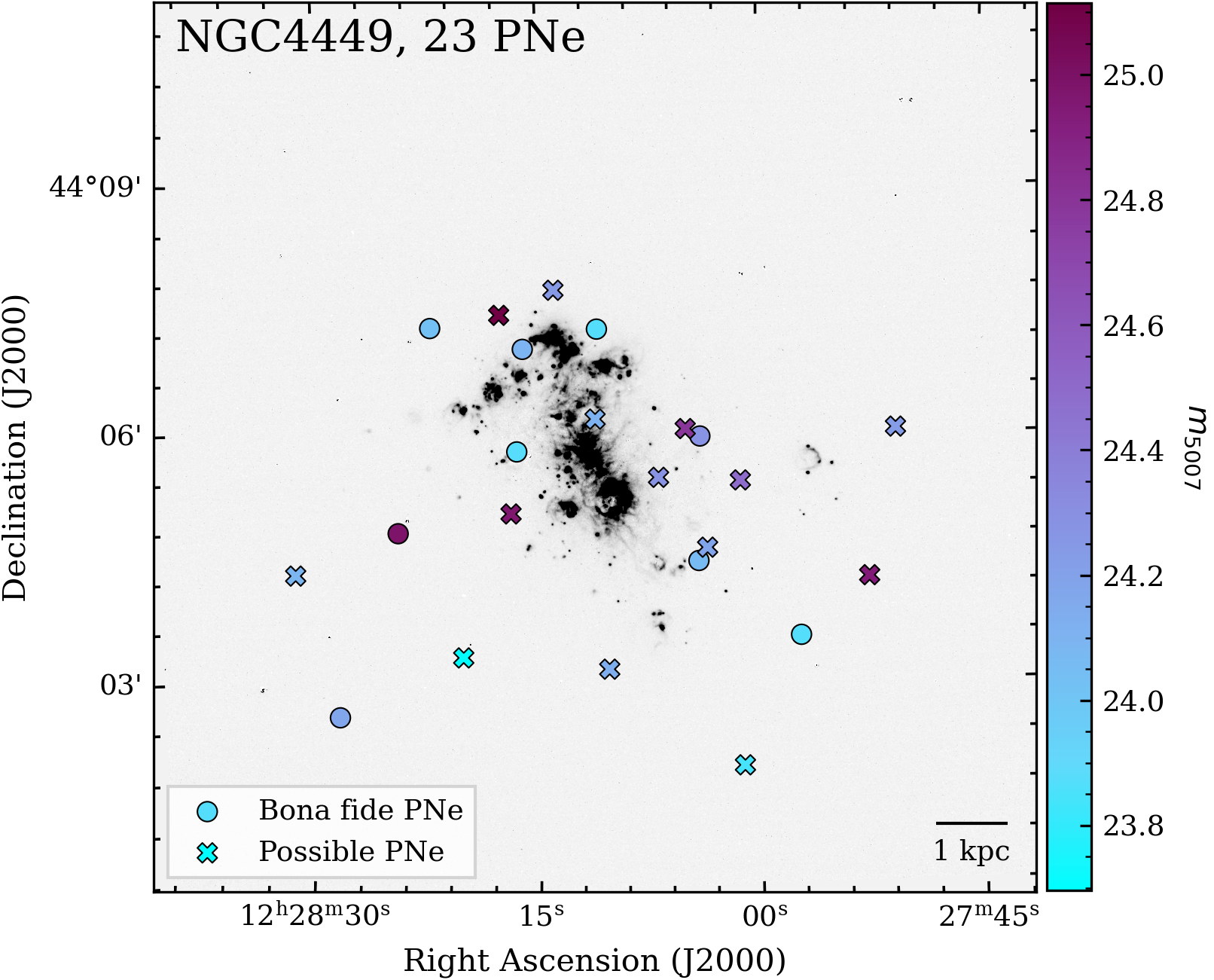}%
    }
    \subfloat[]{%
        \includegraphics[width=.45\linewidth]{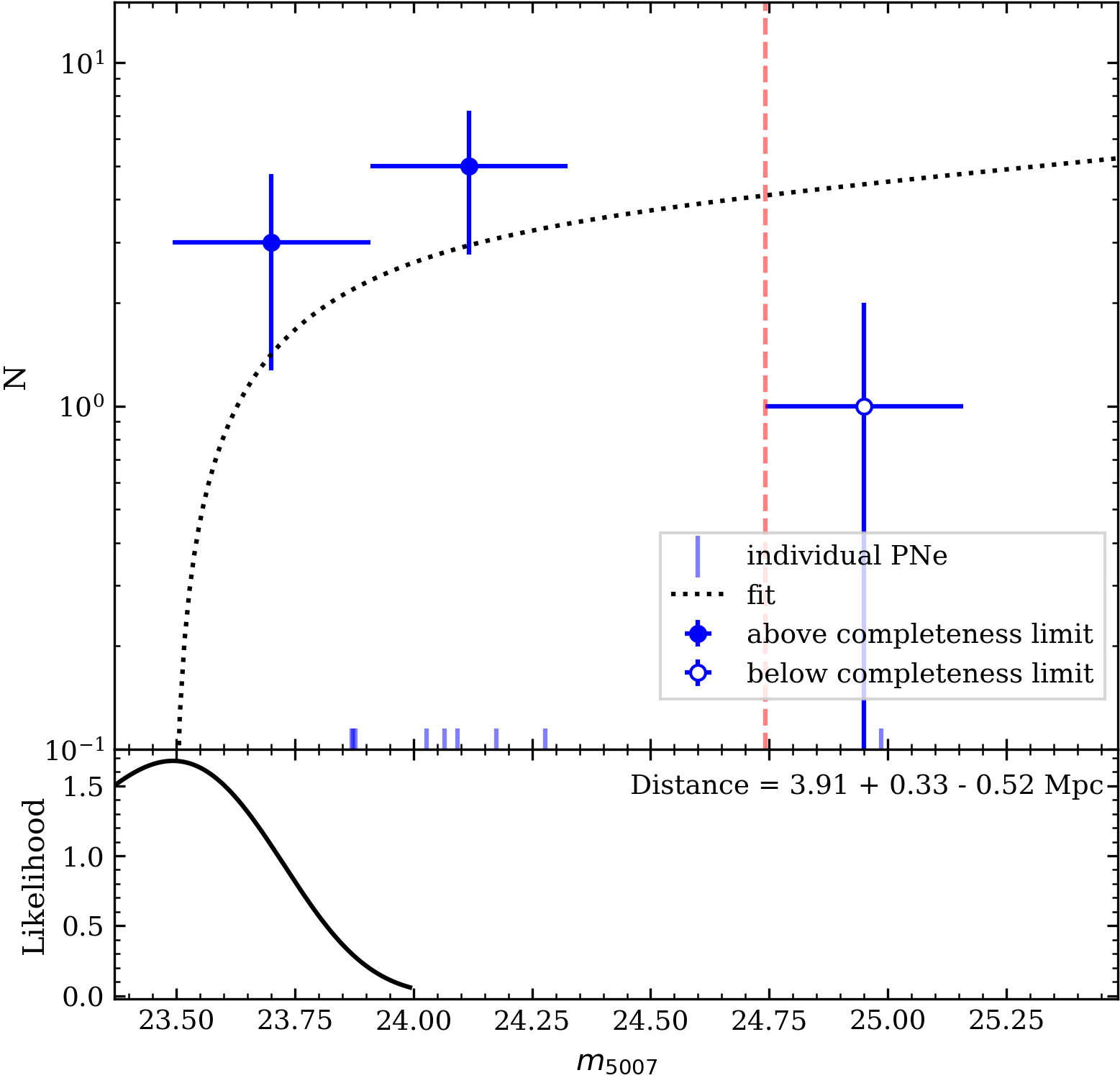}%
    }
    \caption{Same as Figure~\ref{fig:pne_4214} but for NGC~4449.}
    \label{fig:pne_4449}
\end{figure*}

The final \ac{PN} catalogue of NGC~4449 is presented in Table~\ref{tab:NGC4449}, while the \ac{PN} spatial distribution is shown in the left panel of Fig.~\ref{fig:pne_4449}. 

Using \ac{HST} data of the central region of NGC~4449, \citet{annibali_ngc4449_2017} detected $28$ \ac{PN} candidates. Of these, we detect $23$ with reliable \oiii\ emission and $10$ make it into our final catalogue of bona fide and possible PNe. 

From \ac{PNLF} fitting to the bona fide \ac{PN} sample, we derive a distance modulus of $27.96^{+0.18}_{-0.29}$~mag and thus a distance of $3.91^{+0.33}_{-0.52}$~Mpc (see the right panel of Fig.~\ref{fig:pne_4449}). Using the bona fide and possible \acp{PN}, we derive a distance modulus of $28.05^{+0.11}_{-0.19}$~mag and a distance of $4.07^{+0.21}_{-0.35}$~Mpc. Our \ac{PN} sample for NGC~4449 is only complete up to $0.5$ mag from the bright-end cut-off, and we only fit the PNLF to the \acp{PN} with a \mpn\ brighter than the completeness limit. 

\Ac{TRGB} measurements have yielded distances ranging from $3.82\pm0.27$~Mpc \citep{annibali_ngc4449_2008} to $4.30^{+0.44}_{-0.32}$~Mpc \citep{jacobs_distance_2009}. The most recent \ac{TRGB} measurement by the Legacy ExtraGalactic UV Survey \citep{sabbi_legus_2018} yields a distance of $4.01\pm0.30$~Mpc. All are consistent with our PNLF measurement. We list the corresponding $\alpha$ parameters in Table~\ref{tab:alpha}. 

\begin{figure}
    \centering
    \includegraphics[width=1\linewidth]{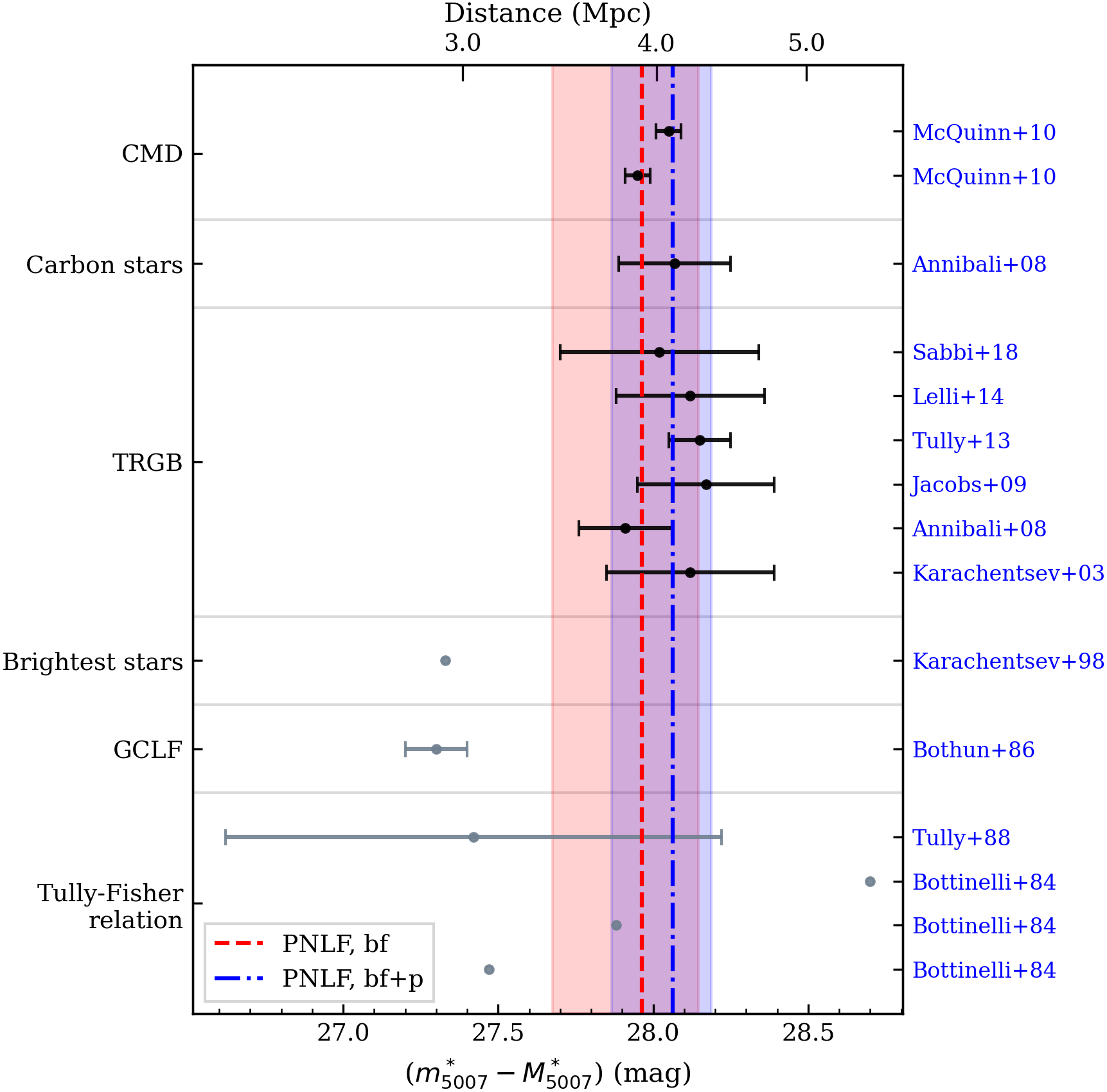}
    \caption{Same as Figure~\ref{fig:distances_4214} but for NGC~4449. GCLF: globular cluster luminosity function.}
    \label{fig:distances_4449}
\end{figure}
\nocite{bottinelli_distance_1984,karachentsev_distance_1998, jacobs_distance_2009, tully_distance_2013,karachentsev_distance_2003, lelli_distances_2014, sabbi_legus_2018, annibali_ngc4449_2008, bothun_distance_1986, mcquinn_distance_2010, tully_distance_1988}


\section{Discussion}
\label{sec:discussion}

\subsection{$\alpha$ parameter comparisons}

\begin{figure}
    \centering
    \includegraphics[width=1\linewidth]{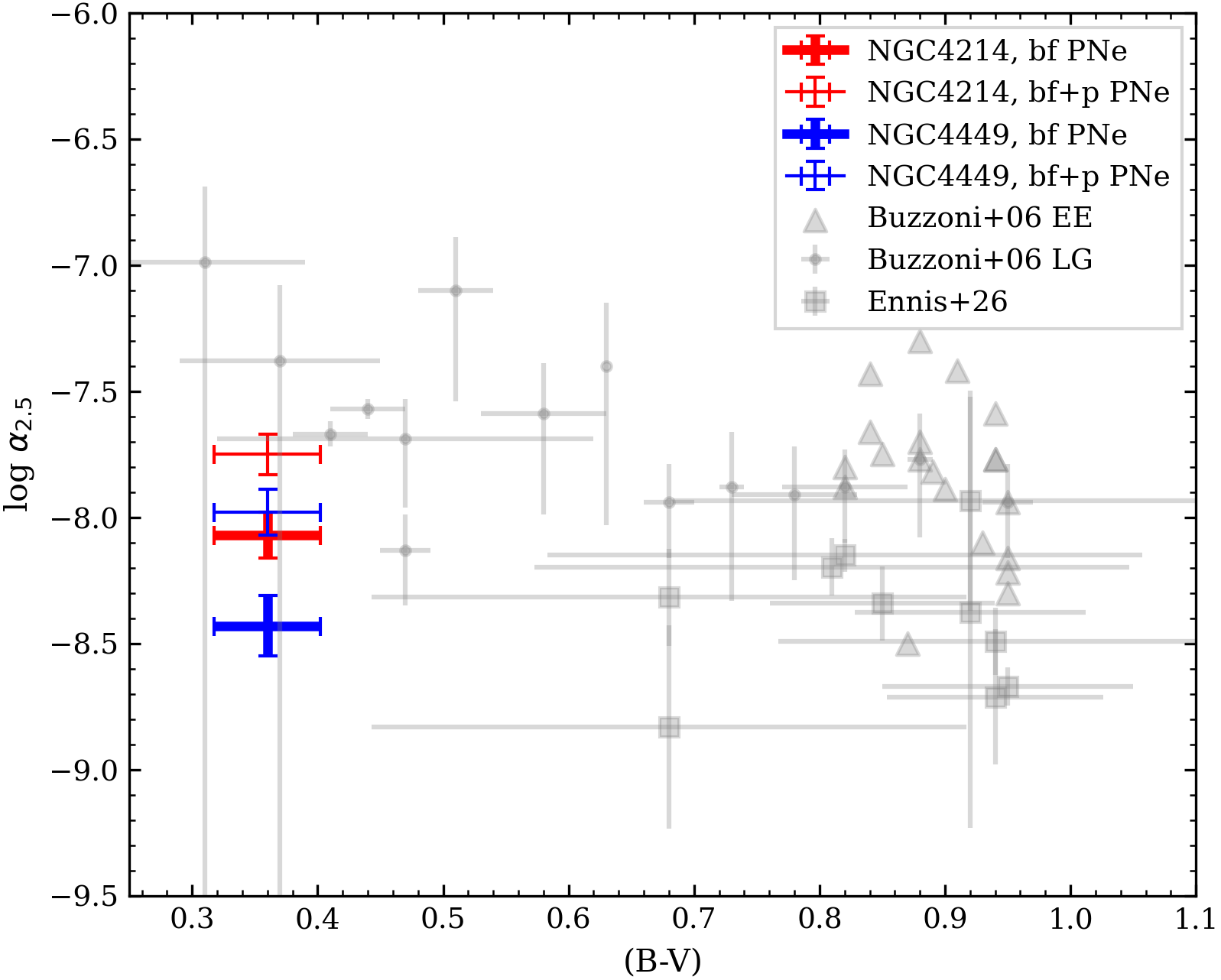}
    \caption{Comparison of the $\alpha$ parameters derived in this work to those of \citet{buzzoni_planetary_2006} and \citet{ennis_muse_2026}. The \citet{buzzoni_planetary_2006} elliptical galaxies and local group galaxies have been scaled using $\log(\alpha_{2.5})=\log(\alpha_8) -1$.}
    \label{fig:alpha_BV}
\end{figure}

We compare our $\alpha_{\mathrm{bol},2.5}$ parameters to those of \citet{buzzoni_planetary_2006} in Fig.~\ref{fig:alpha_BV}. We calculate $(B-V)$ using the magnitudes integrated within the infrared apertures of \citet{cook_spitzerphoto_2014}. The \citet{buzzoni_planetary_2006} elliptical galaxies are not as blue as NGC~4214 and NGC~4449. Compared to their \ac{LG} galaxy sample, the $\alpha$ parameters we derive using our bona fide samples are slightly smaller. This follows the theoretical expectation that bluer galaxies will have smaller $\alpha$ parameters \citep{buzzoni_planetary_2006}, but is in contrast to past observations which have shown the opposite trend \citep{peimbert_alpha_1990, ciardullo_alpha_1991, ciardullo_binaries_2005, buzzoni_planetary_2006}. However, since historically most \ac{PN} surveys were carried out in \acp{ETG}, there are not much data available for relatively blue galaxies. \ac{PN} catalogues for the remaining \ac{SIGNALS} galaxies, which are all highly star-forming, will enable us to populate Fig.~\ref{fig:alpha_BV} further and study the correlation between $\alpha$ and galaxy colour.

\subsection{Flux and \ac{PNLF} uncertainties}

The \ac{PN} \mpn\ listed in Tables~\ref{tab:NGC4214} and \ref{tab:NGC4449} were calculated using the \ac{HST}-derived calibration factors (Section~\ref{sec:flux_cal}). Additional flux uncertainties are introduced by the aperture correction, and as the correction factors were calculated using isolated undistorted objects, they may not be accurate for sources closer to the edges of the \acp{FOV}, where the \acp{PSF} are more distorted. Background subtraction is also very important for star-forming galaxies such as those from \ac{SIGNALS}. There is much contaminant emission from a variety of sources (principally \hii\ regions and \acp{SNR}), including in the \oiii\ line that can in turn be over- or under-subtracted. Additionally, while we do correct for foreground Milky Way extinction, we do not account for extinction within the host galaxy. All these factors can influence the derived \ac{PN} apparent magnitudes, in turn impacting the derived \ac{PNLF} distances.

The \ac{PNLF} distances are also affected by incompleteness. Our mock \ac{PN} catalogues have shown that we are unable to recover most of the \acp{PN} in the central regions of our galaxies (Fig.~\ref{fig:mock_recovery}). If the brightest \acp{PN} of a galaxy remain undetected because of this, the resulting fit to the \ac{PNLF} would overestimate the galaxy distance. Conversely, the presence of contaminants can lead to an underestimation of the galaxy distance, if bright compact \hii\ regions are confused for \acp{PN}.

\subsection{Velocity fitting}

As discussed in Section~\ref{sec:fitting}, the fitting software \texttt{ORCS} requires a rather precise initial guess for the input velocity of a source. The precision required increases with increased spectral resolution. While we do not expect this to have impacted the results of this paper, it could potentially be a problem for galaxies with higher rotational velocities. The way the velocity guess in SN3 is now implemented, three different guesses are compared, and the best one is selected from the reduced $\chi^2$. In galaxies with higher velocities, a larger number of guesses over a wider range of velocities may need to be compared. 

For some noisy spectra or spectra of very faint objects, even with the aforementioned three velocity guesses, a noise peak sometimes gets fitted and an incorrect flux and velocity are returned. These bad fits can be difficult to identify, as the associated uncertainties are relatively small ($S/N\approx2$). We have attempted to minimise the impact on our \ac{PN} catalogue by only quoting $v_\mathrm{H\alpha}$ for \acp{PN} which have $S/N>3$ in the \ha\ line, and using $[\ion{O}{III}]$ velocities otherwise (see Tables~\ref{tab:NGC4214} and \ref{tab:NGC4449}). For each galaxy, there are six (bona fide or possible) \acp{PN} for which this applies. One of these, NGC4214-Y6, still has an outlying $v_{[\ion{O}{III}]}$ (compared to the rest of the catalogue) and may therefore be a contaminant.

The required precision for fitting is only a challenge because we are looking at faint point-sources that have velocities distinct from that of the diffuse gas. For other science done using the \ac{SIGNALS} data, initial velocity guesses can be easily sampled from velocity maps created using diffuse emission and \hii\ regions.

We would like to repeat the caveat that the calibration of $v_{[\ion{O}{III}]}$, as described in Section~\ref{sec:SN2_vel}, relies on the assumption that the velocity offset $v_\mathrm{SN3}-v_\mathrm{SN2}$ varies smoothly across the FOV.


\section{Conclusions}
\label{sec:conclusions}

In this paper we have described in detail a new automated pipeline to detect \acp{PN} in \ac{IFS} data, particularly suited to the \ac{SIGNALS} data obtained with SITELLE. Compared to previous methods which rely on visual inspection of \oiii\ flux maps or similar \citep[e.g.][]{vicens-mouret_planetary_2023}, this pipeline mostly eliminates the need for visual inspection. This speeds up the process considerably, while also making the results more objective and replicable.

We applied our new pipeline to the star-forming galaxies NGC~4214 and NGC~4449. In NGC~4214, we recovered $19$ known \acp{PN} and discovered $6$ (possible) new ones. We derived a PNLF distance of $3.09^{+0.25}_{-0.46}$~Mpc, in agreement with previous distance measurements. In NGC~4449, we confirmed $10$ previously reported \acp{PN} and discovered $13$ (possible) new ones. We derived a PNLF distance of $3.91^{+0.33}_{-0.52}$ Mpc, also in agreement with previous distance measurements. 

As became apparent in the tests we ran using mock \acp{PN} (Fig.~\ref{fig:mock_recovery}), detecting \acp{PN} superposed or close to bright star-forming regions is challenging even with the \ac{SIGNALS} IFS data. This is reflected in the spatial distributions of the recovered \acp{PN}, as shown in the left panels of Figs.~\ref{fig:pne_4214} and \ref{fig:pne_4449}. However, because the mock \ac{PN} samples allow us to robustly quantify the recovery rates, we can correct for this when calculating the \ac{PN} specific frequencies ($\alpha$ parameters). Additionally, SITELLE's wide \ac{FOV} allows us to detect \acp{PN} to greater galactocentric distances, where contaminants are also less prominent. Of course, the \ac{SIGNALS} data also yield the \ac{PN} emission-line ratios, allowing to reject compact contaminants, and to measure velocities. It is therefore highly complementary to narrow-band \ac{HST} \ac{PN} surveys.

Although our \ac{PN} detection pipeline is meant to be generic, small changes may be necessary to apply it to the other \ac{SIGNALS} galaxies, to optimally handle different galaxy morphologies. For example, even though this functionality already exists in the current version of the pipeline, the initial velocity guesses required for the emission-line fits may have to be sampled from wider velocity ranges. Galaxies at larger distances may also prove more challenging, as larger fractions of the contaminants will appear spatially unresolved and only smaller (i.e.\ brighter) portions of the \acp{PNLF} will be probed, resulting in proportionally fewer \acp{PN} detected. This will however be partially compensated for brighter galaxies, as they inherently harbour more \acp{PN}.

Overall, the new method developed in this paper is bound to lead to hundreds of new \ac{PN} discoveries in a range of star-forming galaxies, most of which have never been surveyed for \acp{PN} before.


\section*{Acknowledgements}
J.H.\ and C.S.\ acknowledge the financial support from the Visitor and Mobility program of the Finnish Centre for Astronomy with ESO (FINCA). J.H.\ also acknowledges financial support from the Turku Collegium for Science, Medicine, and Technology (TCSMT) in the form of a starting grant.

This research was based on observations obtained at the CFHT, which is operated from the summit of Mauna Kea by the National Research Council of Canada, the Institut National des Sciences de l’Univers of the Centre National de la Recherche Scientifique of France, and the University of Hawaii. The authors wish to recognize and acknowledge the very significant cultural role that the summit of Mauna Kea has always had within the indigenous Hawaiian community. The authors are most grateful to have the opportunity to conduct observations from this mountain. The observations were obtained with SITELLE, a joint project between Université Laval, ABB-Bomem, Université de Montréal, and the CFHT, with funding support from the Canada Foundation for Innovation (CFI), the National Sciences and Engineering Research Council of Canada (NSERC), Fonds de Recherche du Québec – Nature et Technologies (FRQNT), and CFHT.

The collaboration is grateful to the FRQNT, CFHT, the Canada Research Chair program, the National Science Foundation NSF – 2109124, Natural Sciences and Engineering Research Council of Canada NSERC – RGPIN-2023-03487, the Swedish Research Council, the Swedish National Space Board, the Royal Society, and the Newton Fund via the award of a Royal Society-Newton Advanced Fellowship (NAF\textbackslash R1\textbackslash 180403), FAPESC, CNPq, FAPESP (2014/11156-4), FAPESB (7916/2015), and CONACyT (CB2015-254132).


\section*{Data Availability}
The data used in this work are publicly available, along with all other SIGNALS data. 


\bibliographystyle{mnras}
\bibliography{refs} 

\bsp	
\label{lastpage}

\onecolumn
\appendix

\section{Decision tree for BPT-diagrams} \label{app:tree}
\begin{figure*}
    \centering
    \includegraphics[width=1\linewidth]{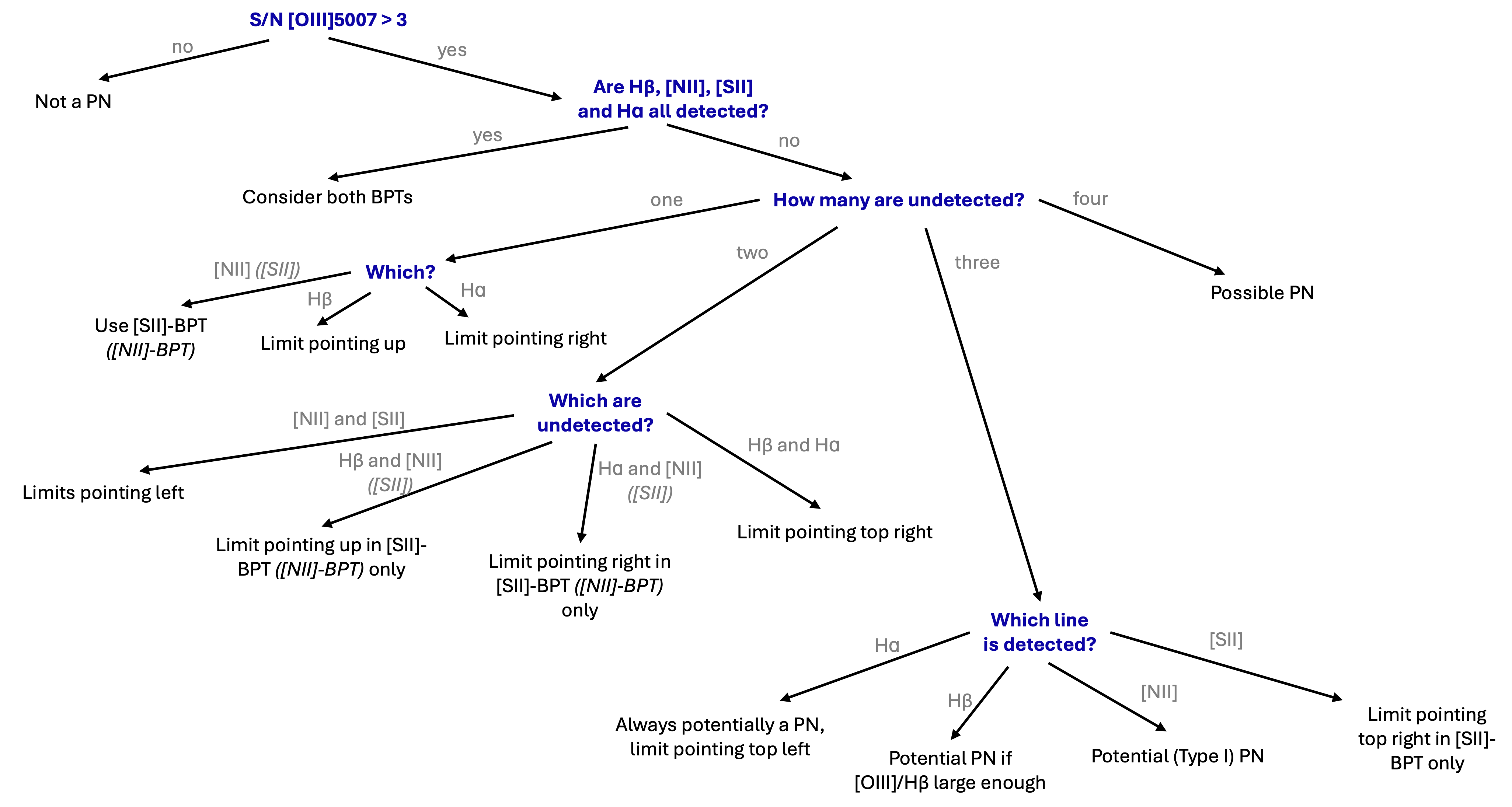}
    \caption{Decision tree showing what combination of diagnostic diagram and limits are used to classify a source as a \ac{PN} candidate or contaminant. Unless stated otherwise, both the \nbpt\ and the \sbpt\ diagram are considered. 'Limit pointing up' would refer to a lower limit on the \oiii/\hb emission line ratio.
    }
    \label{fig:tree}
\end{figure*}

Fig.~\ref{fig:tree} shows the decision tree used to determine which BPT diagrams should be considered for any given source, depending on its combination of detected and undetected emission lines.

\begin{landscape}
\section{PN tables}
Tables~\ref{tab:NGC4214} and~\ref{tab:NGC4449} are our full \ac{PN} catalogues, for both bona fide and possible \acp{PN}. $v_\mathrm{H\alpha}$ is given only for PNe with $S/N>3$ in the \ha\ line; we quote $v_{[\ion{O}{III}]}$ otherwise.
\input{tables/NGC4214_PNe_table}

\input{tables/NGC4449_PNe_table}
\end{landscape}

\end{document}

%% file: tables/obs.tex
\begin{table*}
\caption{Overview of targets and observations.}
\label{tab:obs}
\begin{tabular}{l|ccccccccl}
\toprule
Galaxy & Type & $D$ & $Z$ & Filter & Band & $R$ & Time/step & Steps & Date \\
& & (Mpc) & & & (nm) & & (s) & & \\
\hline
\multirow{2}{*}{NGC~4214} & \multirow{2}{*}{IAB(s)m} & \multirow{2}{*}{$2.93$} & \multirow{2}{*}{$8.20 \pm 0.05$} & SN2 & $482$ -- $513$ & $\phantom{1}943$ & $50\phantom{.5}$ & $219$ & May 2018 \\
     & & & & SN3 & $647$ -- $685$ & $1895$ & $30\phantom{.5}$ & $337$ & May 2018 \\
\hline
\multirow{2}{*}{NGC~4449} & \multirow{2}{*}{IBm} & \multirow{2}{*}{$4.01$} & \multirow{2}{*}{$8.26\pm 0.01$} & SN2 & $482$ -- $513$ & $\phantom{1}943$ & $45.5$ & $219$ & Feb 2020 \\
   & & & & SN3 & $647$ -- $685$ & $4741$ & $13\phantom{.5}$ & $842$ & Feb 2020  \\
   \bottomrule
\end{tabular} \\
Notes: Columns: (1) Galaxy. (2) Morphological type. (3) Uncertainty-weighted mean distance of a selection of distances from the NASA/IPAC Extragalactic Database. (4) Gas-phase metallicity ($12+\log(\mathrm{O}/\mathrm{H})$) from \citet{pilyugin_metallicity_2015}. (5) SITELLE filter. (6) SITELLE band. (7) Spectral resolution. (8) Integration time per scanning mirror step. (9) Number of scanning mirror steps. (10) Observation date.
\end{table*}

%% file: tables/alpha.tex
\begin{table*}
\caption{Number of \acp{PN} and $\alpha$ parameters calculated for different ranges of $m_{5007}$ from the bright-end cut-off. The parameter $\alpha_\mathrm{bol}$ is calculated using the bolometric correction factor from \citet{buzzoni_planetary_2006}. The parameter $\alpha_V$ is calculated using the $V$-band luminosity. The column headers $0.5$ and $2.5$ refer to the magnitude range from the best-fitting bright-end cut-off. The 'bf' sample contains only bona fide PNe, whereas the 'bf+p' sample also contains the possible PNe.}
\begin{tabular}{@{}l|ll|cc|l|cc|l|cc@{}}
\toprule
\multicolumn{1}{l|}{\multirow{2}{*}{Galaxy}} & \multirow{2}{*}{Sample} &  & \multicolumn{2}{c|}{$N$}   &  & \multicolumn{2}{c|}{$\log(\alpha_\mathrm{bol})$} &  & \multicolumn{2}{c}{$\log(\alpha_V)$} \\ \cmidrule(lr){4-5} \cmidrule(lr){7-8} \cmidrule(l){10-11} 
\multicolumn{1}{l|}{}                        &                         &  & $0.5$        & $2.5$        &  & $0.5$          & $2.5$          &  & $0.5$          & $2.5$   \\ \midrule
\multirow{2}{*}{NGC4214}                     & bf                      &  & $1.2\pm0.5$  & $14.8\pm2.4$ &  & $-9.16\pm0.19$ & $-8.07\pm0.09$ &  & $-8.85\pm0.18$ & $-7.75\pm0.08$ \\
                                             & bf+p                    &  & $1.2\pm0.5$  & $30.5\pm4.0$ &  & $-9.16\pm0.19$ & $-7.75\pm0.08$ &  & $-8.85\pm0.18$ & $-7.44\pm0.07$ \\ \midrule
\multirow{2}{*}{NGC4449}                     & bf                      &  & $4.3\pm1.4$ & $15.7\pm3.8$ &  & $-8.99\pm0.15$ & $-8.43\pm0.12$ &  & $-8.68\pm0.14$ & $-8.12\pm0.11$ \\
                                             & bf+p                    &  & $7.2\pm1.8$ & $43.7\pm7.1$ &  & $-8.77\pm0.12$ & $-7.98\pm0.09$ &  & $-8.46\pm0.12$ & $-7.67\pm0.08$ \\ \bottomrule
\end{tabular}

\label{tab:alpha}
\end{table*}

%% file: tables/NGC4214_PNe_table.tex
\begin{table*}
\centering
\caption{PN catalogue of NGC~4214}
\begin{tabular}{@{}lccrrrrrrrrccr@{}}
\toprule
ID & \multicolumn{1}{c}{RA (J2000)} & \multicolumn{1}{c}{Dec (J2000)} & \multicolumn{1}{c}{$m_{5007}$} & \multicolumn{1}{c}{$v_\mathrm{H\alpha}$} & \multicolumn{1}{c}{$v_{[\ion{O}{III}]}$} & \multicolumn{1}{c}{$F([\ion{O}{III}])$} & \multicolumn{1}{c}{$F\mathrm{(H\beta)}$} & \multicolumn{1}{c}{$F\mathrm{(H\alpha)}$} & \multicolumn{1}{c}{$F([\ion{N}{II}])$} & \multicolumn{1}{c}{$F([\ion{S}{II}])$} & [\ion{N}{II}]-BPT & [\ion{S}{II}]-BPT & Final\\
\cline{7-11}
\\
\multicolumn{1}{c}{} & \multicolumn{1}{c}{(hh:mm:ss)} & \multicolumn{1}{c}{(deg:arcmin:arcsec)} & \multicolumn{1}{c}{(mag)} & \multicolumn{1}{c}{(km s$^{-1}$)}  & \multicolumn{1}{c}{(km s$^{-1}$)}   & \multicolumn{5}{c}{($10^{-17}$ erg s$^{-1}$ cm$^{-2}$)} & & & \\

(1) & \multicolumn{1}{c}{(2)} & \multicolumn{1}{c}{(3)} & \multicolumn{1}{c}{(4)} & \multicolumn{1}{c}{(5)} & \multicolumn{1}{c}{(6)} & \multicolumn{1}{c}{(7)} & \multicolumn{1}{c}{(8)} & \multicolumn{1}{c}{(9)} & \multicolumn{1}{c}{(10)} & (11) & (12) & (13)\\
\midrule
VDR2 & 12:15:19.609 & 36:21:27.84 & 24.3 $\pm$ 0.1 &  & 280 $\pm$ 10 & 60 $\pm$\phantom{ 1}8 & 5 $\pm$ 3 & 22 $\pm$ 5 & 4 $\pm$ 3 & < 4 & l &  & b.f. \\ 
VDR3 & 12:15:21.403 & 36:20:34.12 & 25.5 $\pm$ 0.2 &  & 220 $\pm$ 10 & 20 $\pm$\phantom{ 1}4 & 5 $\pm$ 3 & 11 $\pm$ 2 & 4 $\pm$ 2 & < 3 & p &  & p \\ 
VDR5 & 12:15:23.947 & 36:21:28.42 & 25.8 $\pm$ 0.2 & 300 $\pm$ 20 &  & 15 $\pm$\phantom{ 1}3 & 5 $\pm$ 2 & 6 $\pm$ 3 & 9 $\pm$ 4 & < 4 & p &  & p \\ 
VDR6 & 12:15:31.960 & 36:21:10.42 & 25.2 $\pm$ 0.2 &  & 260 $\pm$ 10 & 28 $\pm$\phantom{ 1}5 & 5 $\pm$ 2 & 12 $\pm$ 4 & 7 $\pm$ 3 & < 4 & p &  & p \\ 
VDR7 & 12:15:32.297 & 36:15:27.82 & 24.9 $\pm$ 0.2 &  & 260 $\pm$ 20 & 34 $\pm$\phantom{ 1}6 & 8 $\pm$ 4 & 20 $\pm$ 6 & < 4 & < 5 & p & p & p \\ 
VDR8 & 12:15:35.551 & 36:18:00.19 & 25.0 $\pm$ 0.2 &  & 270 $\pm$ 10 & 33 $\pm$\phantom{ 1}5 & 8 $\pm$ 3 & 17 $\pm$ 4 & 5 $\pm$ 3 & < 4 & p &  & p \\ 
VDR9 & 12:15:37.487 & 36:15:30.52 & 25.4 $\pm$ 0.2 & 240 $\pm$ 20 &  & 22 $\pm$\phantom{ 1}5 & 7 $\pm$ 3 & 11 $\pm$ 5 & 3 $\pm$ 3 & < 4 & p &  & p \\ 
VDR11 & 12:15:38.629 & 36:21:18.35 & 23.6 $\pm$ 0.1 &  & 271 $\pm$\phantom{ 1}7 & 120 $\pm$ 10 & 7 $\pm$ 3 & 33 $\pm$ 5 & 4 $\pm$ 2 & < 4 & l &  & b.f. \\ 
VDR12 & 12:15:42.108 & 36:20:52.23 & 24.9 $\pm$ 0.2 & 230 $\pm$ 20 &  & 36 $\pm$\phantom{ 1}6 & < 3 & 4 $\pm$ 2 & < 2 & < 4 & p & p & p \\ 
VDR13 & 12:15:43.710 & 36:20:26.07 & 24.9 $\pm$ 0.2 &  & 330 $\pm$ 20 & 34 $\pm$\phantom{ 1}6 & < 3 & 10 $\pm$ 3 & < 2 & < 3 & l & l & b.f. \\ 
VDR15 & 12:15:48.130 & 36:22:47.15 & 25.1 $\pm$ 0.2 &  & 326 $\pm$\phantom{ 1}9 & 28 $\pm$\phantom{ 1}5 & 7 $\pm$ 3 & 20 $\pm$ 3 & 5 $\pm$ 2 & 4 $\pm$ 3 & l & p & b.f. \\ 
D1 & 12:15:34.473 & 36:19:36.88 & 24.8 $\pm$ 0.2 & 330 $\pm$ 20 &  & 40 $\pm$\phantom{ 1}6 & < 3 & 8 $\pm$ 4 & < 4 & < 6 & p & p & p \\ 
D3 & 12:15:35.310 & 36:19:14.17 & 24.0 $\pm$ 0.1 & 280 $\pm$ 10 &  & 77 $\pm$ 10 & 6 $\pm$ 3 & 11 $\pm$ 4 & < 3 & < 4 & l & l & b.f. \\ 
D4 & 12:15:36.656 & 36:19:15.05 & 24.0 $\pm$ 0.1 &  & 307 $\pm$\phantom{ 1}9 & 80 $\pm$ 10 & 8 $\pm$ 4 & 28 $\pm$ 5 & 6 $\pm$ 3 & 8 $\pm$ 5 & l & l & b.f. \\ 
D5 & 12:15:36.925 & 36:19:54.82 & 24.0 $\pm$ 0.1 &  & 240 $\pm$ 20 & 80 $\pm$ 10 & < 5 & 26 $\pm$ 8 & 9 $\pm$ 6 & 12 $\pm$ 8 & l & l & b.f. \\ 
D6 & 12:15:37.510 & 36:20:02.05 & 24.1 $\pm$ 0.2 &  & 310 $\pm$ 10 & 70 $\pm$ 10 & < 6 & 47 $\pm$ 10 & 7 $\pm$ 5 & 9 $\pm$ 8 & l & l & b.f. \\ 
D7 & 12:15:38.489 & 36:20:17.83 & 23.3 $\pm$ 0.1 &  & 304 $\pm$\phantom{ 1}7 & 150 $\pm$ 20 & 6 $\pm$ 4 & 60 $\pm$ 9 & 4 $\pm$ 4 & < 5 & l &  & b.f. \\ 
D14 & 12:15:41.002 & 36:20:17.26 & 24.2 $\pm$ 0.1 &  & 291 $\pm$\phantom{ 1}9 & 64 $\pm$\phantom{ 1}8 & 4 $\pm$ 4 & 29 $\pm$ 5 & < 3 & 7 $\pm$ 4 &  & l & b.f. \\ 
D17 & 12:15:46.246 & 36:19:01.22 & 25.7 $\pm$ 0.2 &  & 270 $\pm$ 20 & 16 $\pm$\phantom{ 1}3 & 3 $\pm$ 3 & 14 $\pm$ 4 & < 2 & < 4 & p & p & p \\ 
Y1 & 12:15:41.716 & 36:18:39.97 & 24.0 $\pm$ 0.1 &  & 292 $\pm$\phantom{ 1}6 & 79 $\pm$ 10 & 24 $\pm$ 4 & 55 $\pm$ 7 & < 3 & 17 $\pm$ 5 &  & p & p \\ 
Y2 & 12:15:45.665 & 36:19:36.31 & 24.3 $\pm$ 0.1 &  & 325 $\pm$\phantom{ 1}6 & 60 $\pm$\phantom{ 1}8 & 18 $\pm$ 4 & 63 $\pm$ 8 & < 3 & 18 $\pm$ 4 &  & p & p \\ 
Y3 & 12:15:37.327 & 36:20:39.80 & 24.6 $\pm$ 0.2 &  & 315 $\pm$\phantom{ 1}7 & 44 $\pm$\phantom{ 1}7 & 16 $\pm$ 4 & 45 $\pm$ 7 & < 3 & 14 $\pm$ 5 &  & p & p \\ 
Y4 & 12:15:28.564 & 36:21:55.64 & 25.3 $\pm$ 0.2 &  & 264 $\pm$\phantom{ 1}6 & 24 $\pm$\phantom{ 1}4 & 8 $\pm$ 2 & 53 $\pm$ 7 & 3 $\pm$ 3 & 16 $\pm$ 4 & c & l & b.f. \\ 
Y5 & 12:15:38.372 & 36:22:32.39 & 25.4 $\pm$ 0.2 &  & 300 $\pm$ 10 & 22 $\pm$\phantom{ 1}4 & 10 $\pm$ 3 & 28 $\pm$ 6 & 5 $\pm$ 3 & 14 $\pm$ 5 & p & p & p \\ 
Y6 & 12:15:30.469 & 36:16:19.29 & 26.6 $\pm$ 0.3 & 30 $\pm$ 40 &  & 7 $\pm$\phantom{ 1}2 & < 2 & 4 $\pm$ 2 & 3 $\pm$ 2 & < 3 & p &  & p \\ 

\bottomrule
\end{tabular} \\
Notes: Flux upper limits are listed for non-detections. $1\sigma$ uncertainties are quoted to $1$ significant figure. 
Columns: (1) \ac{PN} ID. (2) -- (3) Coordinates in the FK5 reference frame. (4) \fulloiii\ apparent magnitude. 
(5) -- (6) Velocity as measured from $v_{H\alpha}$ or $v_{[\ion{O}{III}]}$, respectively. (7) -- (11) Emission-line flux. (12) Likely (l) or possible (p) classification based on the \nbpt\ only. 
(13) Likely (l) or possible (p) classification based on the \sbpt\ only. (14) Final classification of bone fide (b.f.) and possible (p) \acp{PN}. A machine readable version of this table has been provided in the online supplementary material.

\label{tab:NGC4214}
\end{table*}

%% file: tables/NGC4449_PNe_table.tex
\begin{table*}
\centering
\caption{PN catalogue of NGC~4449}
\begin{tabular}{@{}lccrrrrrrrrccr@{}}
\toprule
ID & \multicolumn{1}{c}{RA (J2000)} & \multicolumn{1}{c}{Dec (J2000)} & \multicolumn{1}{c}{$m_{5007}$} & \multicolumn{1}{c}{$v_\mathrm{H\alpha}$} & \multicolumn{1}{c}{$v_{[\ion{O}{III}]}$} & \multicolumn{1}{c}{$F([\ion{O}{III}])$} & \multicolumn{1}{c}{$F\mathrm{(H\beta)}$} & \multicolumn{1}{c}{$F\mathrm{(H\alpha)}$} & \multicolumn{1}{c}{$F([\ion{N}{II}])$} & \multicolumn{1}{c}{$F([\ion{S}{II}])$} & [\ion{N}{II}]-BPT & [\ion{S}{II}]-BPT & Final\\
\cline{7-11}
\\
\multicolumn{1}{c}{} & \multicolumn{1}{c}{(hh:mm:ss)} & \multicolumn{1}{c}{(deg:arcmin:arcsec)} & \multicolumn{1}{c}{(mag)} & \multicolumn{1}{c}{(km s$^{-1}$)}  & \multicolumn{1}{c}{(km s$^{-1}$)}   & \multicolumn{5}{c}{($10^{-17}$ erg s$^{-1}$ cm$^{-2}$)} & & & \\

(1) & \multicolumn{1}{c}{(2)} & \multicolumn{1}{c}{(3)} & \multicolumn{1}{c}{(4)} & \multicolumn{1}{c}{(5)} & \multicolumn{1}{c}{(6)} & \multicolumn{1}{c}{(7)} & \multicolumn{1}{c}{(8)} & \multicolumn{1}{c}{(9)} & \multicolumn{1}{c}{(10)} & (11) & (12) & (13)\\
\midrule
A1 & 12:28:04.145 & 44:04:24.70 & 24.1 $\pm$ 0.3 &  & 229 $\pm$ 6 & 80 $\pm$ 20 & 7 $\pm$ 5 & 19 $\pm$\phantom{ 1}5 & < 3 & < 4 & l & l & b.f. \\ 
A2 & 12:28:03.541 & 44:04:34.45 & 24.2 $\pm$ 0.3 & 200 $\pm$ 10 &  & 70 $\pm$ 20 & 15 $\pm$ 6 & 30 $\pm$ 10 & < 5 & < 8 & p & p & p \\ 
A3 & 12:28:03.984 & 44:05:56.47 & 24.3 $\pm$ 0.3 &  & 183 $\pm$ 5 & 60 $\pm$ 20 & < 5 & 22 $\pm$\phantom{ 1}6 & 7 $\pm$ 3 & < 4 & l &  & b.f. \\ 
A5 & 12:28:13.962 & 44:07:44.75 & 24.2 $\pm$ 0.3 &  & 197 $\pm$ 5 & 60 $\pm$ 20 & 8 $\pm$ 4 & 19 $\pm$\phantom{ 1}5 & < 3 & 6 $\pm$ 5 &  & p & p \\ 
A10 & 12:28:16.995 & 44:05:00.13 & 25.0 $\pm$ 0.3 & 310 $\pm$ 20 &  & 33 $\pm$ 10 & < 4 & 11 $\pm$\phantom{ 1}4 & < 3 & < 4 & p & p & p \\ 
A12 & 12:28:01.244 & 44:05:23.80 & 24.5 $\pm$ 0.3 & 240 $\pm$ 10 &  & 50 $\pm$ 10 & 8 $\pm$ 4 & 16 $\pm$\phantom{ 1}7 & < 4 & < 6 & p & p & p \\ 
A14 & 12:28:06.851 & 44:05:26.26 & 24.3 $\pm$ 0.3 & 250 $\pm$ 20 &  & 60 $\pm$ 20 & < 7 & 15 $\pm$\phantom{ 1}9 & < 5 & 9 $\pm$ 9 &  & p & p \\ 
A16 & 12:28:16.566 & 44:05:45.87 & 23.9 $\pm$ 0.3 &  & 179 $\pm$ 5 & 90 $\pm$ 20 & 32 $\pm$ 10 & 80 $\pm$ 20 & 25 $\pm$ 8 & 17 $\pm$ 9 & l & p & b.f. \\ 
A24 & 12:28:11.166 & 44:06:09.62 & 24.1 $\pm$ 0.3 & 140 $\pm$ 30 &  & 70 $\pm$ 20 & < 9 & 18 $\pm$\phantom{ 1}7 & < 6 & < 9 & p & p & p \\ 
A28 & 12:28:16.112 & 44:07:01.45 & 24.1 $\pm$ 0.3 & 200 $\pm$ 40 &  & 70 $\pm$ 20 & < 10 & 12 $\pm$\phantom{ 1}4 & < 4 & 7 $\pm$ 5 &  & l & b.f. \\ 
Y1 & 12:28:20.277 & 44:03:14.77 & 23.7 $\pm$ 0.2 &  & 209 $\pm$ 6 & 110 $\pm$ 20 & 13 $\pm$ 6 & 28 $\pm$\phantom{ 1}8 & < 4 & < 5 & p & p & p \\ 
Y2 & 12:28:01.165 & 44:01:55.14 & 23.8 $\pm$ 0.2 &  & 221 $\pm$ 4 & 90 $\pm$ 20 & 19 $\pm$ 6 & 70 $\pm$ 20 & < 4 & 12 $\pm$ 6 &  & p & p \\ 
Y3 & 12:28:11.009 & 44:07:15.94 & 23.9 $\pm$ 0.3 &  & 216 $\pm$ 4 & 90 $\pm$ 20 & 29 $\pm$ 8 & 60 $\pm$ 10 & 16 $\pm$ 5 & 25 $\pm$ 6 & l & l & b.f. \\ 
Y4 & 12:27:57.214 & 44:03:29.86 & 23.9 $\pm$ 0.2 &  & 207 $\pm$ 6 & 90 $\pm$ 20 & 5 $\pm$ 4 & 24 $\pm$\phantom{ 1}7 & < 4 & < 5 & l & l & b.f. \\ 
Y5 & 12:28:22.422 & 44:07:17.18 & 24.0 $\pm$ 0.3 & 210 $\pm$ 10 &  & 80 $\pm$ 20 & < 5 & 24 $\pm$\phantom{ 1}8 & < 4 & < 6 & l & l & b.f. \\ 
Y6 & 12:28:31.603 & 44:04:16.53 & 24.1 $\pm$ 0.3 &  & 189 $\pm$ 5 & 70 $\pm$ 20 & 13 $\pm$ 5 & 19 $\pm$\phantom{ 1}5 & < 3 & 6 $\pm$ 4 &  & p & p \\ 
Y7 & 12:28:10.330 & 44:03:05.46 & 24.1 $\pm$ 0.3 &  & 149 $\pm$ 6 & 70 $\pm$ 20 & 7 $\pm$ 4 & 21 $\pm$\phantom{ 1}6 & 9 $\pm$ 4 & < 5 & p &  & p \\ 
Y8 & 12:28:28.562 & 44:02:32.75 & 24.2 $\pm$ 0.3 &  & 201 $\pm$ 6 & 70 $\pm$ 20 & < 4 & 13 $\pm$\phantom{ 1}4 & 6 $\pm$ 3 & 4 $\pm$ 4 & l & l & b.f. \\ 
Y9 & 12:27:50.645 & 44:06:02.11 & 24.2 $\pm$ 0.3 & 210 $\pm$ 10 &  & 60 $\pm$ 20 & < 4 & 14 $\pm$\phantom{ 1}5 & 10 $\pm$ 4 & < 4 & p &  & p \\ 
Y10 & 12:28:04.985 & 44:06:02.06 & 24.8 $\pm$ 0.3 &  & 229 $\pm$ 6 & 40 $\pm$ 10 & 8 $\pm$ 5 & 16 $\pm$\phantom{ 1}5 & < 3 & < 5 & p & p & p \\ 
Y11 & 12:27:52.526 & 44:04:13.11 & 24.9 $\pm$ 0.3 & 250 $\pm$ 20 &  & 34 $\pm$\phantom{ 1}8 & 6 $\pm$ 4 & 10 $\pm$\phantom{ 1}3 & < 3 & < 4 & p & p & p \\ 
Y12 & 12:28:24.700 & 44:04:46.42 & 25.0 $\pm$ 0.3 & 160 $\pm$ 20 &  & 32 $\pm$\phantom{ 1}8 & < 3 & 15 $\pm$\phantom{ 1}6 & < 4 & 6 $\pm$ 6 &  & l & b.f. \\ 
Y13 & 12:28:17.701 & 44:07:26.59 & 25.1 $\pm$ 0.3 &  & 199 $\pm$ 5 & 29 $\pm$\phantom{ 1}8 & 7 $\pm$ 4 & 18 $\pm$\phantom{ 1}5 & < 3 & 11 $\pm$ 4 &  & p & p \\ 

\bottomrule
\end{tabular} \\
Notes: Flux upper limits are listed for non-detections. $1\sigma$ uncertainties are quoted to $1$ significant figure. 
Columns: (1) \ac{PN} ID. (2) -- (3) Coordinates in the FK5 reference frame. (4) \fulloiii\ apparent magnitude. 
(5) -- (6) Velocity as measured from $v_{H\alpha}$ or $v_{[\ion{O}{III}]}$, respectively. (7) -- (11) Emission-line flux. (12) Likely (l) or possible (p) classification based on the \nbpt\ only. 
(13) Likely (l) or possible (p) classification based on the \sbpt\ only. (14) Final classification of bone fide (b.f.) and possible (p) \acp{PN}. A machine readable version of this table has been provided in the online supplementary material.

\label{tab:NGC4449}
\end{table*}